\documentclass[paper=A4,11pt,DIV=12,headings=big]{scrartcl}
\pdfoutput=1
\usepackage{graphicx}
\usepackage{tabularx}
\usepackage{array}
\usepackage{amsmath}
\usepackage{amssymb}
\usepackage{xcolor}
\usepackage{longtable}
\usepackage{verbatim}
\usepackage{cite}
\usepackage{bbm}
\usepackage{multirow}
\usepackage{colortbl}
\usepackage[normalem]{ulem}
\definecolor{darkblue}{rgb}{0,0,0.5}
\usepackage[colorlinks,linkcolor=darkblue,citecolor=darkblue,urlcolor=darkblue,%
pdftitle={State of new physics in b->s transitions},%
pdfauthor={Wolfgang Altmannshofer, David M. Straub}]{hyperref}

\allowdisplaybreaks

\addtokomafont{disposition}{\rmfamily\boldmath}
\DeclareMathAlphabet{\mathbfit}{OML}{cmm}{b}{it}

\newcommand{\mailref}[1]{\href{mailto:#1}{#1}}

\newcommand{\beq}{\begin{equation}}
\newcommand{\eeq}{\end{equation}}
\newcommand{\ba}{\begin{array}}
\newcommand{\ea}{\end{array}} 
\newcommand{\beqa}{\begin{eqnarray}}
\newcommand{\eeqa}{\end{eqnarray}}

\newcommand{\eff}{\text{eff}}

\begin{document}
%-------------------------------------------------------------------

\vskip0.5cm

\begin{center}
{\LARGE
\textbf{\boldmath 
New physics in $b\to s$ transitions after LHC run 1
}
}\\[0.8 cm]
{\large
Wolfgang~Altmannshofer$^{a}$ and David M. Straub$^{b}$} \\[0.4 cm]
\small
$^a$ {\em Perimeter Institute for Theoretical Physics, 31 Caroline St. N, Waterloo, Ontario, Canada N2L~2Y5}\\[0.1cm]
$^b$ {\em Excellence Cluster Universe, TUM,
Boltzmannstr.~2, 85748~Garching, Germany} \\[0.4cm]
E-mail: \mailref{waltmannshofer@perimeterinstitute.ca}, \mailref{david.straub@tum.de}%
\end{center}

\medskip
%###################################################################
\begin{abstract}\noindent
We present results of global fits of all relevant experimental data on rare $b \to s$ decays.
We observe significant tensions between the Standard Model predictions and the data.
After critically reviewing the possible sources of theoretical uncertainties, we find that within the Standard Model,
the tensions 
could be explained 
if there are unaccounted hadronic effects much larger than our estimates.
Assuming hadronic uncertainties are estimated in a sufficiently conservative way, we discuss the implications of the experimental results on new physics, both model independently as well as in the context of the minimal supersymmetric standard model and models with flavour-changing $Z'$ bosons.
We discuss in detail the violation of lepton flavour universality as hinted by the current data and make predictions for additional lepton flavour universality tests that can be performed in the future.
We find that the ratio of the forward-backward asymmetries in $B \to K^* \mu^+\mu^-$ and $B \to K^* e^+e^-$ at low dilepton invariant mass is a particularly sensitive probe of lepton flavour universality and allows to distinguish between different new physics scenarios that give the best description of the current data.
\end{abstract}
%###################################################################

\setcounter{tocdepth}{2}
\tableofcontents

%###################################################################
\section{Introduction}
%###################################################################

Rare decays based on the flavour-changing neutral current $b\to s$ transition are sensitive probes of physics beyond the Standard Model (SM).
In recent years, a plethora of observables, including branching ratios,  CP and angular asymmetries in inclusive and exclusive $B$ decay modes, has been measured at the $B$ factories and at LHC experiments.
This wealth of data allows to investigate the helicity structure of flavour-changing interactions as well as possible new sources of CP violation.

In 2013, the observation by LHCb of a tension with the SM in $B\to K^*\mu^+\mu^-$ angular observables \cite{Aaij:2013qta} has received considerable attention from theorists and it was shown that the tension could be softened by assuming the presence of new physics (NP) \cite{Descotes-Genon:2013wba,Altmannshofer:2013foa,Beaujean:2013soa,Hurth:2013ssa}.
In 2014, another tension with the SM has been observed by LHCb, namely a suppression of the ratio $R_K$ of $B\to K\mu^+\mu^-$ and $B\to Ke^+e^-$ branching ratios at low dilepton invariant mass~\cite{Aaij:2014ora}.
Assuming new physics in $B\to K\mu^+\mu^-$ only, a consistent description of these anomalies seems possible \cite{Hiller:2014yaa,Ghosh:2014awa,Hurth:2014vma,Glashow:2014iga}.
In addition, also branching ratio measurements of $B\to K^*\mu^+\mu^-$ and $B_s\to \phi\mu^+\mu^-$ decays published recently~\cite{Aaij:2014pli,Aaij:2013aln} seem to be too low compared to the SM predictions when using state-of-the art form factors from lattice QCD or light-cone sum rules (LCSR) \cite{Ball:2004rg,Horgan:2013hoa,Horgan:2013pva,Straub:2015ica}.
Finally, in the latest update of the LHCb $B\to K^*\mu^+\mu^-$ analysis from 2015~\cite{LHCb-CONF-2015-002}, the tensions in angular observables persist.

While the ratio $R_K$ is theoretically extremely clean, predicted to be 1 to an excellent accuracy in the SM~\cite{Bobeth:2007dw}, the 
other observables mentioned are plagued by sizable hadronic uncertainties. On the one hand, they require the knowledge of the QCD form factors; on the other hand, even if the form factors were known exactly, there would be uncertainties from contributions of the hadronic weak Hamiltonian that violate quark-hadron duality and/or break QCD factorization.
These two sources of theoretical uncertainty have been discussed intensively in the recent literature
\cite{Lyon:2014hpa,Descotes-Genon:2014uoa,Jager:2014rwa,Straub:2015ica} (see 
also the earlier work 
\cite{Khodjamirian:2010vf,Beylich:2011aq,Khodjamirian:2012rm,Jager:2012uw}, as 
well as efforts to design observables with limited sensitivity to hadronic 
uncertainties in various kinematic 
regimes~\cite{Bobeth:2010wg,Bobeth:2011gi,Becirevic:2011bp,Matias:2012xw,
DescotesGenon:2012zf,Bobeth:2012vn,Descotes-Genon:2013vna}).
Understanding how large these hadronic effects could be is crucial to disentangle potential new physics effects from underestimated non-perturbative QCD effects, if significant tensions from the SM expectations are observed in the data. The main aim of our present analysis is thus to perform a global analysis of all relevant experimental data to answer the following questions,
\begin{enumerate}
\item Is there a significant tension with SM expectations in the current data on $b\to s$ transitions?
\item Assuming the absence of NP, which QCD effects could have been underestimated and how large would they have to be to bring the data into agreement with predictions, assuming they are wholly responsible for an apparent tension?
\item Assuming the QCD uncertainties to be estimated sufficiently conservatively, what do the observations imply for NP, both model-independently and in specific NP models?
\end{enumerate}

Our work builds on our previous global analyses of NP in $b\to s$ transitions \cite{Altmannshofer:2011gn,Altmannshofer:2012az,Altmannshofer:2013foa}, but we have built up our analysis chain from scratch to incorporate a host of improvements, including in particular the following.
\begin{itemize}
\item In our global $\chi^2$ fits, we take into account all the correlations of theoretical uncertainties between different observables and between different bins. This has become crucial to assess the global significance of any tension, as the experimental data are performed in more and more observables in finer and finer bins.

\item We assess the impact of different choices for the estimates of theoretical uncertainties on the preferred values for the Wilson coefficients.

\item We model the subleading hadronic uncertainties in exclusive semi-leptonic decays in a different way, motivated by discussions of these effects in the recent literature (see e.g.\ \cite{Khodjamirian:2010vf,Beylich:2011aq,Khodjamirian:2012rm,Jager:2012uw,Lyon:2014hpa,Descotes-Genon:2014uoa,Straub:2015ica}), see sec.~\ref{sec:obs} for details.
\end{itemize}
The novel features of our analysis in comparison to similar recent studies in the literature \cite{Descotes-Genon:2013wba,Beaujean:2013soa,Hurth:2013ssa,Ghosh:2014awa,Hurth:2014vma},
are as follows.
\begin{itemize}
\item We use the information on $B\to K^*$ and $B_s\to \phi$ form factors from the most precise LCSR calculation \cite{Ball:2004rg,Straub:2015ica}, taking into account all the correlations between the uncertainties of different form factors and at different $q^2$ values. This is particularly important to estimate the uncertainties in angular observables that involve ratios of form factors.

\item We include in our analysis the branching ratio of $B_s\to\phi\mu^+\mu^-$, showing that there exists a significant tension between the recent LHCb measurements and our SM predictions.
\end{itemize}

Our paper is organized as follows. In section~\ref{sec:obs}, we define the effective Hamiltonian and discuss the most important experimental observables, detailing our treatment of theoretical uncertainties.
In section~\ref{sec:modelindependent}, we perform the
numerical analysis. We start by investigating which sources of theoretical uncertainties, if underestimated, could account for the tension even within the SM. We then proceed with a model-independent analysis beyond the SM,
studying the allowed regions for the NP Wilson coefficients.
In section~\ref{sec:np}, we discuss what the model-independent findings imply for the Minimal Supersymmetric Standard Model as well as for models with a new heavy neutral gauge boson.
We summarize and conclude in section~\ref{sec:concl}.
Several appendices contain all our SM predictions for the observables of interest, details on our treatment of form factors, and plots of constraints on Wilson coefficients.

\section{Observables and uncertainties}\label{sec:obs}

In this section, we specify the effective Hamiltonian encoding potential new physics contributions and we discuss the most important observables entering our analysis.
The calculation of the observables included in our previous analyses \cite{Altmannshofer:2011gn,Altmannshofer:2012az,Altmannshofer:2013foa} (see also \cite{Altmannshofer:2008dz,Straub:2015ica}) have been discussed in detail there and in references therein; here we only focus on the novel aspects of the present analyses -- like the $B_s\to \phi\mu^+\mu^-$ decay -- and on our refined treatment of theoretical uncertainties.

%################################################################
\subsection{Effective Hamiltonian}\label{sec:Heff}
%################################################################

The effective Hamiltonian for $b\to s$ transitions can be written as
\begin{equation}
\label{eq:Heff}
{\cal H}_{\eff} = - \frac{4\,G_F}{\sqrt{2}} V_{tb}V_{ts}^* \frac{e^2}{16\pi^2}
\sum_i
(C_i O_i + C'_i O'_i) + \text{h.c.}
\end{equation}
and we consider NP effects in the following set of dimension-6 operators,
\begin{align}
O_7 &= \frac{m_b}{e}
(\bar{s} \sigma_{\mu \nu} P_{R} b) F^{\mu \nu},
&
O_7^{\prime} &= \frac{m_b}{e}
(\bar{s} \sigma_{\mu \nu} P_{L} b) F^{\mu \nu},
\\
O_9 &= 
(\bar{s} \gamma_{\mu} P_{L} b)(\bar{\ell} \gamma^\mu \ell)\,,
&
O_9^{\prime} &= 
(\bar{s} \gamma_{\mu} P_{R} b)(\bar{\ell} \gamma^\mu \ell)\,,\label{eq:O9}
\\
O_{10} &=
(\bar{s} \gamma_{\mu} P_{L} b)( \bar{\ell} \gamma^\mu \gamma_5 \ell)\,,
&
O_{10}^{\prime} &=
(\bar{s} \gamma_{\mu} P_{R} b)( \bar{\ell} \gamma^\mu \gamma_5 \ell)\,,\label{eq:O10}
\end{align}
Of the complete set of dimension-6 operators invariant under the strong and electromagnetic gauge groups, this set does not include
\begin{itemize}
\item Four-quark operators (including current-current, QCD penguin, and electroweak penguin operators). These operators only contribute to the observables considered in this analysis through mixing into the operators listed above and through higher order corrections. Moreover, at low energies they are typically dominated by SM contributions. Consequently, we expect the impact of NP contributions to these operators on the observables of interested to be negligible.\footnote{Note that the situation is different when also non-leptonic decays are considered, see e.g.~\cite{Bobeth:2014rra}.}
\item Chromomagnetic dipole operators. In the radiative and semi-leptonic decays we consider, their Wilson coefficients enter at leading order only through mixing with the electromagnetic dipoles and thus enter in a fixed linear combination, making their discussion redundant.
\item Tensor operators.
Our rationale for not considering these operators is that they do not appear in the dimension-6 operator product expansion of the Standard Model \cite{Buchmuller:1985jz,Grzadkowski:2010es,Alonso:2014csa}. Consequently, they are expected to receive only small NP contributions unless the scale of new physics is very close to the electroweak scale, which is in tension with the absence of new light particles at the LHC.
\item Scalar operators of the form $(\bar{s}  P_{A} b)( \bar{\ell} P_{B} \ell)$.
The operators with $AB=LL$ or $RR$ do not appear in the dimension-6 operator product expansion of the Standard Model either.
While the ones with $AB=LR$ and $RL$ do appear at dimension 6, their effects in semi-leptonic decays are completely negligible once constraints from $B_s\to\mu^+\mu^-$ are imposed \cite{Alonso:2014csa}.
The constraints from $B_s\to\mu^+\mu^-$ can only be avoided for a new physics 
scale close to the electroweak scale such that scalar $LL$ and $RR$ operators 
can have non-negligible impact.
\end{itemize}

\subsection{\texorpdfstring{$B\to K\mu^+\mu^-$}{B --> K mu+ mu-}}

\subsubsection{Observables}
The differential decay distribution of $B\to K\mu^+\mu^-$ in terms of the dimuon invariant mass squared $q^2$ and the angle between the $K$ and $\mu^-$ gives access to two angular observables, the so-called flat term $F_H$ and the forward-backward asymmetry $A_\text{FB}$, in addition to the differential decay rate (or branching ratio).
The observables $A_\text{FB}$ and $F_H$ only deviate significantly from zero in the presence of scalar or tensor operators \cite{Bobeth:2007dw}.
Due to the argument given above, we do not consider NP contributions to these operators in semi-leptonic decays.
While the direct CP asymmetry  has been measured recently as well \cite{Aaij:2014bsa}, we do not include it in our analysis since it is suppressed by small strong phases and therefore does not provide constraints on new physics at the current level of experimental accuracy.
Consequently, the only observable we need to consider is the (CP-averaged) differential branching ratio of the charged $B$ decay,
\begin{align}
\frac{d\text{BR}(B^\pm\to K^\pm\mu^+\mu^-)}{dq^2} &= \frac{\tau_{B^+}}{2}
\left(\frac{d\Gamma(B^+\to K^+\mu^+\mu^-)}{dq^2}+\frac{d\Gamma(B^-\to K^-\mu^+\mu^-)}{dq^2}\right).
\end{align}
and analogously for the neutral $B$ decay.

\subsubsection{Theoretical uncertainties}

The theoretical analysis of the $B \to K \mu^+ \mu^-$ observables is complicated not only by the need to know the $B\to K$ form factors, but also by the fact that the ``naive'' factorization of the amplitude into a hadronic and a leptonic part is violated by contributions from the hadronic weak Hamiltonian, connecting to the lepton pair through a photon.
Concretely, in the limit of vanishing lepton mass\footnote{We take the non-zero lepton mass into account in our numerics; the zero-mass limit is taken here just for illustration.}, the decay rate can be written as
\begin{align}
\frac{d\Gamma(B\to K\mu^+\mu^-)}{dq^2} = \frac{G_F^2\alpha^2_\text{em}|V_{tb}V_{ts}^*|^2}{2^{10} \pi^5 m_B^3} \lambda^{3/2}(m_B^2,m_{K^*}^2,q^2)
\left( |F_V|^2+|F_A|^2 \right),
\end{align}
where
\begin{align}
\lambda(a,b,c)&= a^2  +b^2 + c^2 - 2 (ab+ bc + ac)
\,,\\
F_V(q^2) &=
\left(C_{9}^\text{eff}(q^2)+C_{9}'\right) f_+(q^2)
+ \frac{2m_b}{m_B+m_K} \left(C_{7}^\text{eff}+C_{7}'\right) f_T(q^2)
+ h_K(q^2)
\,,
\\
F_A(q^2) &= \left(C_{10}+C_{10}'\right) f_+(q^2)
\,.
\end{align}
Here, $f_+$ and $f_T$ are the full QCD form factors and $h_K$ includes the non-factorizable contributions from the weak effective Hamiltonian. An additional form factor, $f_0$, enters terms that are suppressed by the lepton mass. We now discuss our treatment of these quantities, which represent the main source of theoretical uncertainties in the $B\to K\mu^+\mu^-$ observables.

For the form factors, we perform a combined fit of the recent lattice computation by the HPQCD collaboration \cite{Bouchard:2013eph}, valid at large $q^2$, and form factor values at $q^2=0$ obtained from light-cone sum rules (LCSR)~\cite{Ball:2004ye,Bartsch:2009qp}, to a simplified series expansion. Details of the fit are discussed in appendix~\ref{sec:BKFFs}. The results are 3-parameter (4-parameter) fit expressions for the form factors $f_{+,T}$ ($f_0$) as well as the full $10\times10$ covariance matrix. We retain the correlations among these uncertainties throughout our numerical analysis.

Concerning $h_K(q^2)$, we emphasize the following contributions.
\begin{itemize}
\item Virtual corrections to the matrix elements of the four-quark operators $O_1$ and $O_2$. We include them to NNLL accuracy using the results of ref.~\cite{Greub:2008cy}.
\item Contributions from weak annihilation and hard spectator scattering. These have been estimated in QCD factorization to be below a percent \cite{Bartsch:2009qp} and we neglect them.
\item Soft gluon corrections to the virtual charm quark loop at low $q^2$. This effect was computed recently in LCSR with $B$ meson distribution amplitudes in ref.~\cite{Khodjamirian:2010vf} and was found to be ``unimportant at least up to $q^2\sim5-6~\text{GeV}^2$.'' (See also \cite{Khodjamirian:2012rm}).
\item Violation of quark-hadron duality at high $q^2$, above the open charm threshold, due to the presence of broad charmonium resonances. Employing an OPE in inverse powers of the dilepton invariant mass, this effect has been found to be under control at a few percent in ref.~\cite{Beylich:2011aq}.
\end{itemize}

Concerning the last two items, the uncertainties due to these effects have to be estimated in a consistent and conservative manner to draw robust conclusions about the compatibility of experimental measurements with the SM predictions. We do this by parametrizing our ignorance of sub-leading corrections to $h_K$ in the following way,
\begin{equation}
h_K^\text{subl.} = [C_9^\text{eff}(q^2)]^\text{SM} f_+(q^2)
\times\,\left\lbrace
\begin{array}{ll}
a_K e^{i\phi_a}+ b_K e^{i\phi_b} (q^2/6\,\text{GeV}^2)
& \text{at low }q^2\,,
\\
c_K e^{i\phi_c}
& \text{at high }q^2\,,
\end{array}
\right.
\label{eq:SLK}
\end{equation}
where we used the leading contribution to the amplitude $F_V$ as an overall normalization factor. To obtain the theory uncertainties, we vary the strong phases $\phi_{a,b,c}$ within $(-\pi,\pi]$. At low $q^2$, since the main contribution is expected to come from the soft gluon correction to the charm loop, we vary $a$ within $[0,0.02]$ and $b$ within $[0,0.05]$. In this way, the central value of the effect discussed in \cite{Khodjamirian:2010vf,Khodjamirian:2012rm} is contained within our $1\sigma$ error band.
Although (\ref{eq:SLK}) is just a very crude parametrization of the (unknown) $q^2$ dependence at low $q^2$, we believe it is sufficiently general at the current level of experimental precision.
At high $q^2$, the presence of broad charmonium resonances means that $h_K(q^2)$ 
varies strongly with $q^2$, but since we will only consider observables 
integrated over the whole high-$q^2$ region, we can ignore this fact and the 
parameter $c$ simply parametrizes the violation of the OPE result. We estimate 
it by varying $c$ within $[0,0.05]$, which corresponds to an uncertainty on the 
rate more than twice the uncertainty quoted in \cite{Beylich:2011aq}. This large 
range is chosen to take into account the fact that ref.~\cite{Beylich:2011aq} 
uses a toy model for the charm loop. In section~\ref{sec:sm_comp}, we will also 
discuss the consequences of increasing the ranges for these parameters.

\subsection{\texorpdfstring{$B\to K^*\mu^+\mu^-$}{B --> K* mu+ mu-} and \texorpdfstring{$B\to K^*\gamma$}{B --> K* gamma}}

\subsubsection{Observables}\label{sec:BKsmmobs}

The angular decay distribution of $\bar B^0\to \bar K^{*0}\mu^+\mu^-$ contains in general 12 angular coefficient functions. In the presence of CP violation, the 12 angular coefficients of the CP-conjugate decay $B^0\to K^{*0}\mu^+\mu^-$ represent another 12 independent observables \cite{Altmannshofer:2008dz}. However, since scalar contributions are negligible in our setup and one can neglect the muon mass to a good approximation, there are only 9 independent observables in each decay. Moreover, the absence of large strong phases implies that several of the observables are hardly sensitive to new physics. In practice, the observables that are sensitive to new physics are
\begin{itemize}
\item the CP-averaged differential branching ratio $d\text{BR}/dq^2$,
\item the CP-averaged $K^*$ longitudinal polarization fraction $F_L$ and forward-backward asymmetry $A_\text{FB}$,
\item the CP-averaged angular observables $S_{3,4,5}$,
\item the T-odd CP-asymmetries $A_{7,8,9}$.
\end{itemize}
All of these observables can be expressed in terms of angular coefficients and are functions of $q^2$. Alternative bases have been considered in the literature (see e.g.~\cite{Bobeth:2010wg,Bobeth:2011gi,Becirevic:2011bp,Matias:2012xw,Descotes-Genon:2013vna}). Choosing different normalizations can reduce the sensitivity of the observables to the hadronic form factors, at least in the heavy quark limit and for naive factorization. In our analysis, the choice of basis is irrelevant for the impact of hadronic uncertainties, as we consistently take into account all the correlations between theoretical uncertainties.

In the case of $B\to K^*\gamma$, we consider the following observables: the branching ratio of $B^\pm\to K^{*\pm}\gamma$, the branching ratio of $B^0\to K^{*0}\gamma$, the direct CP asymmetry $A_\text{CP}$ and the mixing-induced CP asymmetry $S_{K^*\gamma}$ in $B^0\to K^{*0}\gamma$. Since we take all known correlations between the observables into account in our numerical analysis, including the branching ratios of the charged and neutral $B$ decays is to a very good approximation equivalent to including one of these branching ratios and the isospin asymmetry.

\subsubsection{Theoretical uncertainties}

Similarly to the $B\to K\mu^+\mu^-$ decay, the main challenges of $B\to K^*\mu^+\mu^-$ are the form factors and the contributions of the hadronic weak Hamiltonian.

For the form factors, we use the preliminary results of a 
a combined fit \cite{Straub:2015ica} to a LCSR calculation of the full set of seven form factors \cite{Ball:2004rg} with correlated uncertainties as well as lattice results for these form factors \cite{Horgan:2013hoa}. This leads to strongly reduced uncertainties in angular observables.

The non-factorizable contributions from the hadronic weak Hamiltonian are more involved in $B\to K^*\mu^+\mu^-$ compared to $B\to K\mu^+\mu^-$ for several reasons. First, it contributes to three helicity amplitudes instead of just one; Second, the presence of the photon pole at $q^2=0$ enhances several of the contributions at low $q^2$; Third, since we do not only consider branching ratios but also a host of angular observables where form factor uncertainties partly cancel, we require a higher theoretical accuracy in the $h_\lambda$. Concretely, we include the following contributions.
\begin{itemize}
\item The NNLL contributions to the matrix elements of $O_{1,2}$ as in the case of $B\to K\mu^+\mu^-$.
\item At low $q^2$, hard spectator scattering at $O(\alpha_s)$ from QCD factorization \cite{Beneke:2001at} including the sub-leading doubly Cabibbo-suppressed contributions \cite{Beneke:2004dp}.
\item At low $q^2$, weak annihilation beyond the heavy quark limit as obtained from LCSR \cite{Lyon:2013gba}.
\item At low $q^2$, contributions from the matrix element of the chromomagnetic operator as obtained from LCSR
\cite{Dimou:2012un}.
\end{itemize}
As in $B\to K\mu^+\mu^-$, there are additional, sub-leading contributions, such as the soft gluon corrections to the charm loop \cite{Ball:2006eu,Khodjamirian:2010vf,Khodjamirian:2012rm,Lyon:2014hpa}. We parametrize them at low $q^2$ by a correction relative to the leading contribution to the helicity amplitudes proportional to $C_7^\text{eff}$,
\begin{equation}
[C_7^\text{eff}]^\text{SM} \to 
[C_7^\text{eff}]^\text{SM} \left[ 1 +a_\lambda e^{i\phi_a^\lambda} + b_\lambda e^{i\phi_b^\lambda} \left(\frac{q^2}{6\,\text{GeV}^2}\right)\right]
\,.
\label{eq:SL}
\end{equation}
The parameters $a_\lambda$ and $b_\lambda$ are allowed to be different for each of the three helicity amplitudes, $\lambda=+,-,0$.
We vary the $a_\lambda$ and $b_\lambda$ in the following ranges,
\begin{align}
a_{+,-} &\in [0,0.05] \,,
&
b_{+,-} &\in [0,0.2] \,,
&
a_{0} &\in [0,0.2] \,,
&
b_{0} &\in [0,0.5] \,,
\end{align}
Again, with this choice the effect discussed in
\cite{Khodjamirian:2010vf,Khodjamirian:2012rm} is within our $1\sigma$ uncertainty band.
Although the normalization of the correction is arbitrary and could have also been written as a relative correction to $C_9$, we choose $C_7$ as normalization in $B\to K^*\mu^+\mu^-$ since the leading contribution proportional to $C_9$ vanishes at $q^2=0$ and does not contribute to $B\to K^*\gamma$.
It is due to this choice that we need to allow for larger $a_0$, $b_0$ since the $C_7^\text{eff}$ contribution is not enhanced in the $\lambda=0$ amplitude.

At high $q^2$, as in the case of $B\to K\mu^+\mu^-$, we do not have to consider a $q^2$ dependent correction as we are only considering observables integrated over the full high $q^2$ region. Analogous to $B\to K\mu^+\mu^-$, we parametrize the sub-leading uncertainties by a relative correction to $C_9$. To be conservative, we allow it to be up to $7.5\%$ in magnitude, independently for the three helicity amplitudes, with an arbitrary strong phase.

\subsubsection{Direct CP asymmetry in $B\to K^*\gamma$}\label{sec:ACPBKsg}

While direct CP asymmetries in the $B$ decays considered by us are suppressed by small strong phases and so typically do not lead to strong constraints on NP, the direct CP asymmetry in $B\to K^*\gamma$ is a special case since the measurements by the $B$ factories and LHCb are so precise that this suppression could be overcome. The world average reads\footnote{Here, we gloss over the fact that the $B$ factories actually measure the direct CP asymmetry in an admixture between charged and neutral $B$ decays. However, the isospin difference between the CP asymmetries generated by an imaginary $C_7$ or $C_7'$ turns out to be negligibly small, so this is not relevant for our purposes.}
\begin{equation}\label{eq:ACPexp}
A_\text{CP}(B^0\to K^{*0}\gamma)_\text{HFAG} = (0.1\pm 1.3)\%.
\end{equation}
Allowing for general NP contributions in $C_7$, we find the following central value for the asymmetry,
\begin{equation} \label{eq:ACPNP}
A_\text{CP}(B^0\to K^{*0}\gamma) 
\approx
\left[0.003-0.45 \, \text{Im}\,C_7(m_b)\right]
\frac{\text{BR}(B^0\to K^{*0}\gamma)_\text{SM}}{\text{BR}(B^0\to K^{*0}\gamma) }\,,
\end{equation}
where we have neglected contributions from NP in $C_7'$ and $C_8$. We observe that the experimental bound (\ref{eq:ACPexp}) can constrain an imaginary part of the Wilson coefficient $C_7$ at the $m_b$ scale at the level of $0.1$, which is still allowed by all other measurements as we will see.

The problem with using this observable as a constraint on NP is that it is 
proportional to a strong phase that appears only at sub-leading order and is 
afflicted with a considerable uncertainty. With our error treatment described 
above, 
taking the subleading contributions from ref.~\cite{Dimou:2012un},
we find an overall relative uncertainty of 20\% in the presence of a large 
imaginary $C_7$.
However, to be conservative, we will not include $A_\text{CP}(B^0\to K^{*0}\gamma)$ in our global fits, but we will discuss the impact of including it separately in section~\ref{sec:1WC}.

\subsection{\texorpdfstring{$B_s\to \phi\mu^+\mu^-$}{Bs --> phi mu+ mu-}}

The decay $B_s\to \phi\mu^+\mu^-$ is very similar to the $B\to K^*\mu^+\mu^-$ decay, so here we only discuss the differences in the calculation of the observables compared to $B\to K^*\mu^+\mu^-$, in addition to the obvious parametric replacements throughout the calculation.

\begin{itemize}
 \item The {\bfseries form factors} are of course different; we use the combined fit of lattice and LCSR results obtained in \cite{Straub:2015ica} including the correlated uncertainties.
 \item The sub-leading {\bfseries non-factorizable corrections} are parametrized as in the case of $B\to K^*\mu^+\mu^-$, and the coefficients $a_\lambda$, $b_\lambda$ and $c_\lambda$ are varied in the same ranges. We assume the uncertainty in these coefficients to be $90\%$ correlated between $B_s\to \phi\mu^+\mu^-$ and $B\to K^*\mu^+\mu^-$ since we do not see a physical reason why they should be drastically different\footnote{In the case of $B\to K^*\mu^+\mu^-$, all known spectator-dependent non-factorizable effects are very small (see e.g.~\cite{Lyon:2013gba}), while e.g.\ the sizable effect discussed in ref.~\cite{Khodjamirian:2010vf} does not depend on the flavour of the spectator quark and we therefore expect it to be very similar between $B_s\to \phi\mu^+\mu^-$ and $B\to K^*\mu^+\mu^-$. We also stress that this guess for the correlation has a small impact on the numerical results as the uncertainty of BR($B_s\to\phi\mu^+\mu^-$) is by far dominated by form factor uncertainties \cite{Straub:2015ica}, 
which we assume to be uncorrelated between $B\to K^*$ and $B_s\to\phi$ to be conservative.}.
\item In contrast to $B\to K^*\mu^+\mu^-$, the $B_s\to \phi\mu^+\mu^-$ decay is 
{\bfseries not self-tagging}. Therefore, the only observables among the ones 
mentioned at the beginning of section~\ref{sec:BKsmmobs} that are experimentally 
accessible  in a straightforward way at a hadron collider are 
\cite{Bobeth:2008ij}
\begin{itemize}
 \item the differential branching ratio $d\text{BR}/dq^2$,
 \item the CP-averaged angular observables $F_L$ and $S_4$,
 \item the angular CP asymmetry $A_9$.
\end{itemize}
\item An additional novelty is the impact of the sizable {\bfseries$B_s$ width difference}. As shown in \cite{Straub:2015ica} (see also \cite{Descotes-Genon:2015hea}), this effect is small in the SM
and we have checked that it is also negligible in the presence of NP at the current level of experimental precision, unless the Wilson coefficients assume extreme values that are already excluded by other constraints. Therefore, we have neglected the effect in our numerical analysis.
\end{itemize}

\section{Global numerical analysis} \label{sec:modelindependent}
\subsection{Fit methodology}

More and more experimental data on $b\to s\mu^+\mu^-$ transitions becomes available and many observables are measured with a fine binning. Therefore, in order to determine the values of the Wilson coefficients preferred by the data
it becomes more and more important to include the correlation of theoretical uncertainties between different observables as well as between different bins of the same observable.
One possibility to achieve this is to perform a global Bayesian analysis where all the uncertainties are parametrized by nuisance parameters that are marginalized over by sophisticated numerical tools like Markov Chain Monte Carlos. This approach has been applied recently e.g.\ in \cite{Beaujean:2013soa}. A drawback of this approach is that it is time-consuming and the computing time increases with the number of parameters. Here, we follow a different approach. We construct a $\chi^2$ function that only depends on the Wilson coefficients and take into account the theoretical and experimental uncertainties in terms of covariance matrices,
\begin{equation}
\chi^2(\vec C^\text{NP})
=
\left[\vec O_\text{exp}-\vec O_\text{th}(\vec C^\text{NP})
\right]^T
\left[C_\text{exp}+C_\text{th}\right]^{-1}
\left[\vec O_\text{exp}-\vec O_\text{th}(\vec C^\text{NP})
\right].
\label{eq:chi2}
\end{equation}
Here, $\vec O_\text{exp}$ are the experimentally measured central values of all observables of interest, $O_\text{th}$ are the corresponding theory predictions that depend on the (NP contributions to the) Wilson coefficients, $C_\text{exp}$ is the covariance matrix of the experimental measurements and $C_\text{th}$ is the covariance matrix of the theory predictions that contains the theory uncertainties and their correlations. In writing (\ref{eq:chi2}), we have made two main approximations. First, we have assumed all the experimental and theoretical uncertainties to be Gaussian. Second, we have neglected the dependence of the theory uncertainties on the new physics contributions to the Wilson coefficients.
This means that the theory uncertainties and their correlations have been evaluated for the Wilson coefficients fixed to their SM values.
We believe that this assumption is well justified in view of the fact that no 
drastic deviations from the SM expectations have been observed so far.
We checked explicitly that changes are small between the covariance matrix of 
the theory predictions in the SM and the one computed at the best-fit point for 
new physics in the Wilson coefficient $C_9$ alone ($C_9^\text{NP}=-1.07$, see 
Sec.~\ref{sec:1WC} below).
The only possible exception are observables that vanish in the SM but could receive NP contributions much larger than the current experimental bounds. As we will discuss below, the only such observable at present is the direct CP asymmetry in $B\to K^*\gamma$.

We determine $C_\text{th}$ by evaluating all observables of interest for a large 
set of the parameters parametrizing the theory uncertainties, randomly 
distributed following normal distributions according to the 
uncertainties and correlations described above. In this way, we retain not only 
correlated uncertainties between different observables, but also between 
different bins of the same observable. We find these correlations to have a 
large impact on our numerical results.
Concerning $C_\text{exp}$, we symmetrize the experimental error bars and include the experimental error correlations provided by the latest LHCb update of the $B \to K^* \mu^+\mu^-$ analysis~\cite{LHCb-CONF-2015-002}. For branching ratio measurements, where no error correlations are available, we include a rough guess of the correlations by assuming the statistical uncertainties to be uncorrelated and the systematic uncertainties to be fully correlated for measurements of the same observable by a single experiment. We have checked that this treatment has only a  small impact on the overall fit at the current level of experimental and theoretical uncertainties on branching ratios.

We use the following experimental input for our global $b \to s \mu^+\mu^-$ fit:
\begin{itemize}
 \item $B \to K^* \mu^+\mu^-$ branching ratios and angular observables from LHCb~\cite{Aaij:2013iag,Aaij:2013qta,Aaij:2014pli,LHCb-CONF-2015-002}, CMS~\cite{Chatrchyan:2013cda}, ATLAS~\cite{ATLAS-CONF-2013-038}, and CDF~\cite{Aaltonen:2011qs,Aaltonen:2011ja,CDFupdate};
 \item $B \to K \mu^+\mu^-$ branching ratios and angular observables from LHCb~\cite{Aaij:2014pli,Aaij:2014tfa} and CDF~\cite{Aaltonen:2011qs,Aaltonen:2011ja,CDFupdate};
 \item $B_s \to \phi \mu^+\mu^-$ branching ratios and angular observables from LHCb~\cite{Aaij:2013aln} and CDF~\cite{Aaltonen:2011qs,CDFupdate};
 \item branching ratios for $B \to K^* \gamma$ and $B \to X_s \gamma$ and the mixing-induced CP asymmetry in $B \to K^* \gamma$ from HFAG~\cite{Amhis:2012bh};
 \item the combined result of the $B_s \to \mu^+\mu^-$ branching ratio from LHCb and CMS~\cite{Aaij:2013aka,Chatrchyan:2013bka,Bsmumu_combination};
 \item the $B \to X_s \mu^+\mu^-$ branching ratio measurement from BaBar~\cite{Lees:2013nxa}.
\end{itemize}
We do not include the additional results on $b \to s \ell\ell$ transitions from BaBar~\cite{Lees:2012tva,Ritchie:2013mx} and Belle~\cite{Wei:2009zv,Sato:2014pjr}, as they are only available as an average of $\mu^+\mu^-$ and $e^+e^-$ modes.   
As already mentioned in section~\ref{sec:obs}, in the fit we do not explicitly include isospin asymmetries, but instead use results on the charged and neutral modes separately. As we take into account all known error correlations, this approach is essentially equivalent.  

We would like to stress that for none of the observables, we use low $q^2$ bins that extend into the region above the perturbative charm threshold $q^2 > 6$~GeV, where hadronic uncertainties cannot be estimated reliably.
This applies in particular to the bin $[4.3,8.68]$~GeV$^2$ that has been used in several fits in the past \cite{Descotes-Genon:2013wba,Hurth:2014vma,Hurth:2013ssa}
as well as the bin $[6,8]$~GeV$^2$ in the recent $B\to K^*\mu^+\mu^-$ angular analysis by LHCb \cite{LHCb-CONF-2015-002}.

For the $B^0 \to K^{*0} \mu^+\mu^-$ observables at low $q^2$, we choose the smallest available bins satisfying this constraint, since they are most sensitive to the non-trivial $q^2$ dependence of the angular observables. For $B_s\to\phi\mu^+\mu^-$, we use the $[1,6]$~GeV$^2$ bin, since the branching ratio does not vary strongly with $q^2$ and since the statistics is limited.
In the high $q^2$ region, we always consider the largest $q^2$ bins available that extend to values close to the kinematical end point.
All the experimental measurements used in our global fits are listed in appendix~\ref{sec:obstables} along with their theory predictions. All theory predictions are based on our own work and on \cite{Straub:2015ica}, except the $B_s \to \mu^+\mu^-$, $B \to X_s \gamma$ and $B \to X_s \ell^+\ell^-$ branching ratios that we take from~\cite{Bobeth:2013uxa},~\cite{Misiak:2015xwa} and~\cite{Huber:2007vv}\footnote{Note also the recent update \cite{Huber:2015sra} which appeared after our analyses had been completed. We expect the changes to be much smaller than the experimental uncertainty.}, respectively. In the case of the SM prediction for BR$(B_s \to \mu^+\mu^-)$ we rescale the central value and uncertainty obtained in~\cite{Bobeth:2013uxa}, to reflect our choice of $V_{cb}$ (see section~\ref{sec:Vcb} below).

\subsection{Compatibility of the data with the SM} \label{sec:sm_comp}

\begin{table}[tb]
\renewcommand{\arraystretch}{1.5}
\centering
\begin{tabular}{ccccccc}
\hline\hline
Decay & obs. & $q^2$ bin & SM pred. & \multicolumn{2}{c}{measurement} & pull\\
\hline
$\bar B^0\to\bar K^{*0}\mu^+\mu^-$ & $10^{7}~\frac{d\text{BR}}{dq^2}$ & $[2,4.3]$ & $0.44 \pm 0.07$ & $0.29 \pm 0.05$ & LHCb & {$+1.8$}\\
$\bar B^0\to\bar K^{*0}\mu^+\mu^-$ & $10^{7}~\frac{d\text{BR}}{dq^2}$ & $[16,19.25]$ & $0.47 \pm 0.06$ & $0.31 \pm 0.07$ & CDF & {$+1.8$}\\
$\bar B^0\to\bar K^{*0}\mu^+\mu^-$ & $F_L$ & $[2,4.3]$ & $0.81 \pm 0.02$ & $0.26 \pm 0.19$ & ATLAS & {$+2.9$}\\
$\bar B^0\to\bar K^{*0}\mu^+\mu^-$ & $F_L$ & $[4,6]$ & $0.74 \pm 0.04$ & $0.61 \pm 0.06$ & LHCb & {$+1.9$}\\
$\bar B^0\to\bar K^{*0}\mu^+\mu^-$ & $S_5$ & $[4,6]$ & $-0.33 \pm 0.03$ & $-0.15 \pm 0.08$ & LHCb & {$-2.2$}\\
$B^-\to K^{*-}\mu^+\mu^-$ & $10^{7}~\frac{d\text{BR}}{dq^2}$ & $[4,6]$ & $0.54 \pm 0.08$ & $0.26 \pm 0.10$ & LHCb & {$+2.1$}\\
$\bar B^0\to\bar K^{0}\mu^+\mu^-$ & $10^{8}~\frac{d\text{BR}}{dq^2}$ & $[0.1,2]$ & $2.71 \pm 0.50$ & $1.26 \pm 0.56$ & LHCb & {$+1.9$}\\
$\bar B^0\to\bar K^{0}\mu^+\mu^-$ & $10^{8}~\frac{d\text{BR}}{dq^2}$ & $[16,23]$ & $0.93 \pm 0.12$ & $0.37 \pm 0.22$ & CDF & {$+2.2$}\\
$B_s\to\phi\mu^+\mu^-$ & $10^{7}~\frac{d\text{BR}}{dq^2}$ & $[1,6]$ & $0.48 \pm 0.06$ & $0.23 \pm 0.05$ & LHCb & {$+3.1$}\\
$B\to X_se^+e^-$ & $10^{6}~\text{BR}$ & $[14.2,25]$ & $0.21 \pm 0.07$ & $0.57 \pm 0.19$ & BaBar & {$-1.8$}\\
\hline\hline
\end{tabular}
\caption{Observables where a single measurement deviates from the SM by 
$1.8\sigma$ or more. The full list of observables is given in 
appendix~\ref{sec:obstables}.
Differential branching ratios are given in units of GeV$^{-2}$.}
\label{tab:largepulls}
\end{table}

Evaluating (\ref{eq:chi2}) with the Wilson coefficients fixed to their SM values, we obtain the total $\chi^2$ of the SM.
Including both $b\to s\mu^+\mu^-$ and $b\to se^+e^-$ observables, we find $\chi^2_\text{SM}\equiv\chi^2(\vec 0)=125.8$ for 91 independent measurements.
This corresponds to a $p$-value of $0.9\%$. Including only $b\to s\mu^+\mu^-$ observables, we find $\chi^2_\text{SM}=116.9$ for 88 independent measurements, corresponding to a $p$-value of $2.1\%$.
In table~\ref{tab:largepulls}, we list the observables with the largest deviation from the SM expectation.
The full list of observables entering the $\chi^2$, together with the SM predictions and experimental measurements, is given in appendix~\ref{sec:obstables}.
We note that some of these observables have strongly correlated uncertainties and that for two of the observables, $A_\text{FB}$ and $F_L$, there is some tension between 
different experiments.
Still, there does seem to be a systematic suppression of branching ratios in different decay modes and we will see in section~\ref{sec:1WC} that the quality of the fit can be improved substantially in the presence of new physics. An important questions is whether these tensions could be due to underestimated theory uncertainties and we will investigate this question in the following paragraphs. It should be kept in mind that none of these sources of uncertainties can account for violation of lepton flavour universality.

\subsubsection{Underestimated hadronic effects?}

We will see in section~\ref{sec:1WC} that the agreement of the theory predictions with the experimental data is improved considerably assuming non-standard values for the Wilson coefficient $C_9$. Since this coefficient corresponds to a left-handed quark current and a leptonic vector current, it is conceivable that a NP effect in $C_9$ is mimicked by a hadronic SM effect that couples to the lepton current via a virtual photon, e.g.\ the charm loop effects at low $q^2$ and the resonance effects at high $q^2$ as discussed in section~\ref{sec:obs} (see e.g.~\cite{Lyon:2014hpa}). In our numerical analysis, in addition to the known non-factorizable contributions taken into account as described in section~\ref{sec:obs}, sub-leading effects of this type are parametrized by the parameters 
$a_i, b_i, c_i$ in (\ref{eq:SLK}), (\ref{eq:SL}), and analogously for $B_s\to \phi\mu^+\mu^-$.
Since they parametrize unknown sub-leading uncertainties, the central values of these parameters are 0 in our SM predictions.

%%%%%%%%%%%%%%%%%%%%%%%%%%%%%%%%%%%%
\begin{figure}[tbp]
\centering
\includegraphics[width=0.43\textwidth]{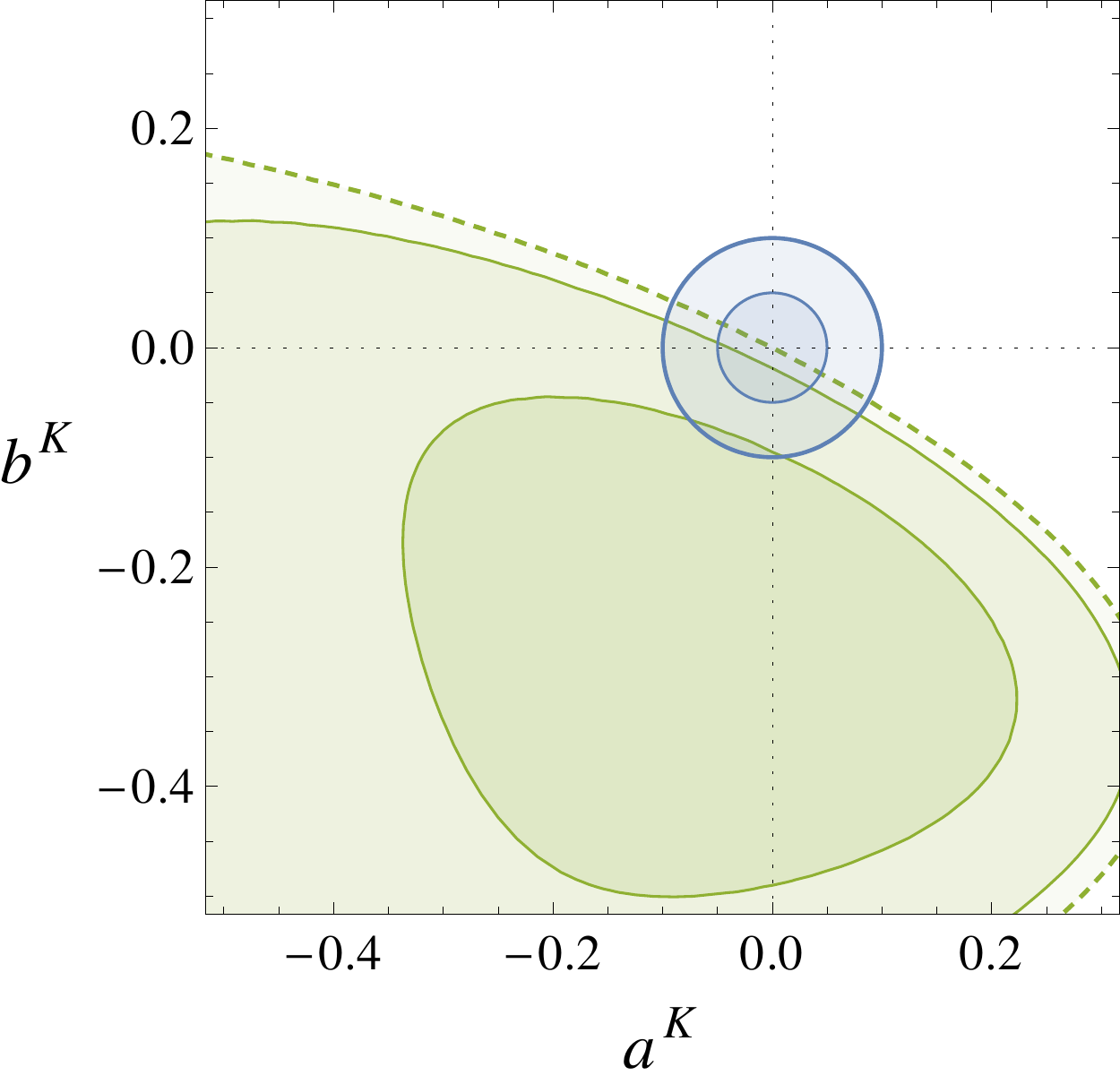} ~~~~~~
\includegraphics[width=0.43\textwidth]{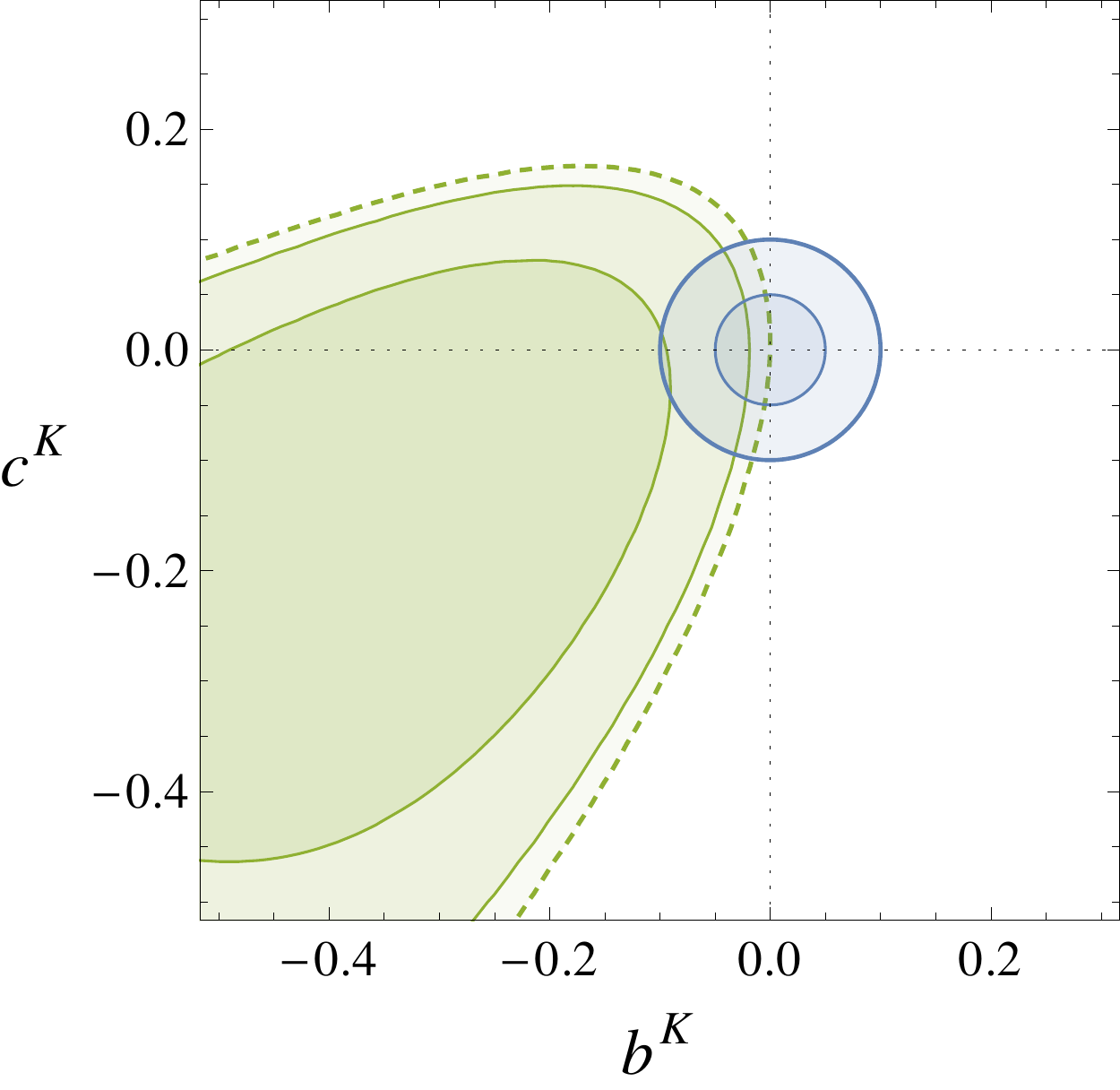} \\[20pt]
\includegraphics[width=0.43\textwidth]{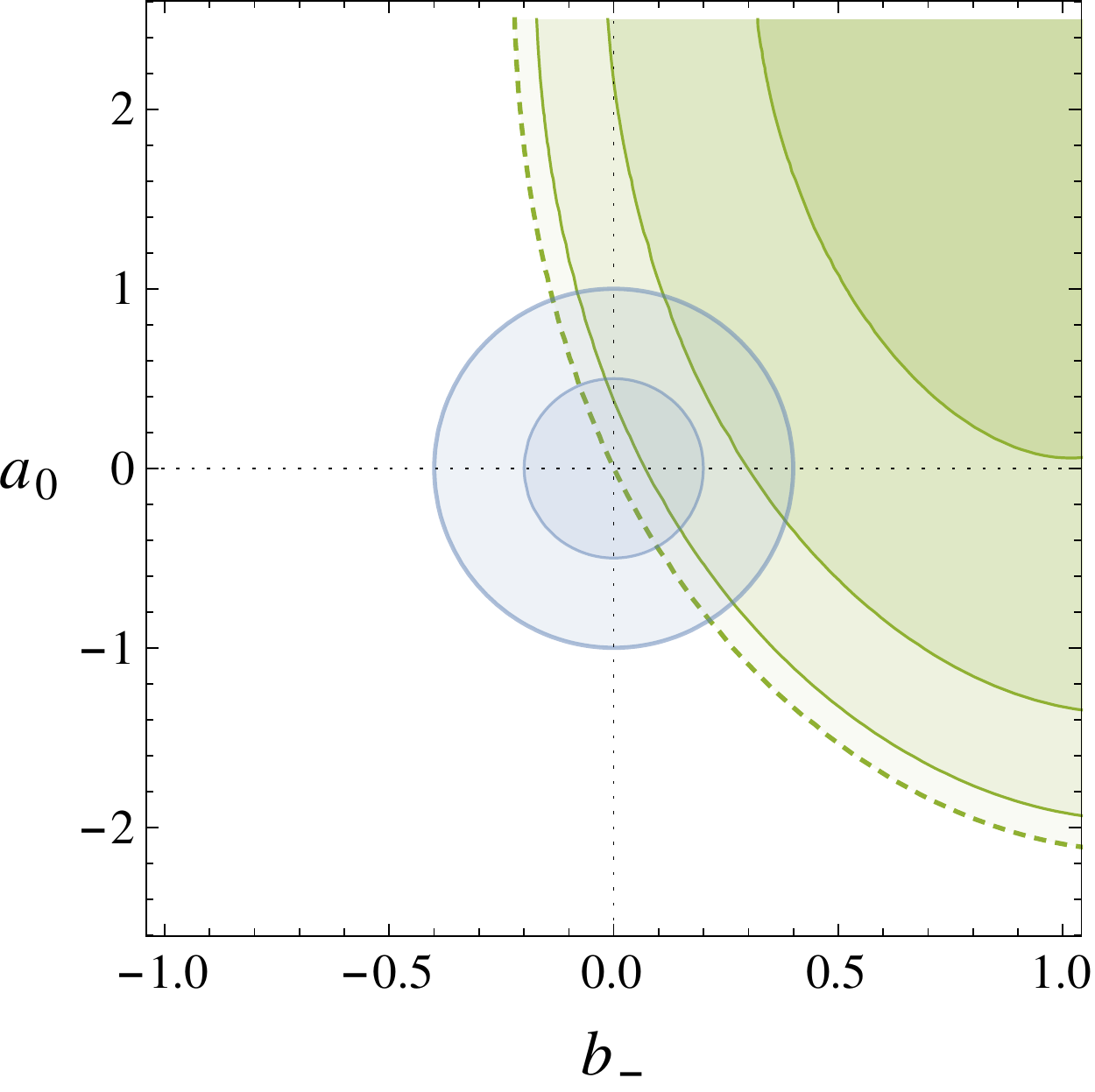} ~~~~~~
\includegraphics[width=0.43\textwidth]{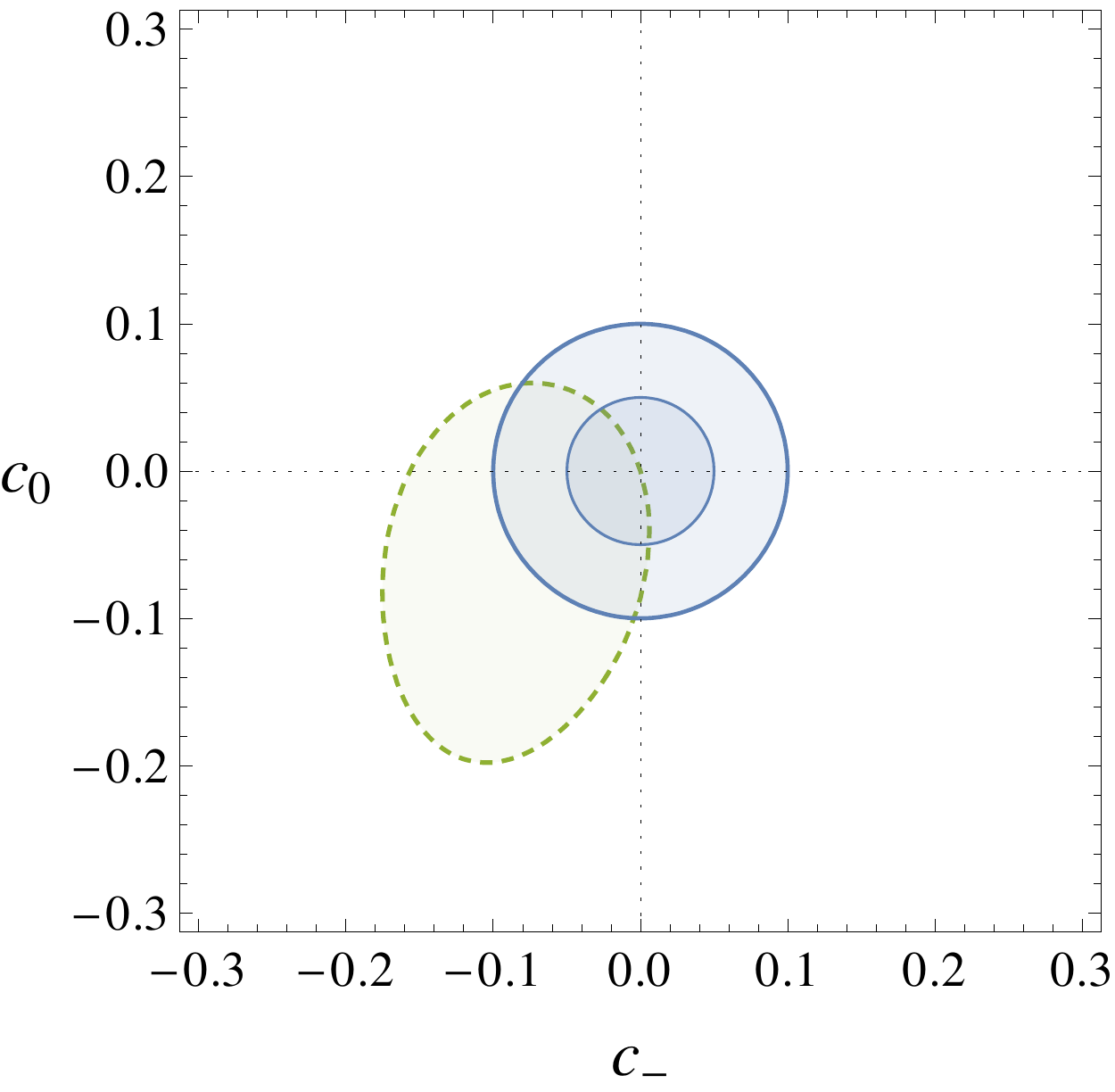}
\caption{Change of $\chi^2$ compared to the SM central value in the planes of pairs of coefficients that parameterize the size of unknown sub-leading non-perturbative QCD effects. Coefficients entering the $B \to K \ell^+ \ell^-$ amplitude at low and high $q^2$ (top); coefficients entering the $B \to K^* \ell^+ \ell^-$ amplitudes at low $q^2$ (bottom left) and high $q^2$ (bottom right). Along the dashed line the $\chi^2$ remains unchanged. In the shaded green region the $\chi^2$ is improved, with the solid lines indicating contours of $\Delta\chi^2=1,4,9$. The blue circles show our 1 and $2\sigma$ assumptions for the uncertainties on the shown parameters.}
\label{fig:SLplots}
\end{figure}
%%%%%%%%%%%%%%%%%%%%%%%%%%%%%%%%%%%% 

Any underestimation of a non-perturbative QCD effect (not related to form factors) should then manifest itself as a drastic reduction of the $\chi^2$ for a sizable value of one of the parameters, when treating them as completely free.
To investigate this question, we have constructed a $\chi^2$ function analogous to (\ref{eq:chi2}), but writing the central values $\vec O_\text{th}$ as functions of the parameters $a_i, b_i, c_i$ instead of the Wilson coefficients.

In fig.~\ref{fig:SLplots}, we show the reduction of the $\chi^2$ compared to our SM central value under variation of pairs of these parameters, while treating two of them at a time as free parameters and fixing all the others to 0. 
We show the cases of varying the coefficients entering the $B \to K \ell^+ \ell^-$ amplitude at low and high $q^2$ (top); the coefficients entering the $\lambda = -$ and $\lambda = 0$  $B \to K^* \ell^+ \ell^-$ helicity amplitudes at low $q^2$ (bottom left) and high $q^2$ (bottom right). Corrections to the $\lambda = +$ helicity amplitude are expected to be suppressed~\cite{Jager:2012uw} and we checked explicitly that they have a weak impact.
On the green dashed contours, the $\chi^2$ is the same as for the central value, so there is no improvement of the fit. In the green shaded area, the fit is improved, with the solid contours showing $\Delta \chi^2\equiv\chi^2-\chi^2_\text{SM}=1, 4, 9$, etc. In the unshaded region to the other side of the dashed contour, the fit is worsened compared to the central value. The blue circles show our 1 and $2\sigma$ assumptions for the uncertainties on the parameters in question, as discussed in section~\ref{sec:obs}. We stress that these assumptions have {\em not} been used as priors to determine the green contours. We make the following observations.
\begin{itemize}
\item The $\chi^2$ can be reduced by up to 4 when pushing the parameter $b_K$, parametrizing sub-leading corrections in $B\to K\mu^+\mu^-$ at low $q^2$, to the border of our estimated uncertainty. The fit does not improve significantly when changing the parameter $c_K$ from 0, i.e.\ when assuming large violations of quark-hadron duality in the global (integrated) high $q^2$ observables in $B\to K\mu^+\mu^-$, unless $b_K$ is shifted at the same time.
\item A simultaneous positive shift in the sub-leading corrections to the $\lambda=-$ and $0$ helicity amplitudes in $B\to K^*\mu^+\mu^-$ can significantly reduce the $\chi^2$ as well.
$\Delta \chi^2=9$ requires a shift in both parameters that is four times larger than our error estimate.
\item Corrections to quark-hadron duality in the global high $q^2$ observables in $B\to K^*\mu^+\mu^-$ do not lead to any significant reduction of the $\chi^2$.
\end{itemize}
We conclude that the agreement of the data with the predictions cannot be improved by assuming (unexpectedly) large violations of quark-hadron duality in integrated observables at high $q^2$ alone, while sizable corrections to $B\to K\mu^+\mu^-$ and $B\to K^*\mu^+\mu^-$ at low $q^2$ could improve the agreement with the data.
We stress however that fig.~\ref{fig:SLplots} should not be misinterpreted as a determination of the size of subleading QCD effects from the data. Indeed, the regions where the $\chi^2$ is significantly reduced correspond to values that are larger than any known hadronic effect.

We will see in section~\ref{sec:1WC} that a good fit to the data can be obtained assuming a large negative NP contribution to the Wilson coefficient $C_9$. We find it instructive to consider the size of the sub-leading parameters that would make them ``mimic'' a NP effect. Experimentally, it would be difficult to distinguish between the cases  $i)$ where $C_9 = C_9^\text{SM} + \Delta_9$ and all $a_i= b_i=c_i=0$ or $ii)$ where $C_9 = C_9^\text{SM}$ as well as
\begin{align}
a_K &\approx 0.25\,\Delta_9 \,,&
c_K &\approx 0.25\,\Delta_9 \,,\\
b_- &\approx -0.6\,\Delta_9 \,,&
c_- &\approx 0.25\,\Delta_9 \,,\\
a_0 &\approx -2\,\Delta_9 \,,&
c_0 &\approx 0.25\,\Delta_9 \,,
\end{align}
and all other $a_i, b_i,c_i$ equal to zero.
This pattern of effects is indeed similar to  what is seen in fig.~\ref{fig:SLplots}. Distinguishing such a scenario from a NP effect is straightforward if the NP effect is not lepton-flavour universal. If it is lepton-flavour universal, a correlated analysis of exclusive and inclusive observables, of the $q^2$ dependence, and of consistency relations among observables valid in the SM (see e.g.~\cite{Mandal:2014kma}) could help to disentangle QCD and NP.

\subsubsection{Underestimated parametric uncertainties?} \label{sec:Vcb}

While the angular observables in $B\to K^*\mu^+\mu^-$ are almost free from parametric uncertainties\footnote{%
By ``parametric'' here we refer to uncertainties that are not due to the form factors or other non-perturbative QCD effects.}, the apparent systematic suppression of branching ratios could also be due to an underestimated overall parametric uncertainty. The uncertainties of the $B_{u,d,s}$ meson lifetimes quoted by the PDG \cite{Agashe:2014kda} are well below $1\%$ and are therefore very unlikely to be responsible. The dominant parametric uncertainty is the CKM factor $|V_{tb}V_{ts}^*|^2$ to which all branching ratios are proportional and which itself is dominated by the uncertainty of the measurement of $|V_{cb}|$. The relative uncertainty of all $b\to s$ branching ratios due to $|V_{cb}|$ is twice the relative uncertainty of $|V_{cb}|$. In our numerical analysis, we use
\begin{equation}
 |V_{cb}| = (4.09 \pm 0.10) \times 10^{-2},
\label{eq:Vcb}
\end{equation}
which leads to an uncertainty of $4.9\%$ on the branching ratios.
In fact there is a long standing tension between determinations of $|V_{cb}|$ from inclusive and exclusive decays. The PDG \cite{Agashe:2014kda}  quotes
\begin{align}
 |V_{cb}|^\text{PDG}_\text{incl.} &= (4.22 \pm 0.07) \times 10^{-2},
 &
 |V_{cb}|^\text{PDG}_\text{excl.} &= (3.95 \pm 0.08) \times 10^{-2},
 \label{eq:VcbPDG}
\end{align}
which are at a $2.5\sigma$ tension with each other.
Choosing the inclusive value instead of (\ref{eq:Vcb}) would increase the central values of all our branching ratios by $6.5\%$ and would worsen the agreement with the data.
Choosing the exclusive value instead would lead to a reduction of the branching ratios by $6.7\%$.

To see whether this has an impact on the significance of the tensions, we multiply all branching ratios by a scale factor $\eta_\text{BR}$ and fit this scale factor to the data. We find $\eta_\text{BR}=0.79\pm0.08$, i.e. a $21\%$ reduction of the branching ratios with respect to our central values is preferred. 
The $\chi^2$ is improved by $7.0$ with respect to the SM.
The obtained central value for $\eta_\text{BR}$ would correspond to $|V_{cb}| 
\simeq 3.6 \times 10^{-2}$, which is in tension with both the inclusive and 
exclusive determinations.

We conclude that underestimated parametric uncertainties are unlikely to be responsible for the observed tensions in the branching ratio measurements.
Needless to say, the angular observables and $R_K$ would be unaffected by a shift in $|V_{cb}|$ anyway.

\subsubsection{Underestimated form factor uncertainties?}

The tensions between data and SM predictions could also be due to underestimated uncertainties in the form factor predictions from LCSR, lattice, or both.
A first relevant observation in this respect is that the tensions in table~\ref{tab:largepulls} include observables in decays involving $B\to K$, $B\to K^*$, and $B_s\to \phi$ transitions, both at low $q^2$ (where LCSR calculations are valid) and at high $q^2$ (where the lattice predictions are valid).
Explaining all of them would imply underestimated uncertainties in several completely independent theoretical form factor determinations.

In the case of $B\to K\mu^+\mu^-$ and $B_s\to \phi\mu^+\mu^-$, tensions are present only in branching ratios, which seem to be systematically below the SM predictions. This could be straightforwardly explained if the form factor predictions were systematically too high.
Note that the largest tensions in the $B \to K \mu^+\mu^-$ branching ratios appear in the neutral mode. The branching ratio of the charged mode, $B^+ \to K^+ \mu^+\mu^-$, is measured with considerably smaller statistical uncertainty and agrees better with the SM predictions (see Appendix~\ref{sec:obstables}). Nevertheless, also the charged mode seems to be systematically below the SM prediction and would profit from a reduction of the form factors.

%%%%%%%%%%%%%%%%%%%%%%%%%%%%%%%%%%%%
\begin{figure}[t]
\centering
\includegraphics[width=0.43\textwidth]{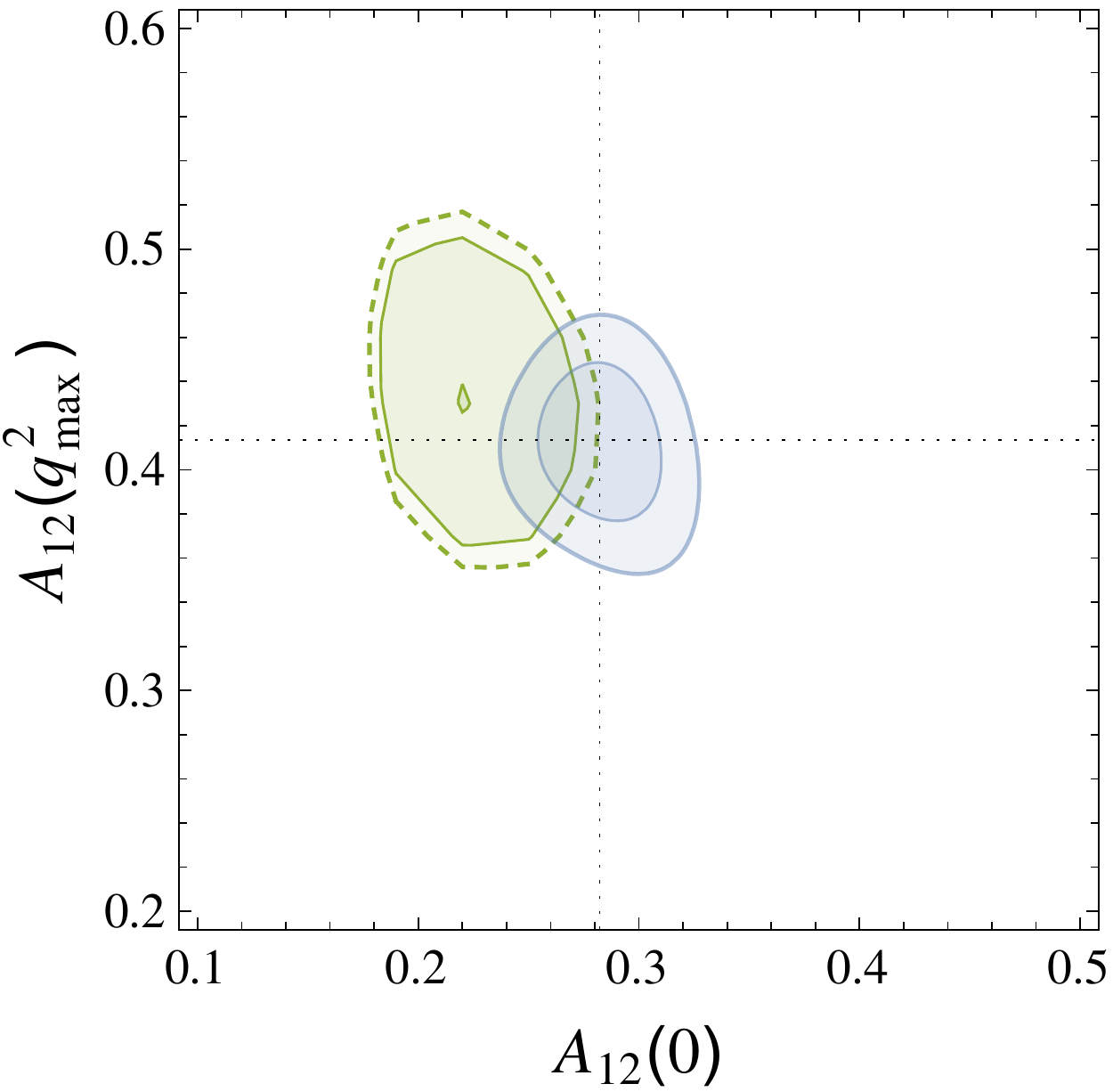}
\caption{Change of $\chi^2$ compared to the SM central value when changing the central value of the form factor $A_{12}$ at minimal or maximal $q^2$, while fixing the central values of all other form factors to their nominal values. Colours are as in fig.~\ref{fig:SLplots}.
}
\label{fig:FFplots}
\end{figure}
%%%%%%%%%%%%%%%%%%%%%%%%%%%%%%%%%%%% 

The case of $B\to K^*\mu^+\mu^-$ is less trivial due to the tensions in angular observables, which cannot simply be due to an overall rescaling of the form factors. To investigate this case, we have parametrized all seven $B\to K^*$ form factors by a two-parameter $z$ expansion\footnote{%
For our global numerical analysis, we use a three-parameter $z$ expansion as in \cite{Straub:2015ica}. The two-parameter expansion is only used in this case for simplicity. Note that two of the 14 parameters are redundant due to two exact kinematical relations at $q^2=0$.}
and constructed a $\chi^2$ function analogous to (\ref{eq:chi2}), but writing the central values $\vec O_\text{th}$ as functions of the 12 $z$ expansion parameters instead of the Wilson coefficients.
Varying the expansion parameters, we have found that the most significant 
shift, i.e.\ preference for a non-standard value, is obtained by modifying 
the form factor\footnote{
Here we use the transversity basis of form factors, cf.~\cite{Horgan:2013hoa}.} $A_{12}$.
In fig.~\ref{fig:FFplots}, we show the improvement in the $\chi^2$ obtained when changing the $A_{12}$ form factor, while fixing all the other form factors to their central values.
Instead of the two $z$ expansion coefficients, we present it in terms of the values of the form factor at the borders of the kinematical region, 0 and $q^2_\text{max}=(m_B-m_{K^*}^2)$.
The colours are analogous to fig.~\ref{fig:SLplots}.
We observe that an improvement of $\Delta \chi^2\sim 4$ can be obtained if the value at $q^2=0$ is significantly lower than what is obtained from LCSR.
This improvement is quite limited compared to the improvement obtained in the presence of NP discussed below or in the presence of large non-form factor corrections discussed above.

Finally, an important observation in the case of $B\to K^*\mu^+\mu^-$ angular 
observables is that the tensions are only present at low $q^2$, where the seven 
form factors can be expressed in terms of two independent ``soft'' form factors 
up to power corrections of naive order $\Lambda_\text{QCD}/m_b$. It is then 
possible to construct angular observables that do not depend on the soft form 
factors, but only on the power corrections \cite{Descotes-Genon:2013vna}. The 
tensions can then be seen by estimating the power corrections by dimensional 
analysis \cite{Descotes-Genon:2014uoa}. This shows that an explanation of the 
tensions by underestimated form factor uncertainties would imply that 
the values of the power corrections are very different from what LCSR 
calculations predict for them.

\subsection{New physics in a single Wilson coefficient} \label{sec:1WC}

We now investigate whether new physics could account for the tension of the data with the SM predictions.
We start by discussing the preferred ranges for individual Wilson coefficients assuming our nominal size of hadronic uncertainties. We determine the $1\sigma$ ($2\sigma$) ranges by computing $\Delta\chi^2=1~(4)$ while fixing all the other coefficients to their SM values. We also set the imaginary part of the respective coefficient to 0. 
In addition to the Wilson coefficients $C_{7,9,10}^{(\prime)}$, we also consider the case where the NP contributions to $C_9^{(\prime)}$ and $C_{10}^{(\prime)}$ are equal up to a sign, since this pattern of effects is generated by $SU(2)_L$-invariant four fermion operators in the dimension-6 SM effective theory.

\begin{table}[tb]
\renewcommand{\arraystretch}{1.5}
\centering
\begin{tabular}{cccccc}
\hline\hline
Coeff. & best fit & $1\sigma$ & $2\sigma$ & $\chi^2_\text{SM}-\chi^2_\text{b.f.}$ & pull \\
\hline
$C_7^\text{NP}$ & $-0.04$ & $[-0.07,-0.01]$ & $[-0.10,0.02]$ & $2.0$ & $1.4$ \\ 
 $C_7'$ & $0.01$ & $[-0.04,0.07]$ & $[-0.10,0.12]$ & $0.1$ & $0.2$ \\ 
 $C_9^\text{NP}$ & $-1.07$ & $[-1.32,-0.81]$ & $[-1.54,-0.53]$ & $13.7$ & $3.7$ \\ 
 $C_9'$ & $0.21$ & $[-0.04,0.46]$ & $[-0.29,0.70]$ & $0.7$ & $0.8$ \\ 
 $C_{10}^\text{NP}$ & $0.50$ & $[0.24,0.78]$ & $[-0.01,1.08]$ & $3.9$ & $2.0$ 
\\ 
 $C_{10}'$ & $-0.16$ & $[-0.34,0.02]$ & $[-0.52,0.21]$ & $0.8$ & $0.9$ \\ 
 $C_9^\text{NP}=C_{10}^\text{NP}$ & $-0.22$ & $[-0.44,0.03]$ & $[-0.64,0.33]$ & $0.8$ & $0.9$ \\ 
 $C_9^\text{NP}=-C_{10}^\text{NP}$ & $-0.53$ & $[-0.71,-0.35]$ & $[-0.91,-0.18]$ & $9.8$ & $3.1$ \\ 
 $C_9'=C_{10}'$ & $-0.10$ & $[-0.36,0.17]$ & $[-0.64,0.43]$ & $0.1$ & $0.4$ \\ 
 $C_9'=-C_{10}'$ & $0.11$ & $[-0.01,0.22]$ & $[-0.12,0.33]$ & $0.9$ & $0.9$ \\ 
\hline\hline
\end{tabular}
\caption{Constraints on individual Wilson coefficients, assuming them to be 
real. The pull in the last column is defined as $\sqrt{\chi^2_\text{SM} - 
\chi^2_\text{b.f.}}$.}
\label{tab:1Dbounds}
\end{table}

Our results are shown in table~\ref{tab:1Dbounds}.
We summarize the most important points.
\begin{itemize}
 \item A negative NP contribution to $C_9$, approximately $-25\%$ of $C_9^\text{SM}$, leads to a sizable decrease in the $\chi^2$. The best fit point corresponds to a $p$-value of $11.3\%$, compared to $2.1\%$ for the SM.
This was already found in fits of low-$q^2$ angular observables only \cite{Descotes-Genon:2013wba} and in global fits not including data released this year \cite{Altmannshofer:2013foa,Beaujean:2013soa,Descotes-Genon:2014uoa,Hurth:2013ssa}, as well as in a recent fit to a subset of the available data \cite{Hurth:2014vma}.
We find that the significance of this solution has increased substantially. 
This is due in part to the reduced theory uncertainties, in particular the form factors, as well as due to the new measurements by LHCb.
\item A significant improvement is also obtained in the $SU(2)_L$ invariant direction $C_9^\text{NP}=-C_{10}^\text{NP}$, corresponding to an operator with left-handed muons.
\item A positive NP contribution to $C_{10}$ alone can also improve the fit, although to a lesser extent.
\item NP contributions to individual right-handed Wilson coefficients hardly lead to improvements of the fit.
\end{itemize}

While table~\ref{tab:1Dbounds} assumed the Wilson coefficients to be real, i.e.\ aligned in phase with the SM, in general the NP contributions to the Wilson coefficients are complex numbers. Since measurements in semi-leptonic decays are currently restricted to CP-averaged observables or direct CP asymmetries that are suppressed by small strong phases\footnote{The only exception is the measurement of the T-odd CP asymmetry $A_9$ by LHCb~\cite{Aaij:2013iag} and CDF~\cite{CDFupdate} that, however, still has sizable experimental uncertainties.}, the constraints on the imaginary parts are generally weaker than on the real parts, since they do not interfere with the SM contribution.

\begin{figure}[tbp]
\centering
\includegraphics[width=0.43\textwidth]{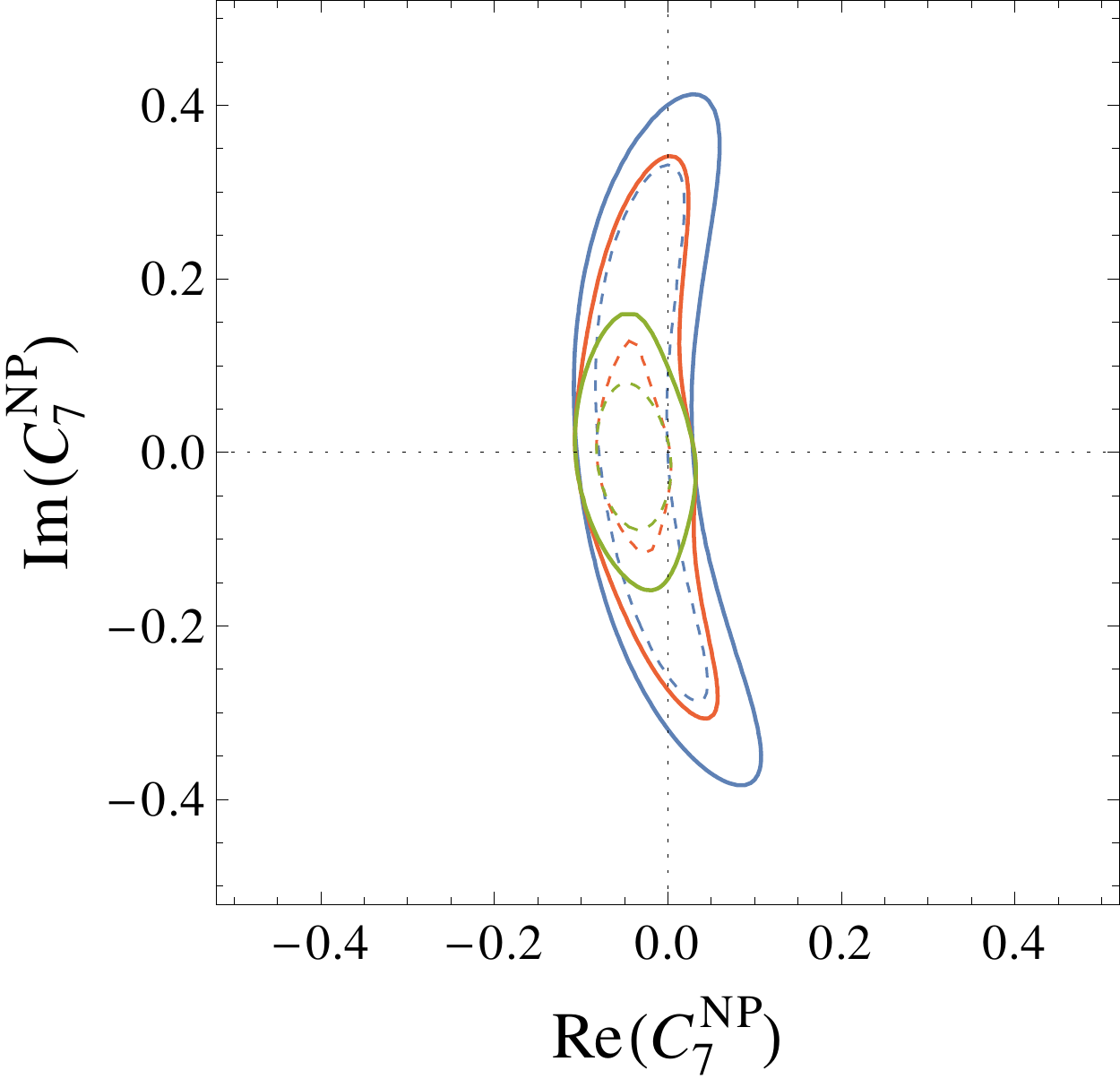}% ~~~~%
\caption{Allowed region in the Re$(C_7^\text{NP})$-Im$(C_7^\text{NP})$ plane.
The blue contours correspond to the 1 and $2\sigma$ best fit region without including the $A_\text{CP}(B\to K^* \gamma)$ measurement. The red (green) contours show the impact of including $A_\text{CP}(B\to K^* \gamma)$ with a relative theoretical uncertainty of 50\% (25\%).}
\label{fig:2DACP}
\end{figure}

An interesting special case is the direct CP asymmetry in $B\to K^*\gamma$. As discussed in section~\ref{sec:ACPBKsg}, this observable is precisely measured and very sensitive to the imaginary part of $C_7$, but we do not include it in our default $\chi^2$ since it is proportional to a strong phase that is afflicted with a considerable uncertainty. 
In fig.~\ref{fig:2DACP}, we show how the allowed region for the NP contribution to $C_7$ would change by including this observable.
The red (green) contours correspond to the 1 and $2\sigma$ regions ($\Delta \chi^2=2.3$ and $6$ while fixing all other coefficients to their SM values) allowed by the global fit including $A_\text{CP}(B^0\to K^{*0}\gamma)$ with a relative uncertainty of 50\% (25\%), while the blue contours correspond to the fit without the CP asymmetry. We observe that the constraint on the imaginary part of $C_7$ improves by a factor of $\sim 2$ even with our conservative estimate for the theory error.
In any case, a more detailed study of the theoretical uncertainties in this observable and a combined analysis with other observables sensitive to $C_7$ -- e.g.\ $B\to K^*e^+e^-$ at very low $q^2$ \cite{Aaij:2015dea} or $B_s\to\phi\gamma$ \cite{LHCb:2012ab,Dutta:2014sxo} -- would be interesting and we leave this to a future study.

The global constraints in the complex planes of all Wilson coefficients are shown in fig.~\ref{fig:2Dconstraints1} of appendix~\ref{sec:2Dplots}.

\subsection{Constraints on pairs of Wilson coefficients}  \label{sec:2WC}

\begin{figure}[tb]
\includegraphics[width=0.45\textwidth]{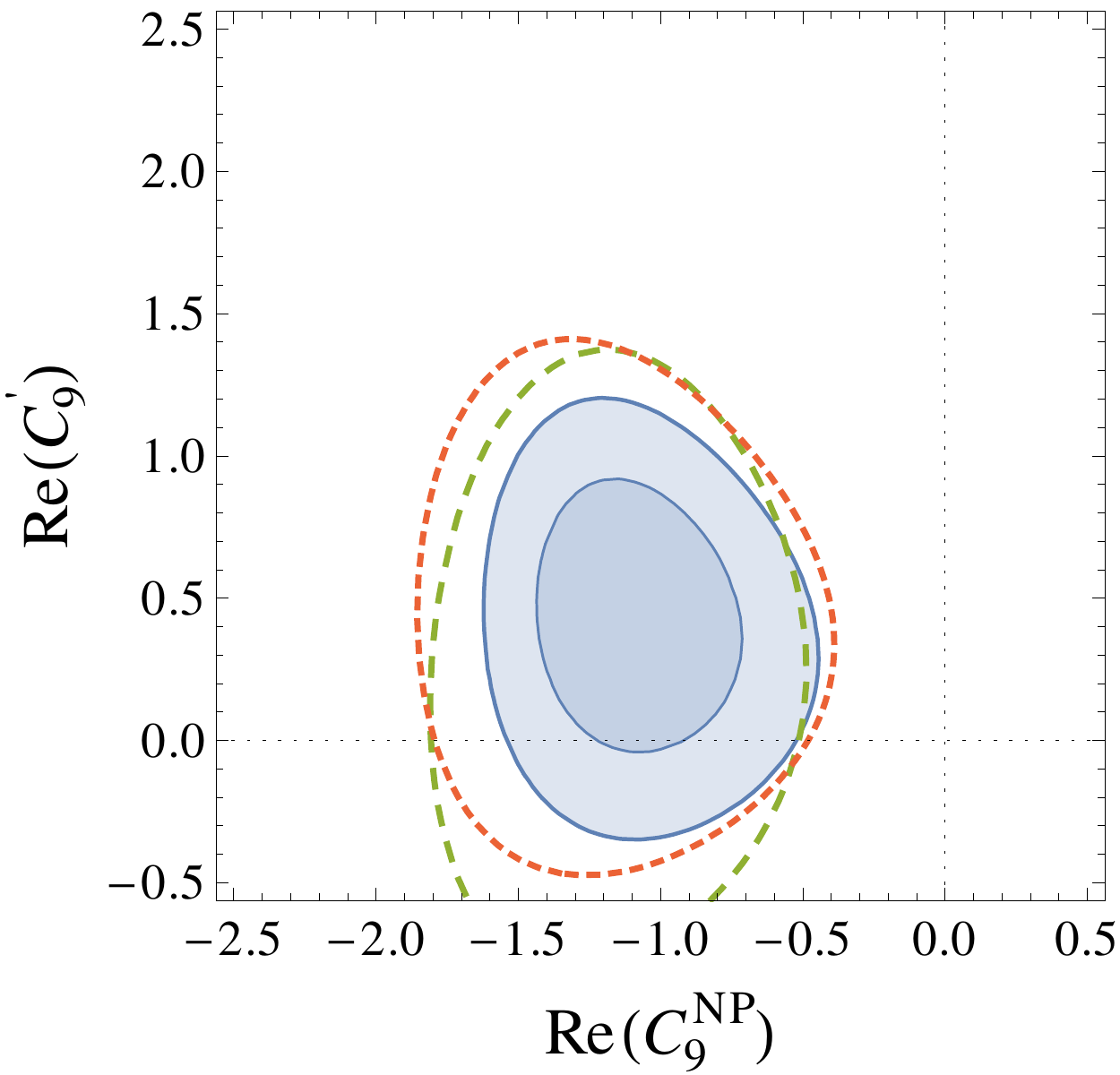}%
\hspace{0.05\textwidth}%
\includegraphics[width=0.45\textwidth]{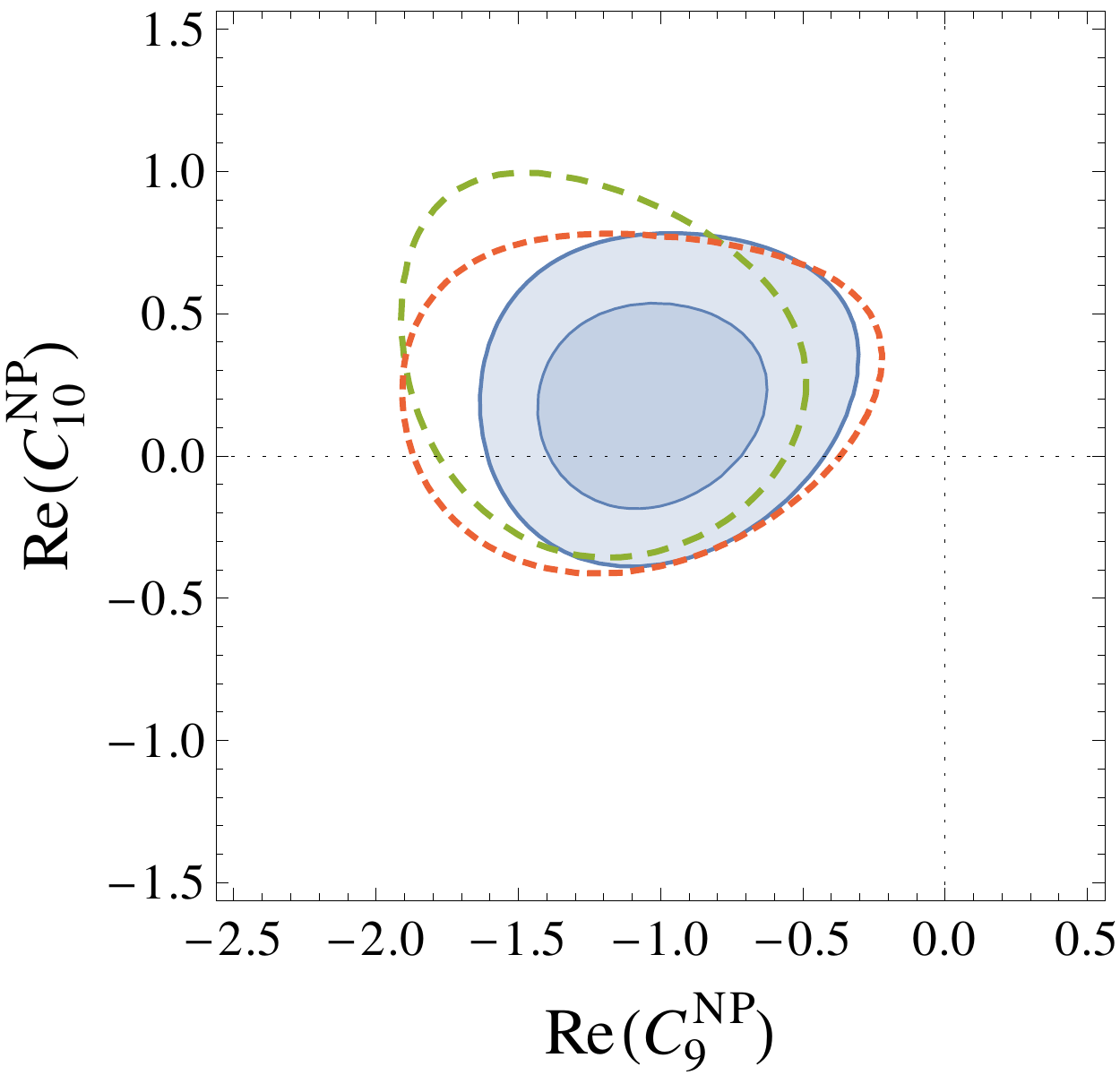}%
\caption{Allowed regions in the Re$(C_9^\text{NP})$-Re$(C_9')$ plane (left) and the Re$(C_9^\text{NP})$-Re$(C_{10}^\text{NP})$ plane (right). The blue contours correspond to the 1 and $2\sigma$ best fit regions. The green and red short-dashed contours correspond to the $2\sigma$ regions in scenarios with doubled form factor uncertainties and doubled uncertainties from sub-leading non-factorizable corrections, respectively.}
\label{fig:2D-zoom}
\end{figure} 

We proceed by analysing the constraints in scenarios where two Wilson coefficients are allowed to differ from their SM values.  
In this section we exemplarily allow for real NP in either $C_9$ and $C_9'$ or $C_9$ and $C_{10}$.
With our nominal values for the theory uncertainties, the best fit values for the Wilson coefficients and the corresponding $\Delta\chi^2$ read in the two cases
\begin{align}
 (C_9^\text{NP})_\text{b.f.} &= -1.10 \,, &(C_9')_\text{b.f.} &= +0.45 ,&&& \chi^2_\text{SM} - \chi^2_\text{b.f.} &= 15.6 \,, \\%[8pt]   
 (C_9^\text{NP})_\text{b.f.} &= -1.06 \,, &(C_{10}^\text{NP})_\text{b.f.} &= +0.16 \,, &&& \chi^2_\text{SM} - \chi^2_\text{b.f.} &= 14.2 \,.
\end{align}
The best fit points correspond to $p$-values of $12.4\%$ and $10.6\%$, respectively. This is comparable to the $11.3\%$ obtained in section~\ref{sec:1WC} in the scenario with new physics only in $C_9$.
In fig.~\ref{fig:2D-zoom}, we show the allowed regions in the Re$(C_9^\text{NP})$-Re$(C_9')$ and Re$(C_9^\text{NP})$-Re$(C_{10}^\text{NP})$ planes. The blue contours correspond to the 1 and $2\sigma$ regions ($\Delta \chi^2=2.3$ and $6$ while fixing all other coefficients to their SM values) allowed by the global fit. In addition, we also show the $2\sigma$ allowed regions for two scenarios with inflated theory uncertainties. For the green short-dashed contours, we have {\em doubled} all the form factor uncertainties. For the red short-dashed contours, we have doubled all the hadronic uncertainties {\em not} related to form factors, i.e.\ the ones that are parametrized as in (\ref{eq:SLK}) and (\ref{eq:SL}). We observe that the negative value preferred for $C_9^\text{NP}$ is above the $2\sigma$ level even for these conservative assumptions. We also observe that $C_9'$ and $C_{10}^\text{NP}$ are preferentially positive, 
although they deviate from 0 less significantly than $C_9^\text{NP}$.
The corresponding plots for all interesting combinations of real Wilson coefficients are collected in fig.~\ref{fig:2Dconstraints2} of appendix~\ref{sec:2Dplots}, together with the $\Delta \chi^2$ values of the corresponding best fit points.

\begin{figure}[tb]
\includegraphics[width=0.45\textwidth]{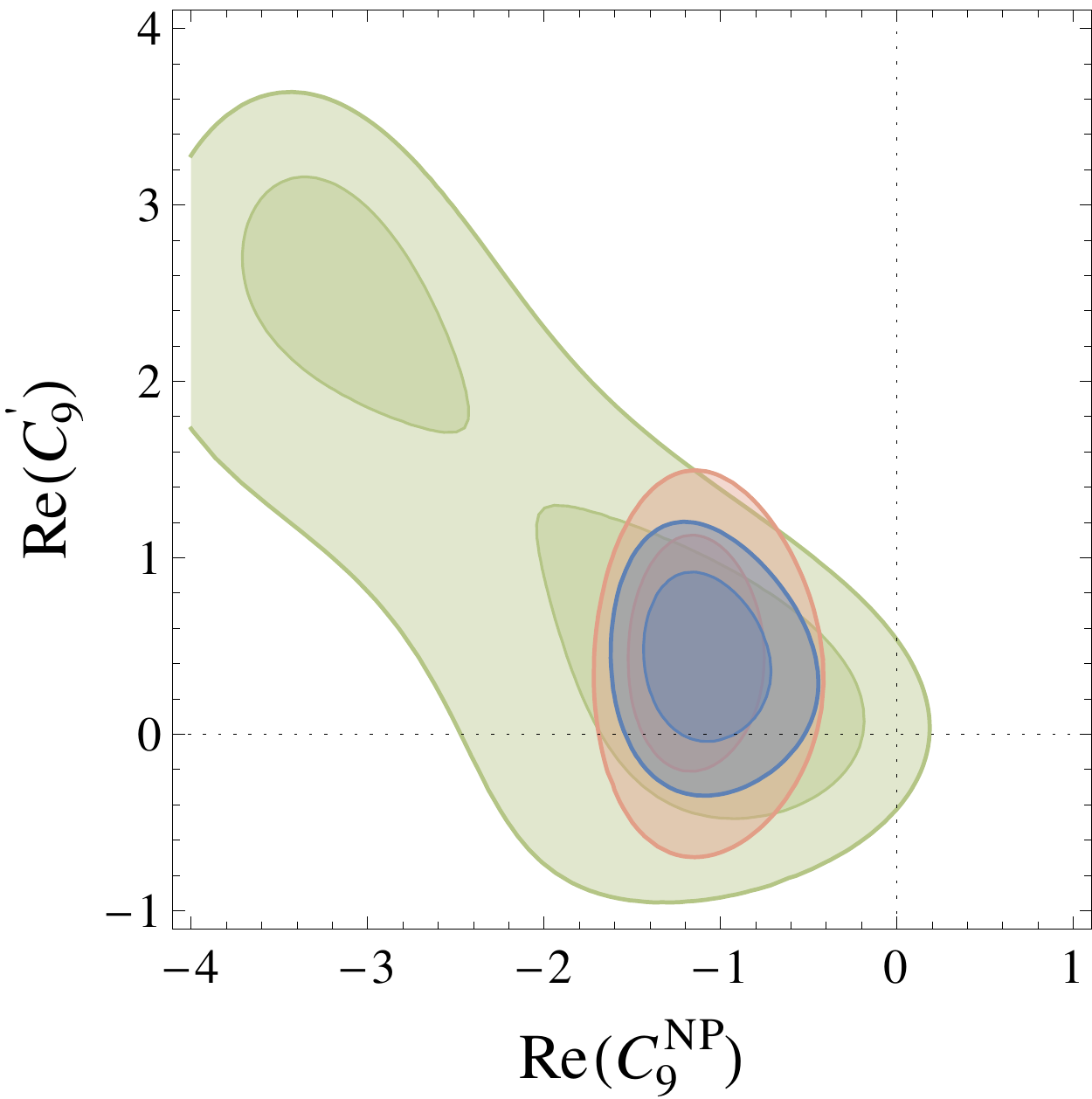}%
\hspace{0.05\textwidth}%
\includegraphics[width=0.45\textwidth]{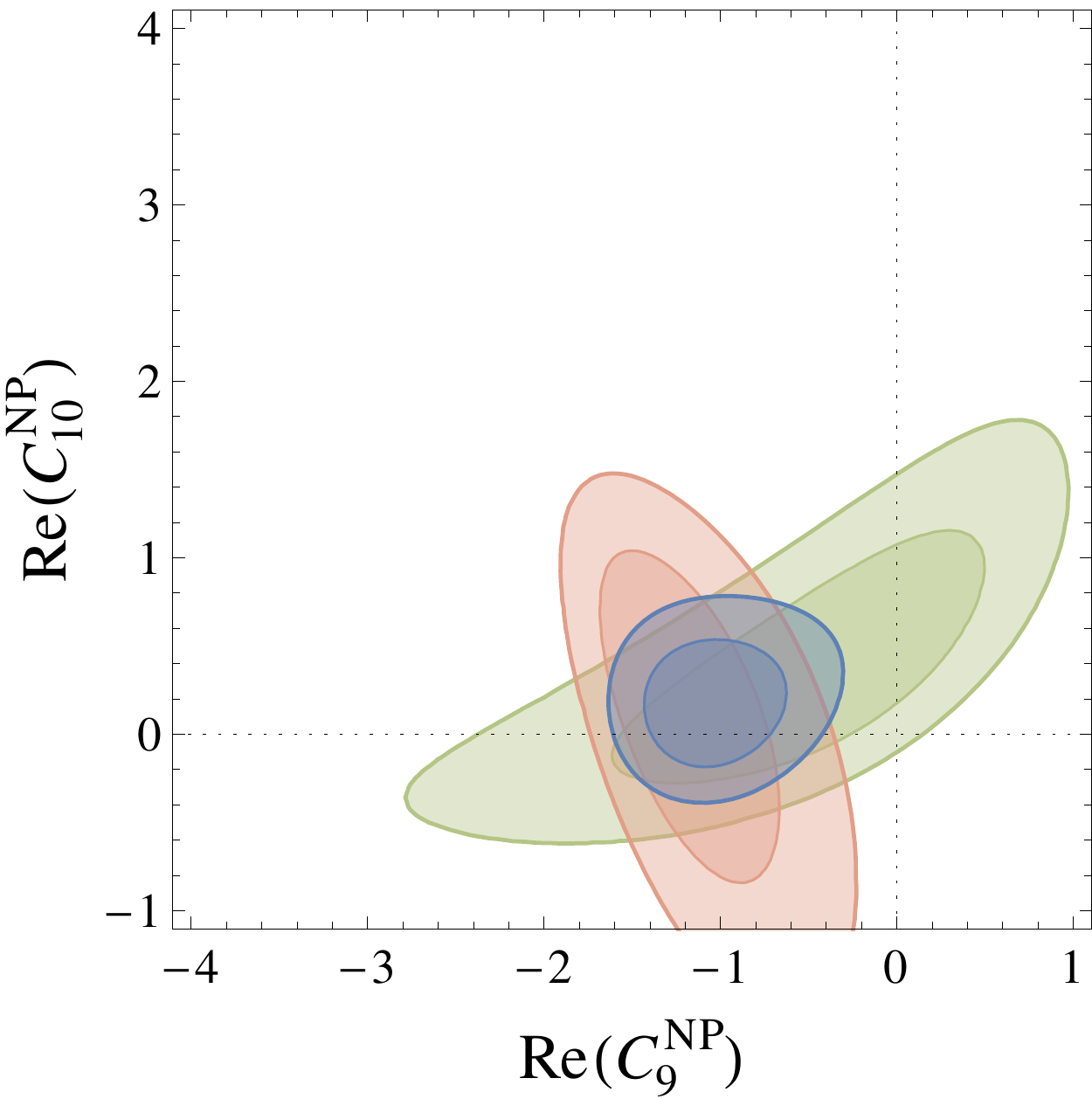}%
\caption{Allowed regions in the Re$(C_9^\text{NP})$-Re$(C_9')$ plane (left) and the Re$(C_9^\text{NP})$-Re$(C_{10}^\text{NP})$ plane (right). The blue contours correspond to the 1 and $2\sigma$ best fit regions from the global fit. The green and red contours correspond to the 1 and $2\sigma$ regions if only branching ratio data or only data on $B \to K^* \mu^+\mu^-$ angular observables is taken into account.}
\label{fig:obs}
\end{figure}

It is also interesting to investigate which observables drive the tensions. In fig.~\ref{fig:obs}, we compare the global constraints in the Re$(C_9^\text{NP})$-Re$(C_9')$ and Re$(C_9^\text{NP})$-Re$(C_{10}^\text{NP})$ planes to the constraints one gets only using branching ratios (green) or only using $B\to K^*\mu^+\mu^-$ angular observables (red). We observe that the angular observables strongly prefer a negative $C_9$ but are not very sensitive to $C_9'$ or $C_{10}$. The branching ratio constraints have an approximate flat direction $C_9^\text{NP}\sim -C_9'$ and show a preference for $C_{10}^\text{NP}>0$ in particular if $C_{9}^\text{NP} > 0$. In fact, from branching ratios alone, one could get a good fit to the data with SM-like $C_9$ and $C_{10}^\text{NP}>0$.

\subsection{Minimal Flavour Violation}

In models with Constrained Minimal Flavour Violation (CMFV) \cite{Buras:2000dm}, only the Wilson coefficients $C_7$, $C_9$ and $C_{10}$ receive new physics contributions and they are aligned in phase with the SM, i.e.\ real in our convention. Since these NP contributions interfere with the SM contributions, they are the most strongly constrained ones at present. In fact, in this simple case, it is a reasonable approximation to expand the $\chi^2$ to quadratic order around the best-fit point,
\begin{equation}
\chi^2_\text{CMFV}(\vec C^\text{NP}) \approx
\chi^2_\text{b.f., CMFV}+
\left(\vec C^\text{NP}- \vec C^\text{NP}_\text{b.f.}\right)^T
C_\text{CMFV}^{-1}
\left(\vec C^\text{NP}- \vec C^\text{NP}_\text{b.f.}\right)
\label{eq:chi2CMFV}
\end{equation}
where the best fit has $\chi^2_\text{b.f., CMFV} = 102.4$. The covariance matrix is given in terms of the variances $\sigma_i$ and correlations $\rho_{ij}$ as $C_\text{CMFV}^{ij} = \sigma_i\sigma_j\rho_{ij}$ (no sum). The central values and variances of the Wilson coefficients read
\begin{equation}
\vec C^\text{NP} =
\begin{pmatrix}
C_7^\text{NP} \\
C_9^\text{NP} \\
C_{10}^\text{NP}
\end{pmatrix}
=
\begin{pmatrix}
-0.017 \pm 0.030 \\
-1.02 \pm 0.27 \\
 0.16 \pm 0.24
\end{pmatrix}\,
\end{equation}
and the correlation matrix reads
\begin{equation}
 \begin{pmatrix}
 1 & -0.28 & 0.06 \\
 -0.28 & 1 & 0.06 \\
 0.06 & 0.06 & 1 \\
 \end{pmatrix}\,.
\end{equation}
The expression \eqref{eq:chi2CMFV} can be used to easily impose the combined fit constraints in phenomenological analyses of models satisfying CMFV.
For scenarios with non-standard CP violation or right-handed currents, it can be understood from figs.~\ref{fig:2Dconstraints1} and~\ref{fig:2Dconstraints2} that at present the constraints are not stringent enough to allow a quadratic expansion of the $\chi^2$ and we cannot provide a comparably simple expression in general.

\subsection{Testing lepton flavour universality}

So far, in our numerical analysis we have only considered the muonic $b \to s \mu^+\mu^-$ modes and the lepton flavour independent radiative $b \to s \gamma$ modes to probe the Wilson coefficients $C_7^{(\prime)}$, $C_9^{(') \mu}$ and $C_{10}^{(\prime) \mu}$, where the superscript $\mu$ indicates that in the semileptonic operators~(\ref{eq:O9}) and~(\ref{eq:O10}) only muons are considered. In this section we will extend our analysis and include also semileptonic operators that contain electrons. In particular, we will allow new physics in the Wilson coefficients $C_9^e$ and $C_{10}^e$ and confront them with the available data on $B \to K e^+e^-$ from LHCb~\cite{Aaij:2014ora} and $B \to X_s e^+e^-$ from BaBar~\cite{Lees:2013nxa}. 

As mentioned already in the introduction, the recent measurement of the ratio $R_K$ of $B\to K \mu^+\mu^-$ and $B \to K e^+e^-$ branching ratios in the $q^2$ bin $[1,6]$~GeV$^2$ by LHCb~\cite{Aaij:2014ora} shows a $2.6\sigma$ tension with the SM prediction
\begin{equation}
 R_K = \frac{\text{BR}(B \to K\mu^+\mu^-)_{[1,6]}}{\text{BR}(B \to Ke^+e^-)_{[1,6]}} = 0.745^{+0.090}_{-0.074} \pm 0.036 ~,~~~ R_K^\text{SM} \simeq 1.00 ~.
\end{equation}
The theoretical error of the SM prediction is completely negligible compared to the current experimental uncertainties.
The tension between the SM prediction and the experimental data is driven by the reduced $B\to K \mu^+\mu^-$ branching ratio, while the measured $B \to K e^+e^-$ branching ratio is in good agreement with the SM. In our extended global fit we do not use the $R_K$ measurement directly but instead include the $B \to K\mu^+\mu^-$ and $B \to Ke^+e^-$ branching rations separately, taking into account the correlations of their theory uncertainties. As the theory uncertainties of BR$(B \to K\mu^+\mu^-)$ and BR$(B \to Ke^+e^-)$ are essentially 100\% correlated, our approach is to a good approximation equivalent to using $R_K$.

\begin{figure}[tb]
\includegraphics[width=0.45\textwidth]{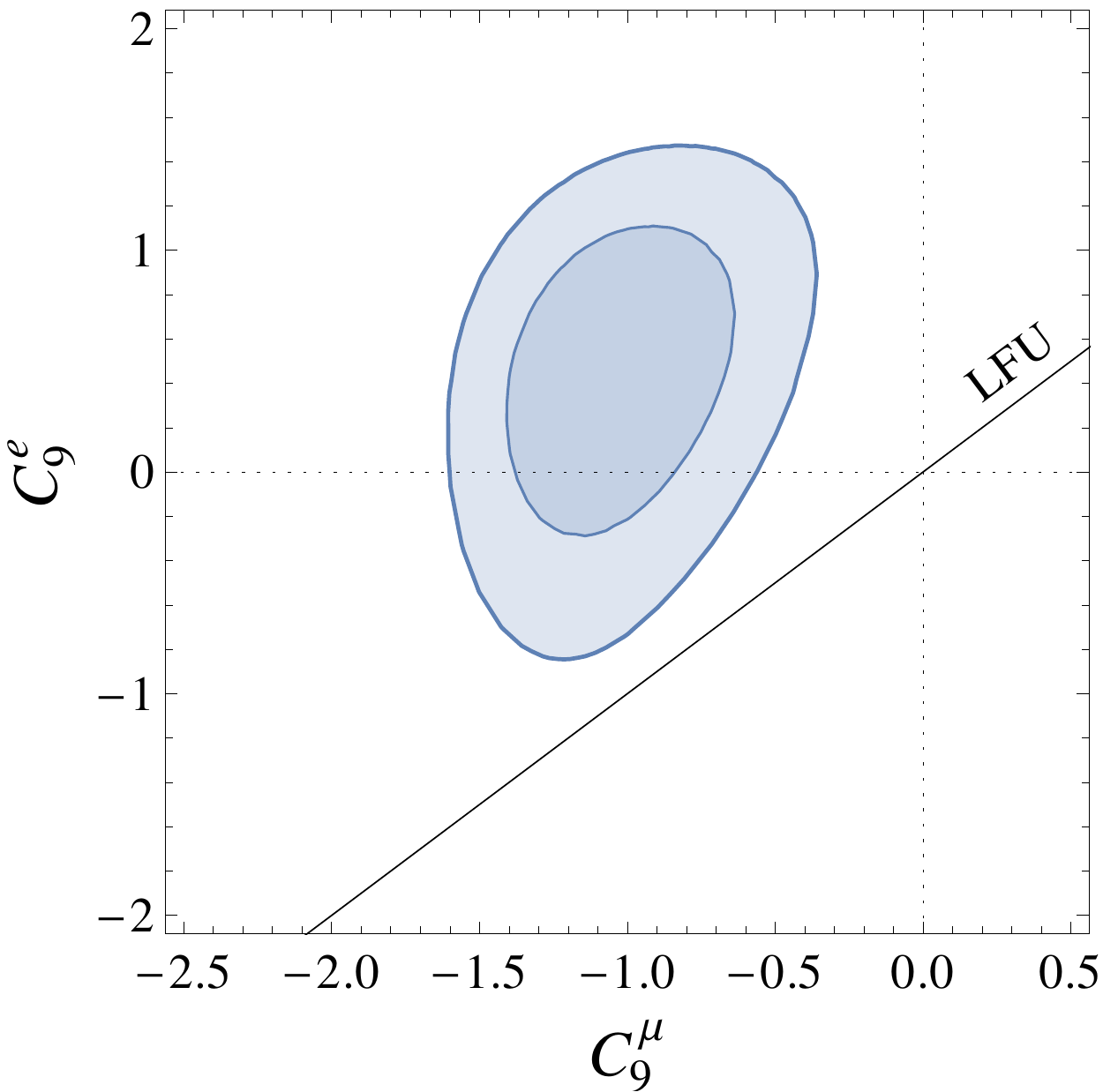}%
\hspace{0.05\textwidth}%
\includegraphics[width=0.46\textwidth]{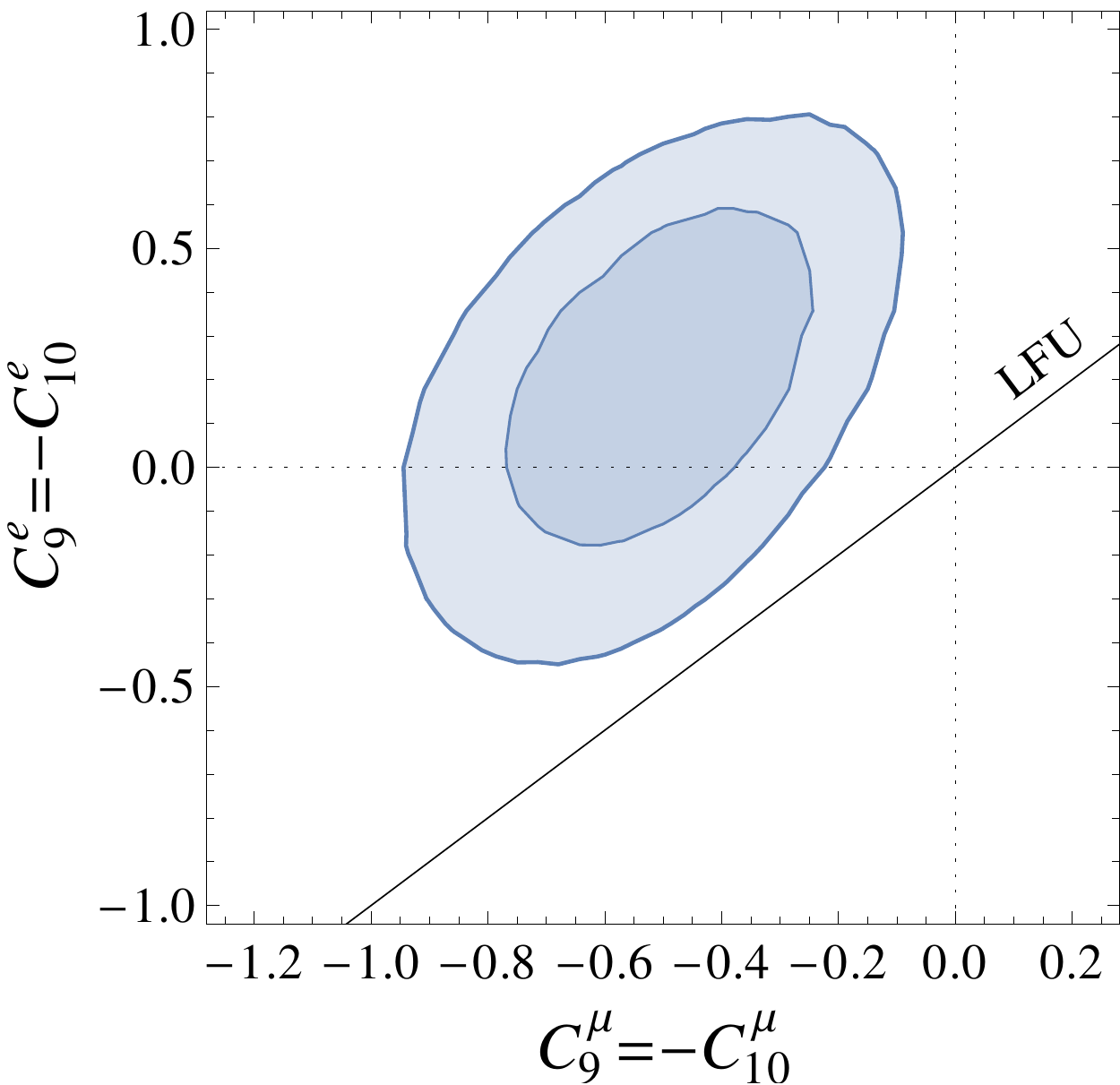}%
\caption{Allowed regions in the plane of new physics contributions to the Wilson coefficients $C_9^\mu$ vs. $C_9^e$ (left) and the plane of the $SU(2)_L$ invariant combinations of Wilson coefficients $C_9^\mu=-C_{10}^\mu$ vs. $C_9^e=-C_{10}^e$ (right). The blue contours correspond to the 1 and $2\sigma$ best fit regions. The diagonal line corresponds to lepton flavour universality.}
\label{fig:muvse-zoom}
\end{figure}

In fig.~\ref{fig:muvse-zoom} we show the result of two fits that allow for new physics in $C_9^\mu$ and $C_9^e$ (left plot) and new physics along the $SU(2)_L$ invariant directions $C_9^\mu = -C_{10}^\mu$ and $C_9^e = -C_{10}^e$. 
Recall that in section~\ref{sec:1WC} we found that new physics in these scenarios gives the by far best description of the experimental $b\to s \mu^+\mu^-$ data. As expected, we again find that a $C_9^\mu$ significantly smaller than in the SM is clearly preferred by the fits.
The best fit regions for $C_9^\mu$ and $C_9^\mu = - C_{10}^\mu$ approximately coincide with the regions found for $C_9$ and $C_9 = - C_{10}$ in section~\ref{sec:1WC}.
The Wilson coefficients $C_9^e$ and $C_9^e = - C_{10}^e$ on the other hand are perfectly consistent with the SM prediction. Lepton flavour universality, i.e. $C_9^\mu = C_9^e$ and $C_{10}^\mu = C_{10}^e$ as indicated by the diagonal line in the plots is clearly disfavoured by the data.
Our results are consistent with similar findings in recent fits to part of the available experimental data \cite{Ghosh:2014awa,Hurth:2014vma}.

\begin{table}[tb]
\renewcommand{\arraystretch}{1.5}
\centering
\begin{tabular}{ccccc}
\hline\hline
Observable & \multicolumn{4}{c}{Ratio of muon vs.\ electron mode} \\
\hline
 &$C_9^\text{NP}=-1.07$ & ~~$-1.10$~~~~  & ~~~$-0.53$~~~ & ~~~$-1.06$~~~ \\
 &$C_9'=0$ & $0.45$ & $0$ & $0$ \\
 &$C_{10}^\text{NP}=0$ & $0$ & $0.53$ & $0.16$ \\
\hline
$10^{7}~\frac{d\text{BR}}{dq^2}(\bar B^0\to\bar K^{*0}\ell^+\ell^-)_{[1,6]}$ & {$0.83$} & {$0.77$} & {$0.77$} & {$0.79$} \\
$10^{7}~\frac{d\text{BR}}{dq^2}(\bar B^0\to\bar K^{*0}\ell^+\ell^-)_{[15,19]}$ & {$0.78$} & {$0.72$} & {$0.75$} & {$0.74$} \\
$F_L(\bar B^0\to\bar K^{*0}\ell^+\ell^-)_{[1,6]}$ & {$0.93$} & {$0.90$} & {$0.98$} & {$0.93$} \\
$F_L(\bar B^0\to\bar K^{*0}\ell^+\ell^-)_{[15,19]}$ & {$1.00$} & {$0.97$} & {$1.00$} & {$1.00$} \\
$A_\text{FB}(\bar B^0\to\bar K^{*0}\ell^+\ell^-)_{[4,6]}$ & {\boldmath$0.33$} & {\boldmath$0.33$} & {$0.74$} & {\boldmath$0.35$} \\
$A_\text{FB}(\bar B^0\to\bar K^{*0}\ell^+\ell^-)_{[15,19]}$ & {$0.90$} & {$0.96$} & {$0.99$} & {$0.92$} \\
$S_5(\bar B^0\to\bar K^{*0}\ell^+\ell^-)_{[4,6]}$ & {$0.73$} & {$0.77$} & {$0.93$} & {$0.74$} \\
$S_5(\bar B^0\to\bar K^{*0}\ell^+\ell^-)_{[15,19]}$ & {$0.91$} & {$0.97$} & {$0.99$} & {$0.92$} \\
$10^{8}~\frac{d\text{BR}}{dq^2}(B^+\to K^+\ell^+\ell^-)_{[1,6]}$ & {$0.77$} & {$0.85$} & {$0.76$} & {$0.74$} \\
$10^{8}~\frac{d\text{BR}}{dq^2}(B^+\to K^+\ell^+\ell^-)_{[15,22]}$ & {$0.78$} & {$0.86$} & {$0.76$} & {$0.74$} \\
$10^{6}~\text{BR}(B\to X_s\ell^+\ell^-)_{[1,6]}$ & {$0.83$} & {$0.83$} & {$0.77$} & {$0.79$} \\
$10^{6}~\text{BR}(B\to X_s\ell^+\ell^-)_{[14.2,25]}$ & {$0.78$} & {$0.78$} & {$0.75$} & {$0.74$} \\
\hline\hline
\end{tabular}
\caption{Predictions for ratios of observables with muons vs.\ electrons for 
four different scenarios with NP only in one or two Wilson coefficients with 
muons. Ratios deviating from the SM prediction $1.00$ by more than 30\% are 
highlighted in boldface.}
\label{tab:LVU}
\end{table}

Working under the assumption that the electron modes are indeed SM like, we can make predictions for ratios of observables that test lepton flavour universality using the best fit regions for the muonic Wilson coefficients from our global fit.
We consider ratios of branching ratios of the exclusive $B\to K^* \ell^+\ell^-$ and $B\to K \ell^+\ell^-$ decays and the inclusive $B \to X_s \ell^+\ell^-$ decays, both at low and high $q^2$. Moreover, we also predict ratios of the $B \to K^* \ell \ell$ angular observables $F_L$, $A_\text{FB}$ and $S_5$ at low and high $q^2$. The results are shown in table~\ref{tab:LVU}. The four columns correspond to the following scenarios: 
\begin{itemize}
 \item new physics only in $C_9^\mu$;
 \item new physics in $C_9^\mu$ and $C_9^{\prime~\mu}$;
 \item new physics along the $SU(2)_L$ invariant direction $C_9^\mu = - C_{10}^\mu$;
 \item new physics independently in $C_9^\mu$ and $C_{10}^\mu$.
\end{itemize}
The Standard Model prediction for all the shown ratios is 1, with negligible 
uncertainties\footnote{We do not quote uncertainties in table~\ref{tab:LVU} 
since any significant deviation from 1 would constitute a clear sign of NP. 
However, it should be noted that for a fixed value of the NP contributions to 
the Wilson coefficients, there are non-zero uncertainties in the observables.}. 
In all scenarios all branching ratio ratios are predicted around $0.8$ both at 
low and high dimuon invariant mass. A similar ratio is seen for $S_5$ at low 
$q^2$. Only very small deviations from the SM are predicted for $S_5$ and 
$A_\text{FB}$ at high $q^2$ as well as $F_L$ at low and high 
$q^2$.\footnote{Note that at high $q^2$, $F_L$ is indeed to a large extent 
insensitive to new physics and largely determined by form factor 
ratios~\cite{Bobeth:2012vn,Hiller:2013cza}}
The most interesting observable turns out to be the ratio of the forward-backward asymmetries in $B\to K^* \mu^+\mu^-$ and $B\to K^* e^+e^-$ in the $q^2$ bin $[4,6]$~GeV
\begin{equation}
 R_{A_\text{FB}} \equiv \frac{A_\text{FB}(B \to K^* \mu^+\mu^-)_{[4,6]}}{A_\text{FB}(B \to K^* e^+e^-)_{[4,6]}} ~.
\end{equation}
Assuming that the electron mode is SM like, $R_{A_\text{FB}}$ is extremely sensitive to the value of $C_9^\mu$. For the considered values of $C_9^\mu$ it deviates drastically from the SM prediction and a precise measurement would even allow to distinguish between the considered scenarios.

%%%%%%%%%%%%%%%%%%%%%%%%%%%%%%%%%%%%%%%%%%%%%%%%%%
\section{Constraints on new physics models}\label{sec:np}
%%%%%%%%%%%%%%%%%%%%%%%%%%%%%%%%%%%%%%%%%%%%%%%%%%

The results from the model-independent fit of the Wilson coefficients in the effective Hamiltonian can be interpreted in the context of new physics models. Here we discuss implications for the minimal supersymmetric standard model (MSSM) and models that contain massive $Z^\prime$ gauge bosons with flavour-changing couplings.

%%%%%%%%%%%%%%%%%%%%%%%%%%%%%%%%%%%%%%%%%%%%%%%%%%
\subsection{SUSY models with generic flavour violation}

Recently, the $B \to K^* \mu^+\mu^-$ decay has been studied in MSSM scenarios that do not contain sources of flavour violation beyond the CKM matrix~\cite{Mahmoudi:2014mja}. We do not find sizable SUSY contributions to $C_9$ and $C_{10}$ in such scenarios. In the following, we will therefore allow for generic flavour violation.

Experimental data on flavour-changing neutral current processes lead to strong constraints on new sources of flavour violation that can be present in the MSSM~\cite{Gabbiani:1996hi,Altmannshofer:2009ne}.
In particular, the experimental information on rare $b \to s \mu^+\mu^-$ decays can be used to put constraints on flavour-violating trilinear couplings in the up squark sector, that are only poorly constrained otherwise~\cite{Cho:1996we,Lunghi:1999uk,Buchalla:2000sk,Altmannshofer:2009ma,Behring:2012mv}.
In principle, the general MSSM also allows for lepton-flavour non-universality effects and we will comment to which extend the $R_K$ measurement can be accommodated. 
 
%%%%%%%%%%%%%%%%%%%%%%%%%%%%%%%%%%%%
\begin{figure}[t]
\centering
\includegraphics[width=0.3\textwidth]{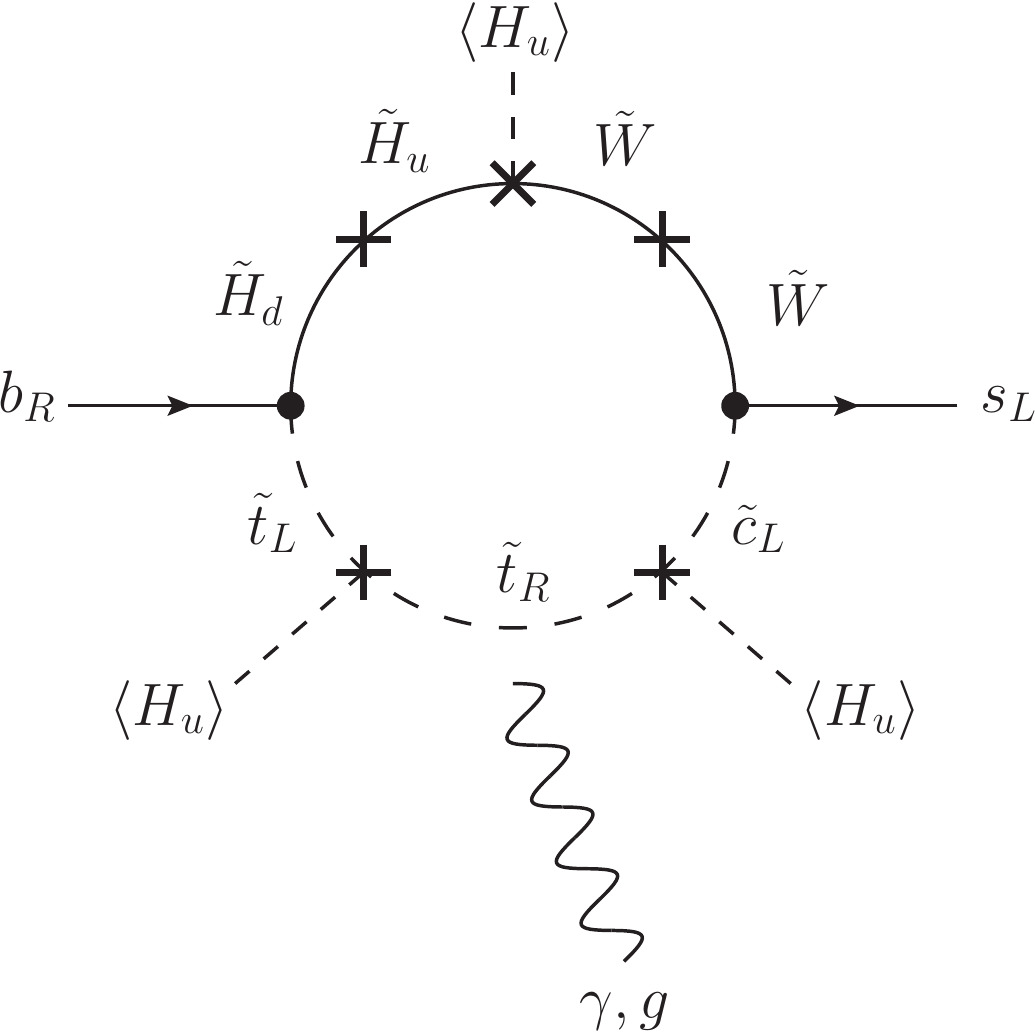} ~~~
\raisebox{6pt}{\includegraphics[width=0.3\textwidth]{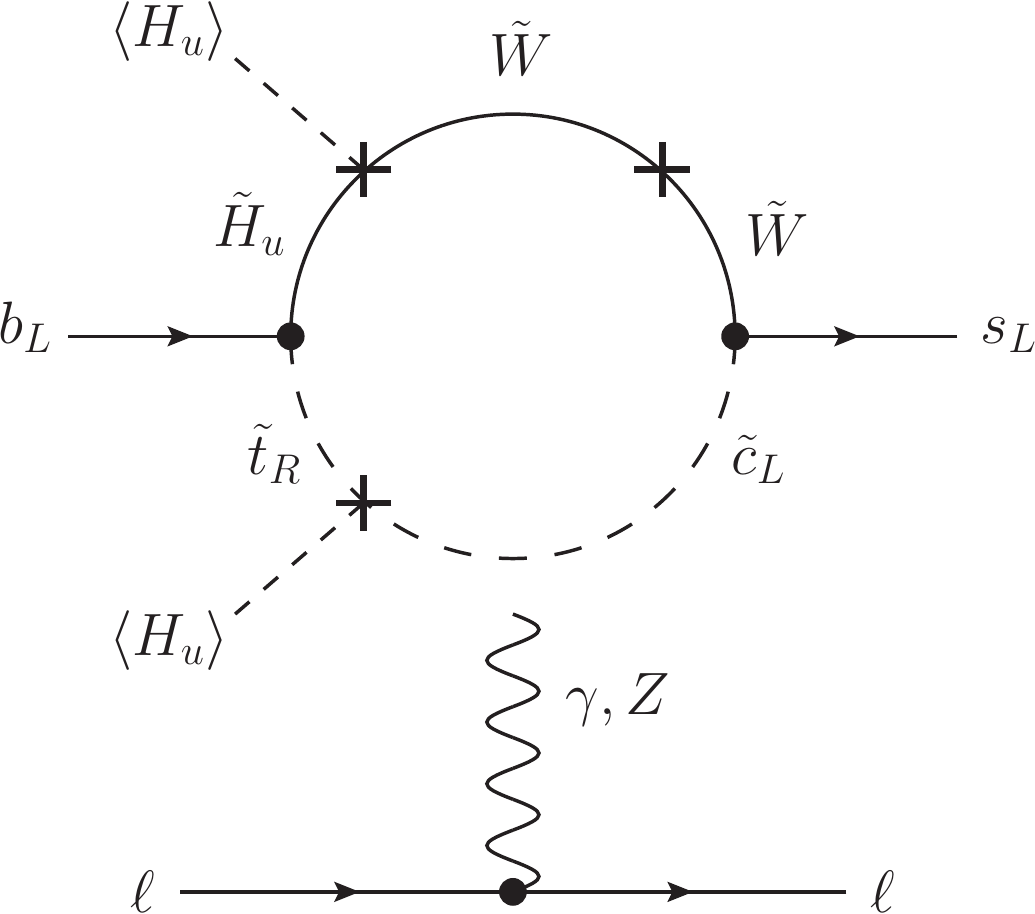}} ~~~
\raisebox{6pt}{\includegraphics[width=0.3\textwidth]{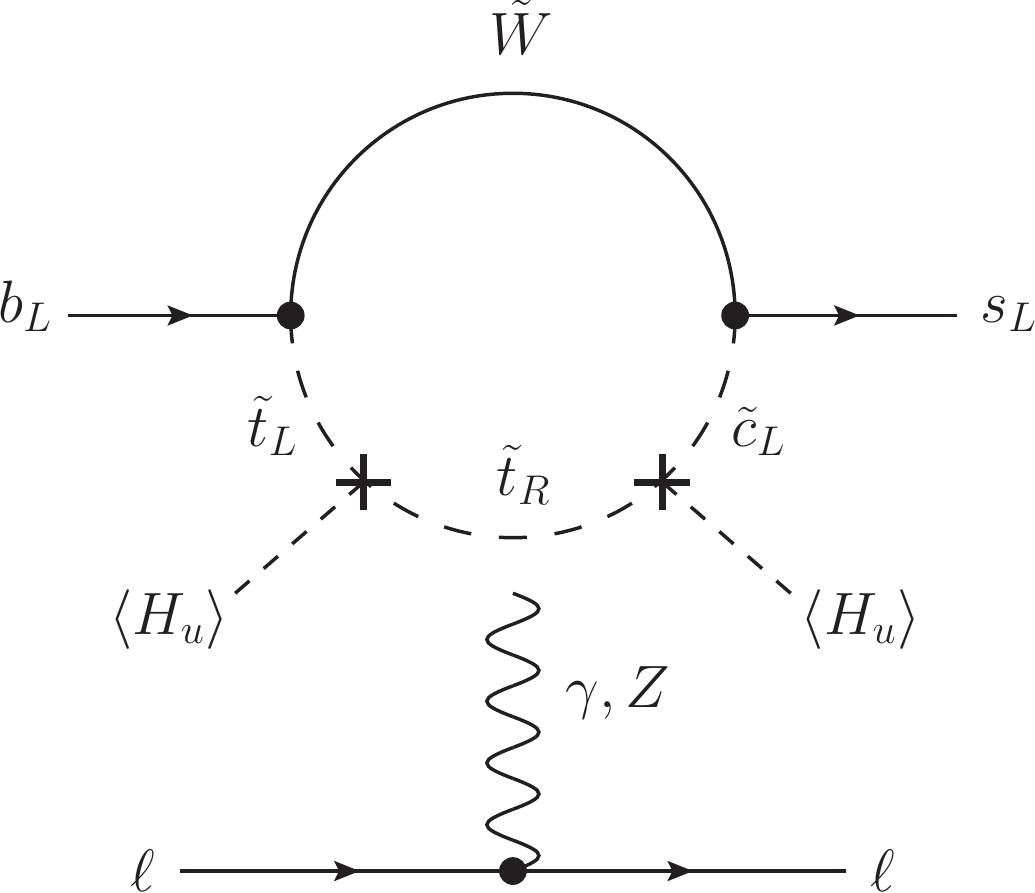}} \\[24pt]
\raisebox{6pt}{\includegraphics[width=0.46\textwidth]{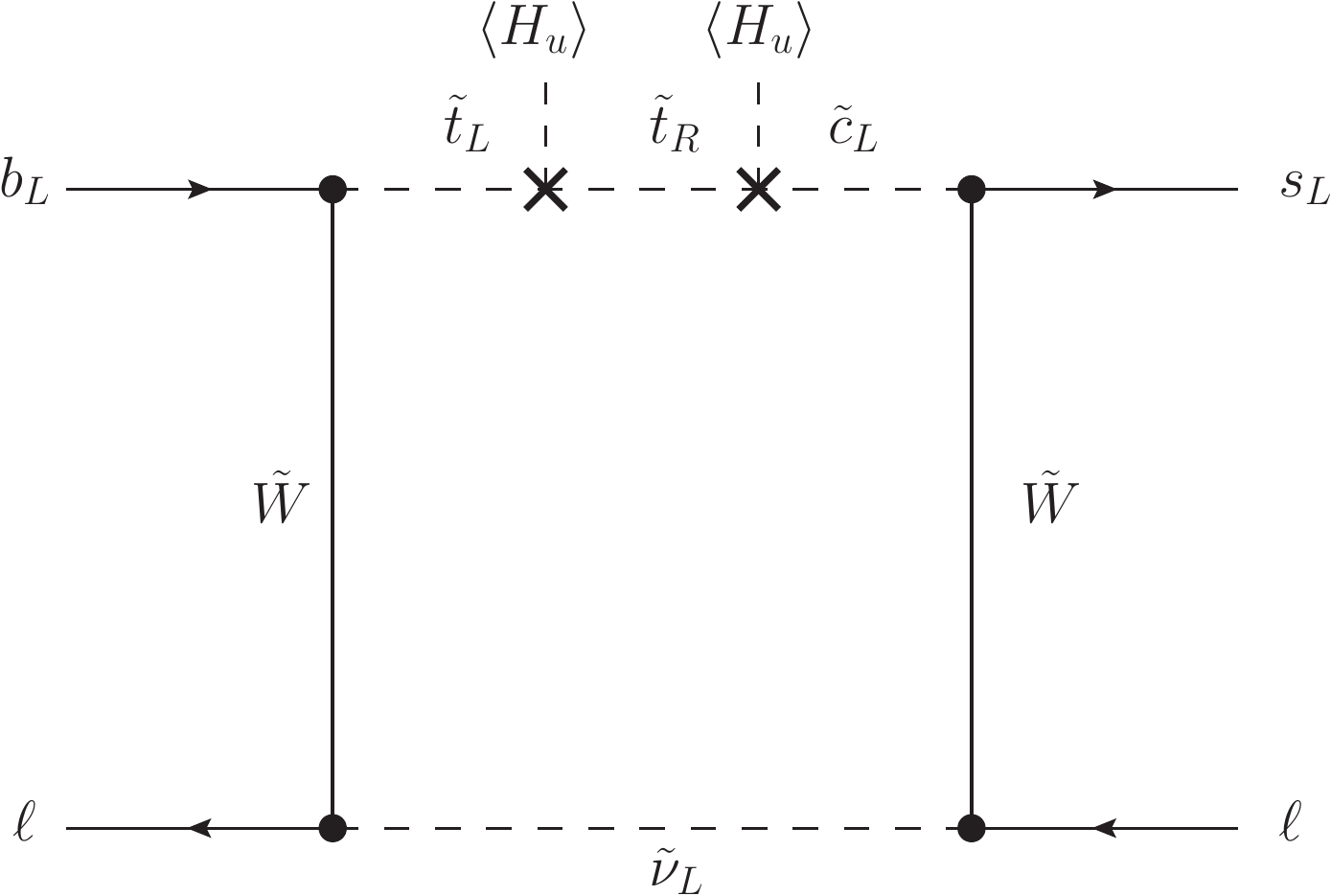}} ~~~~~~
\raisebox{6pt}{\includegraphics[width=0.46\textwidth]{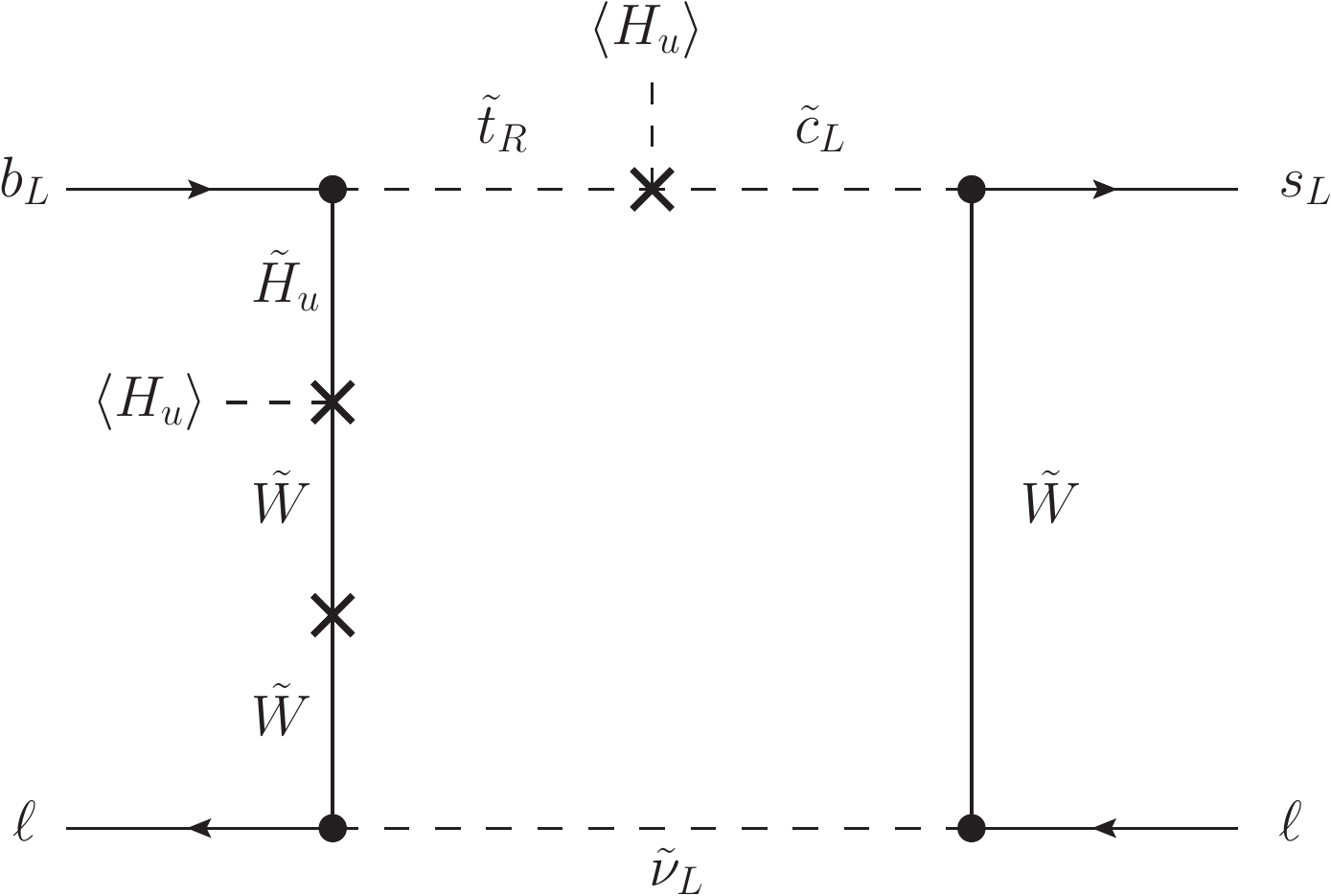}}
\caption{Example Feynman diagrams that correspond to MSSM contributions to the effective Hamiltonian for $b\to s \ell \ell$ transitions proportional to flavour-changing trilinear couplings. In the penguin diagrams, the photon, gluon and $Z$ propagators need to be attached to the loop in all possible ways.}
\label{fig:diagrams}
\end{figure}
%%%%%%%%%%%%%%%%%%%%%%%%%%%%%%%%%%%% 

\subsubsection{Bounds on flavour-changing trilinear couplings}

In addition to the usual flavour diagonal trilinear couplings, the soft SUSY breaking Lagrangian can contain flavour-changing trilinear couplings of the left and right handed top and charm squarks with the up-type Higgs
\begin{equation}
 \mathcal{L}_\text{trilinear} \supset A_t Y_t \tilde t_L^* \tilde t_R H_u + A_{tc} Y_t \tilde t_L^* \tilde c_R H_u + A_{ct} Y_t \tilde c_L^* \tilde t_R H_u  ~~ +~\text{h.c.} ~.
\end{equation}
The flavour-changing trilinears give contributions to the effective Hamiltonian in~(\ref{eq:Heff}) at the one loop level.
Contributions can arise from boxes, photon penguins, and Z penguins and example Feynman diagrams are shown in fig.~\ref{fig:diagrams}.
A straightforward flavour spurion analysis shows the following points:
\begin{itemize}
\item contributions to $C_{7,8}^\prime$, are suppressed by $m_s/m_b$ with respect to contributions to $C_{7,8}$;
\item contributions to $C_{9,10}^\prime$ are suppressed by $m_s m_b / m_t^2$ with respect to contributions to $C_{9,10}$;
\item contributions proportional to $A_{tc}$ are suppressed by $m_c / m_t$ compared to contributions proportional to $A_{ct}$.
\end{itemize}
We therefore concentrate on the Wilson coefficients $C_7$, $C_8$, $C_9$, and $C_{10}$ in the presence of a non-zero $A_{ct}$. 
To illustrate the main parameter dependence, in the following we give simple approximate expressions for the Wilson coefficients that are obtained at leading order in an expansion in $m_\text{EW}^2/m_\text{SUSY}^2$. The most important SUSY masses involved are the Wino mass $M_2$, the Higgsino mass $\mu$, the left-handed slepton mass $m_{\tilde \ell}$, the stop masses $m_{\tilde t_L}$ and $m_{\tilde t_R}$, as well as the left-handed charm squark mass $m_{\tilde c_L}$. The largest effects in $b \to s$ transitions can obviously be achieved if the SUSY spectrum is as light as possible. To keep the expressions compact, we set for simplicity $M_2 = \mu = m_{\tilde \ell} \equiv M$, $m_{\tilde t_L} = m_{\tilde c_L} \equiv m_L$, $m_{\tilde t_R} \equiv m_R$. We also work in the limit $M \ll m_R \ll m_L$ which is least constrained by collider searches and therefore allows to maximize the new physics contributions to the Wilson coefficients. Note also that a light Higgsino and light stops are well motivated by 
naturalness arguments~\cite{Dimopoulos:1995mi,Papucci:2011wy,Brust:2011tb}. For the dipole coefficients we find in a leading log approximation
\begin{subequations}
\begin{eqnarray}
 C_7 &=& \frac{V_{cs}^*}{V_{ts}^*} \left( \frac{A_{ct}}{A_t} \right) \tan\beta \frac{m_W^2 m_t^2}{m_R^4} \frac{\mu M_2  |A_t|^2}{m_L^4} \left[ \frac{m_R^2}{M^2} - \frac{7}{3} \log\left(\frac{m_R^2}{M^2} \right) \right] \nonumber \\
 && - \frac{m_t^2}{m_R^2} \frac{\mu A_t}{m_L^2} \tan\beta \frac{1}{2}\log\left(\frac{m_R^2}{M^2} \right)~, \label{eq:C7} \\
 C_8 &=& \frac{V_{cs}^*}{V_{ts}^*} \left( \frac{A_{ct}}{A_t} \right) \tan\beta \frac{m_W^2 m_t^2}{m_R^4} \frac{\mu M_2  |A_t|^2}{m_L^4} \log\left(\frac{m_R^2}{M^2} \right) - \frac{m_t^2}{m_R^2} \frac{\mu A_t}{m_L^2} \tan\beta \frac{1}{4} ~.  \label{eq:C8}
\end{eqnarray}
\end{subequations}
The contributions to $C_7$ and $C_8$ from $A_{ct}$ arise first at the dimension 8 level, i.e. they are suppressed by $m_\text{EW}^4/m_\text{SUSY}^4$. The last terms in~(\ref{eq:C7}) and~(\ref{eq:C8}) are the leading irreducible MFV contributions to $C_7$ and $C_8$ from Higgsino stop loops. They arise already at dimension 6 and are typically much larger than the contributions proportional to $A_{ct}$. 
%Contributions arising from flavour violation in the squark masses or flavour-violating trilinears in the down sector can also contribute already at the dimension 6 level.    

For the box contributions, $C_{9,10}^\text{box}$, and the photon penguin contribution, $C_9^\gamma$, to the semileptonic operators we find 
\begin{subequations}
\begin{eqnarray}
 C_{10}^\text{box} &=& \frac{1}{s_W^2} \frac{V_{cs}^*}{V_{ts}^*} \left( \frac{A_{ct}}{A_t} \right) \frac{m_W^2 m_t^2}{m_L^2 m_R^2} \left[ \frac{|A_t|^2}{4m_L^2} \log\left(\frac{m_R^2}{M^2}\right) - \frac{A_t}{12M^*} + \frac{M A_t}{4m_R^2} \log\left(\frac{m_R^2}{M^2}\right)   \right]~, \\
 C_9^\text{box} &=& - C_{10}^\text{box} ~, \\
 C_9^\gamma &=& \frac{V_{cs}^*}{V_{ts}^*} \left( \frac{A_{ct}}{A_t} \right) \frac{m_W^2 m_t^2}{m_L^2 m_R^2} \left[ \frac{2|A_t|^2}{3m_L^2} \log\left(\frac{m_R^2}{M^2}\right) - \frac{2A_t}{3M^*} + \frac{5M A_t}{3m_R^2} \log\left(\frac{m_R^2}{M^2}\right)   \right]~,
\end{eqnarray}
\end{subequations}
where $s_W = \sin\theta_W$ and $\theta_W$ is the weak mixing angle. Again we find that these contributions arise at the dimension 8 level. For a TeV scale SUSY spectrum, they are completely negligible.

In the considered scenario, only the Z penguin contributions, $C_{9,10}^Z$, arise already at the dimension 6 level. We find 
\begin{subequations}
\begin{eqnarray}
 C_{10}^Z &=& \frac{1}{s_W^2} \frac{V_{cs}^*}{V_{ts}^*} \left( \frac{A_{ct}}{A_t} \right) \frac{m_t^2}{m_L^2} \left[ \frac{|A_t|^2}{2m_L^2} \log\left(\frac{m_L^2}{m_R^2}\right) + \frac{M A_t}{8m_R^2} \log\left(\frac{m_R^2}{M^2}\right)   \right]~, \\
 C_9^Z &=& (4s_W^2 -1) C_{10}^Z ~.
\end{eqnarray}
\end{subequations}
This suggests that there are regions of MSSM parameter space, where a contribution to $C_{10}^Z$ of $O(1)$ is indeed possible. 
MSSM contributions to $C_9^Z$ on the other hand are suppressed by the accidentally small vector coupling of the Z boson to leptons, $(4s_W^2 -1) \sim -0.08$, and therefore negligible.

Recalling the model independent results from section~\ref{sec:modelindependent}, a positive new physics contribution to the Wilson coefficient $C_{10}^\text{NP} \simeq O(1)$, can improve the agreement with the current experimental $b \to s \mu^+\mu^-$ data significantly (albeit to a lesser extent than NP in $C_9$). Negative NP contributions to $C_{10}$ on the other hand are strongly disfavoured with the current data. We use these results to probe regions of MSSM parameter space with sizable flavour-changing trilinear couplings.

Bounds on flavour-changing trilinear couplings can also be obtained from vacuum stability considerations. As is well known, sizable trilinear couplings can lead to charge and color breaking minima in the MSSM scalar potential~\cite{Casas:1995pd,Casas:1996de}. Requiring that the electro-weak minimum be the deepest gives upper bounds on the trilinear couplings. Taking into account non-zero expectation values for the left and right handed stops, the left handed charm squark, as well as the up-type Higgs, we find the following necessary condition to ensure absolute stability of the electro-weak vacuum~\cite{Altmannshofer:2014qha,Dedes:2014asa}
\begin{equation} \label{eq:Actbound}
 \left( |A_t| + |A_{ct}| \tan\theta  \right)^2 \lesssim \left(3+ \tan^2\theta \right) \left( m_{\tilde t_L}^2 \cos^2\theta + m_{\tilde c_L}^2 \sin^2\theta + m_{\tilde t_R}^2  + m_{H_u}^2 + |\mu|^2 \right) ~.
\end{equation}
This inequality has to hold for all values of $\theta$, that parametrizes the angle in field space between the left handed top and charm squarks. In the limit $\theta = 0$ one recovers a well known bound on $A_t$ given e.g. in~\cite{Casas:1995pd}; for $\theta = \pi/2$ one recovers the bound on $A_{ct}$ found in~\cite{Casas:1996de}.

In principle, additional constraints on $A_{ct}$ can be obtained from the experimental bounds on electric dipole moments (EDMs). In particular, if $A_{ct}$ and $A_t$ contain a relative phase, a strange quark EDM and chromo EDM will be induced analogous to the new physics contributions to $C_7$ and $C_8$. However, predicting an experimentally accessible EDM of a hadronic system, like the neutron, given a strange quark EDM or chromo EDM involves large theoretical uncertainties~\cite{Hisano:2012sc,Fuyuto:2012yf}. Due to these uncertainties, existing EDM bounds do not give appreciable constraints in our setup.
Note also that bounds on the charm quark chromo EDM~\cite{Sala:2013osa} do not constrain the parameter space of our scenario. A sizable charm quark chromo EDM would be generated in the presence of both $A_{ct}$ and $A_{tc}$ couplings, but here we only consider a non-zero $A_{ct}$.   
     
%%%%%%%%%%%%%%%%%%%%%%%%%%%%%%%%%%%%
\begin{figure}[t]
\centering
\includegraphics[width=0.46\textwidth]{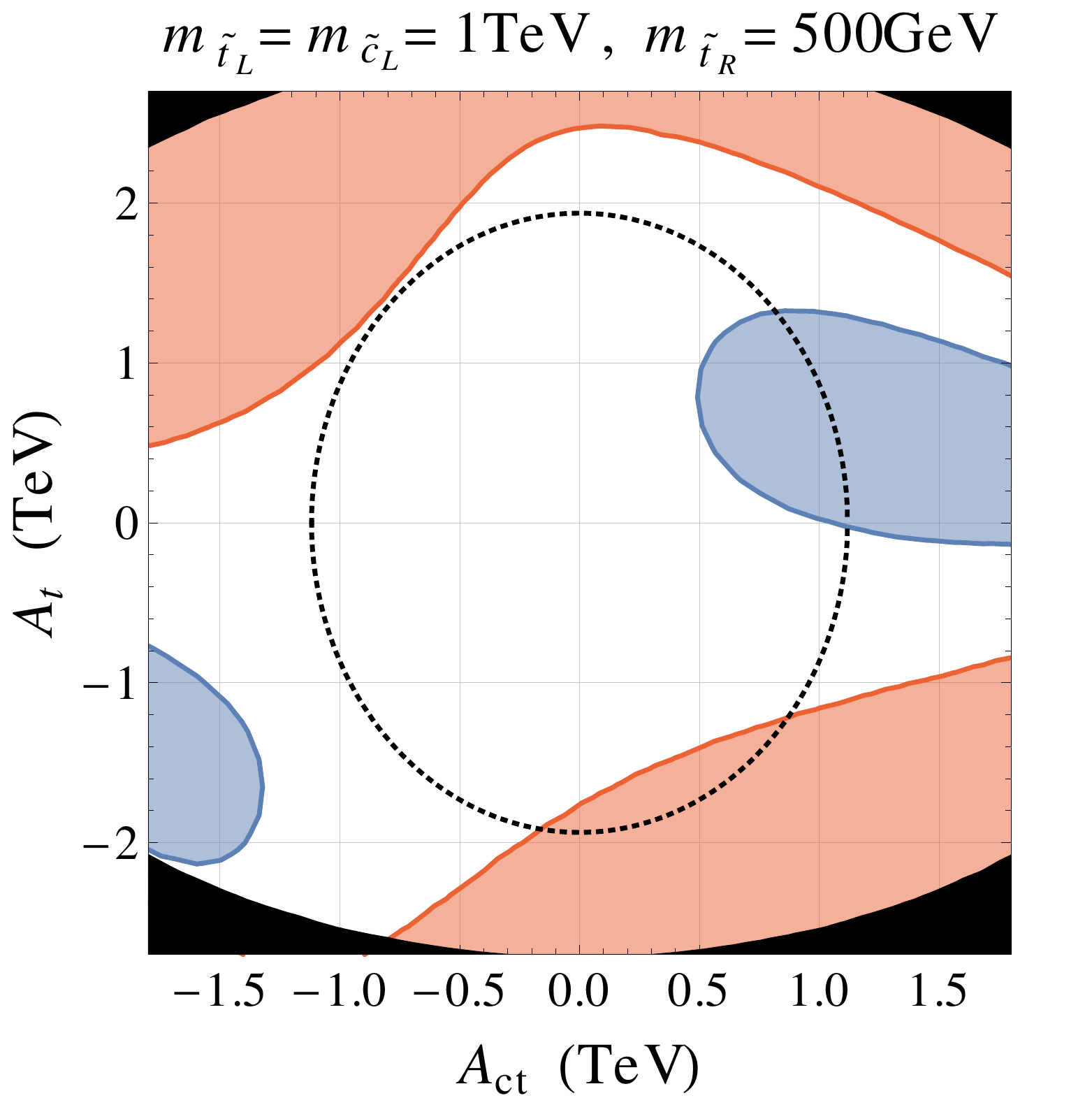} ~~~~~
\includegraphics[width=0.48\textwidth]{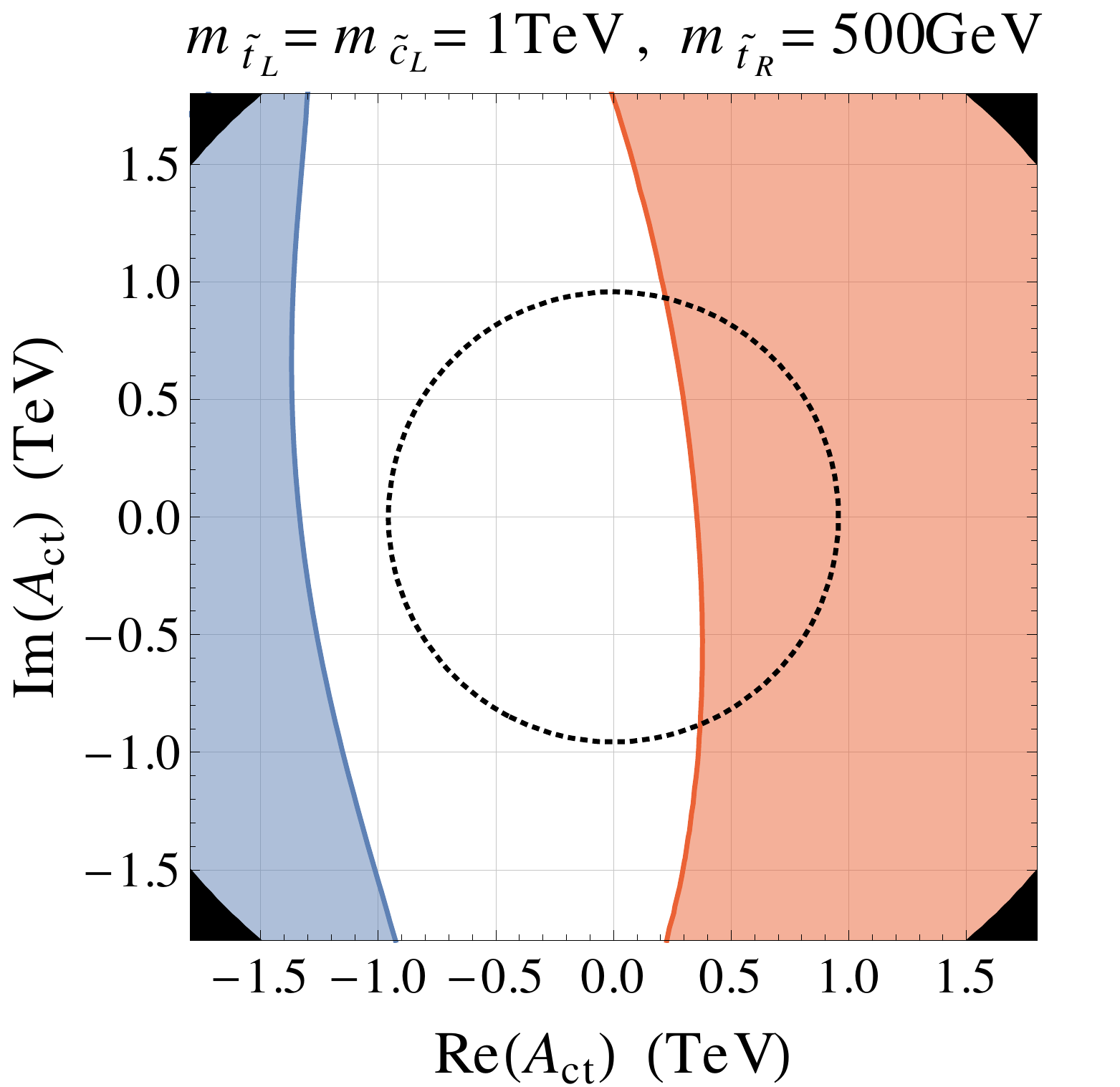}
\caption{Bound on trilinear couplings for an example MSSM scenario defined in the text. Left: bounds in the $A_t$ -- $A_{ct}$ plane, assuming real trilinears. Right: bounds in the Re$(A_{ct})$ -- Im$(A_{ct})$ plane, assuming a fixed $A_t = -1.5$~TeV. The red region is excluded by the $b\to s \mu^+\mu^-$ data by more than $2\sigma$ with respect to the SM. In the blue region the agreement between the theory predictions and the experimental $b\to s \mu^+\mu^-$ data is improved by more then $1\sigma$. Outside the dashed contours there exist charge and color breaking minima in the MSSM scalar potential that are deeper than the electro-weak minimum. In the black corners, the lightest up-squark mass eigenstate is the LSP.}
\label{fig:Abounds}
\end{figure}
%%%%%%%%%%%%%%%%%%%%%%%%%%%%%%%%%%%% 

We now describe the SUSY spectrum that we chose to illustrate the bounds on the trilinear couplings from the $b\to s \mu^+\mu^-$ data. The soft masses for the left-handed stop and charm squark are set to a common value $m_{\tilde t_L} = m_{\tilde c_L} = 1$~TeV. The soft mass of the right-handed stop is set to $m_{\tilde t_R} = 500$~GeV. All other squarks and sleptons as well as the gluino are assumed to be heavy, with masses of 2~TeV. Concerning the trilinear couplings, we only consider non-zero $A_t$ and $A_{ct}$. Due to these trilinear couplings, the lightest up-squark mass eigenstate can have a mass $m_{\tilde t_1} < 500$~GeV and is potentially subject to strong bounds from direct stop searches. Higgsinos, Winos and Binos are assumed to have mass parameters $m_{\tilde B} = 250$~GeV, $m_{\tilde W} = 300$~GeV, $\mu = 350$~GeV. In that way the mass of the lightest neutralino is given by $m_{\tilde \chi_1^0} \simeq 225$~GeV and the mass of the lightest chargino is $m_{\tilde \chi_1^\pm} \simeq 250$~GeV. Such 
a chargino-neutralino spectrum is heavy enough to avoid the bounds from the direct stop 
searches~\cite{Chatrchyan:2013xna,Aad:2014bva,Aad:2014kra,Aad:2014qaa}\footnote{Note that the most important bounds from~\cite{Chatrchyan:2013xna,Aad:2014bva,Aad:2014kra,Aad:2014qaa} assume 100\% branching ratio to either $\tilde t_1 \to t \tilde \chi_1^0$ or $\tilde t_1 \to b \tilde \chi_1^\pm$. In our scenario, both decay modes will compete with each other, weakening the bounds slightly. In addition, in our scenario there is significant 2nd-3rd generation mixing and the lightest stop can also have a sizable branching ratio $\tilde t_1 \to c \tilde \chi_1$. Thus the actual bounds from direct searches are further loosened, see e.g.~\cite{Blanke:2013uia,Agrawal:2013kha}.} as well as bounds from electro-weakino searches~\cite{Aad:2014vma,Khachatryan:2014mma}. 
Finally, we set $\tan\beta = 3$ to minimize contributions to the dipole Wilson coefficients.

In fig.~\ref{fig:Abounds} we show bounds on the trilinear couplings that can be derived from the $b\to s \mu^+\mu^-$ data in the described scenario. We evaluate all MSSM 1-loop contributions to the Wilson coefficients $C_{7,8,9,10}^{(\prime)}$ and compute the $\chi^2$ as defined in (\ref{eq:chi2}) as a function of the trilinear couplings. For the numerical evaluation of the Wilson coefficients in the MSSM, we use an adapted version of the \verb|SUSY_FLAVOR| code~\cite{Rosiek:2010ug,Crivellin:2012jv,Rosiek:2014sia}.
The plot on the left hand side of Fig.~\ref{fig:Abounds} shows constraints in the $A_t$ -- $A_{ct}$ plane, assuming real trilinears. The plot on the right hand side shows constraints in the Re$(A_{ct})$ -- Im$(A_{ct})$ plane, for a fixed $A_t = -1.5$~TeV\footnote{In the MSSM not all the parameter space shown in Fig.~\ref{fig:Abounds} would be compatible with a lightest Higgs mass of 125~GeV. However, there exist various extensions of the MSSM Higgs sector that allow to treat the Higgs mass independently from the stop sector. As the considered SUSY effects in $b \to s \ell \ell$ do not depend on the details of the Higgs sector, we do not consider the Higgs mass constraint in the plots of Fig.~\ref{fig:Abounds}.}. The red region is excluded by the $b\to s \mu^+\mu^-$ data by more than $2\sigma$ with respect to the SM ($\chi^2 > \chi^2_\text{SM} + 6$). In the blue region the agreement between the theory predictions and the experimental $b\to s \mu^+\mu^-$ data is improved by more than $1\sigma$ with respect to 
the SM ($\chi^2 < \chi^2_\text{SM} - 2.3$). In the best fit point in the left 
plot of Fig.~\ref{fig:Abounds}, the $\chi^2$ is reduced by $4.2$ compared to the 
SM. This improvement is rather moderate compared to the results of the model 
independent fits and also compared to the $Z^\prime$ scenarios discussed 
below.
In the black corners, the lightest up-squark mass eigenstate is lighter than the lightest neutralino.
Outside the dashed contours there exist charge and color breaking minima in the MSSM scalar potential that are deeper than the electro-weak minimum.
Inside the contours, the NP effects in the Wilson coefficients are rather moderate. In particular, we find that in this region of parameter space the SUSY contribution to $C_{10}$ does not exceed 0.3; the SUSY contribution to $C_9$ is smaller by approximately one order of magnitude, as expected.

Note that the regions outside of the vacuum stability contours are not necessarily excluded. Even though a deep charge and color breaking minimum exists in these regions, the electro-weak vacuum might be meta-stable with a live time longer than the age of the universe. Studies show that requiring only meta-stability, relaxes the stability bounds on the trilinear couplings slightly~\cite{Kusenko:1996jn,Park:2010wf,Blinov:2013fta,Chowdhury:2013dka,Camargo-Molina:2014pwa}.
A detailed analysis of vacuum meta-stability is beyond the scope of the present work.

\subsubsection{Lepton flavour non-universality in the MSSM}

The $Z$ penguin effects discussed above are lepton flavour universal, i.e. they lead to the same effects in $b \to s e^+e^-$ and $b \to s \mu^+\mu^-$ decays. Breaking of $e$-$\mu$ universality as hinted by the $R_K$ measurement can only come from box contributions as they involve sleptons of different flavours. If there are large mass splittings between the first and second generations of sleptons, or more precisely, if the selectrons are decoupled but smuons are kept light, Wino box diagrams (and to a lesser extent also Bino box diagrams) can contribute to $C_9^\mu$ and $C_{10}^\mu$ but not to $C_9^e$ and $C_{10}^e$. 

Box contributions are, however, typically rather modest in size. 
As discussed above, boxes that are induced by flavour-changing trilinears arise only at the dimension 8 level and are completely negligible. Non-negligible box contributions (at the dimension 6 level) are only possible in the presence of flavour violation in the squark soft masses. However, even allowing for maximal mixing of left-handed bottom and strange squarks, it was found in~\cite{Altmannshofer:2013foa} that Winos and smuons close to the LEP bound of $\sim 100$~GeV as well as bottom and strange squarks with masses of few hundred GeV would be required to obtain contributions to $C_9^\mu$ and $C_{10}^\mu$ of $\gtrsim 0.5$, that could give $R_K \sim 0.75$.
A careful collider analysis would be required to ascertain if there are holes in the LHC searches for stops~\cite{Chatrchyan:2013xna,Aad:2014bva,Aad:2014kra,Aad:2014qaa}, sbottoms~\cite{ATLAS-CONF-2012-165,Aad:2013ija,CMS-PAS-SUS-13-018}, sleptons~\cite{Khachatryan:2014qwa,Aad:2014vma,ATLAS-CONF-2013-049} and electro-weakinos~\cite{Aad:2014vma,Khachatryan:2014mma} that would allow such an extremely light spectrum. We also note that a sizable splitting between the left-handed smuon and selectron masses required to break $e$-$\mu$ universality is only possible if the slepton mass matrix is exactly diagonal in the same basis as the charged lepton mass matrix, since even a tiny misalignment would lead to an excessive $\mu \to e \gamma$ decay rate.

\subsection{Flavour changing \texorpdfstring{$Z'$}{Z'} bosons}

A massive $Z^\prime$ gauge boson with flavour-changing couplings to quarks is an obvious candidate that can lead to large effects in $b \to s \ell \ell$ decays~\cite{Altmannshofer:2013foa,Gauld:2013qba,Buras:2013qja,Gauld:2013qja,Buras:2013dea,Altmannshofer:2014cfa,Buras:2014fpa,Crivellin:2015mga,Crivellin:2015lwa,Niehoff:2015bfa,Sierra:2015fma,Crivellin:2015era}. Instead of discussing a complete model that contains such a $Z'$ boson, we will take a bottom up approach and ask which properties a $Z'$ has to have in order to explain the discrepancies observed in the $b \to s \ell \ell$ data.
To this end we treat the mass of the $Z'$ as well as its couplings to SM quarks and leptons as free parameters. 
Following the notation of~\cite{Buras:2012jb}, we parametrize the $Z'$ couplings as
\begin{equation}
\mathcal L \supset \bar f_i \gamma^\mu \left[ \Delta^{f_if_j}_L(Z') P_L + \Delta^{f_if_j}_R(Z') P_R \right] f_j \,Z'_\mu
\,.
\end{equation}
In the presence of $\Delta_{L/R}^{bs}$ and $\Delta_{L/R}^{\mu\mu}$ couplings, the $Z'$ boson will contribute to the Wilson coefficients $C_9^{(\prime)}$ and $C_{10}^{(\prime)}$ at tree level. As the primed Wilson coefficients hardly improve the agreement of the experimental $b \to s \mu^+\mu^-$ data with the theory predictions, we will not consider them here and set the right-handed $bs$ couplings to zero, $\Delta_R^{bs}=0$.

The $Z'$ couplings $\Delta_{L}^{bs}$ and $\Delta_{L/R}^{\mu\mu}$ are subject to various constraints that bound the maximal effect a $Z'$ prime can have in $C_9$ and $C_{10}$.
In particular, a $Z'$ boson with flavour-changing $b \leftrightarrow s$ couplings will inevitably also contribute to $B_s$-$\bar B_s$ mixing at the tree level. One finds the following modification of the mixing amplitude
\begin{equation}
 \frac{M_{12}}{M_{12}^\text{SM}} -1 = \frac{v^2}{M_{Z'}^2} (\Delta_L^{bs})^2 \left( \frac{g_2^2}{16\pi^2} (V_{tb}V_{ts}^*)^2 S_0 \right)^{-1} ~,
\end{equation}
where $v = 246$~GeV is the Higgs vev, and the SM loop function is given by $S_0 \simeq 2.3$.
We obtain the following stringent bound on the $Z'$ mass and the flavour-changing coupling,
\begin{equation}
\frac{M_{Z'}}{|\Delta^{bs}_L|}
\gtrsim 244~\text{TeV}  \times \left(\frac{10\%}{|M_{12}/M_{12}^\text{SM}-1|} 
\right)^{1/2}
\approx \frac{10~\text{TeV}}{|V_{tb}V_{ts}^*|}  \times 
\left(\frac{10\%}{|M_{12}/M_{12}^\text{SM}-1|} \right)^{1/2}
\,.
\label{eq:Bsbound}
\end{equation}
In the following, we will allow for maximally 10\% new physics 
contribution to the mixing amplitude, which is approximately the size of 
non-standard effects that are currently probed in $B_s$ 
mixing~\cite{Charles:2015gya}.
Concerning the couplings of the $Z'$ to leptons,
we will start with 
the least constrained case, where the $Z'$ only couples to muons, but not to electrons
and consider a coupling to left-handed muons only.
Subsequently, we will discuss how our conclusions change if we assume a vector-like coupling to muons
or a lepton-flavour universal coupling.

\subsubsection{\texorpdfstring{$Z'$}{Z'} with coupling to left-handed muons}

The only non-zero coupling to charged leptons we consider here is $\Delta_L^{\mu\mu}$. Such a $Z^\prime$ is very poorly constrained. Over a very broad range of $Z'$ masses, the strongest constraint on $\Delta_L^{\mu\mu}$ comes from neutrino trident production~\cite{Altmannshofer:2014cfa,Altmannshofer:2014pba}, i.e. the production of a muon pair in the scattering of a muon-neutrino in the Coulomb field of a heavy nucleus\footnote{The only exception relevant in the context of NP in $b\to s\mu^+\mu^-$ is a very low mass window between 10~GeV $\lesssim M_{Z'} \lesssim$ 50~GeV, where the $Z \to 4\mu$ branching ratio measured at the LHC gives a constraint that is slightly stronger than the one obtained from neutrino tridents~\cite{Altmannshofer:2014pba}.}.
The relative correction of the trident cross section in the presence of the considered $Z'$ is given by
\begin{equation}
 \frac{\sigma}{\sigma_\text{SM}} = \frac{1}{1+(1+4s_W^2)^2}\left[ \left( 1 + \frac{v^2 (\Delta_L^{\mu\mu})^2}{M_{Z^\prime}^2} \right)^2 + \left( 1 + 4s_W^2 + \frac{v^2 (\Delta_L^{\mu\mu})^2}{M_{Z^\prime}^2} \right)^2 \right] ~.
\end{equation}
We use the CCFR measurement of the trident cross section, $\sigma_\text{CCFR}/\sigma_\text{SM} = 0.82 \pm 0.28$~\cite{Mishra:1991bv}, to set bounds on the $Z'$ mass and its coupling to muons.
At the $2\sigma$ level we find
\begin{equation}
\frac{M_{Z'}}{|\Delta^{\mu\mu}_L|}
\gtrsim 0.47~\text{TeV}
\,. \label{eq:muonbound1}
\end{equation}
Combining this result with the bound on the flavour-changing quark coupling from $B_s$ mixing, eq.~(\ref{eq:Bsbound}), we can derive an upper bound on the possible size of new physics contributions to the Wilson coefficients $C_9$ and $C_{10}$ that can be achieved in the considered setup. For the Wilson coefficients we have 
\begin{equation}
C_9^\text{NP} = -C_{10}^\text{NP} = -\frac{\Delta^{bs}_L\Delta^{\mu\mu}_L}{V_{tb}V_{ts}^*}\left[\frac{\Lambda_v}{M_{Z'}}\right]^2 ~,~~\text{with}~~ \Lambda_v = \left[\frac{\pi}{\sqrt{2}G_F\alpha_\text{em}}\right]^{1/2}\approx 4.94\,\text{TeV} \,.
\end{equation}
This implies
\begin{equation}
|C_9^\text{NP}| = |C_{10}^\text{NP}| <5.4 ~. 
\end{equation}
The best fit values in the $C_9^\text{NP} = - C_{10}^\text{NP}$ scenario found in section~\ref{sec:1WC} are well within this bound.

Although the explanation of the tensions in $b\to s\mu^+\mu^-$ transitions does not require a coupling of the $Z'$ to first generation quarks, it is interesting to investigate what happens in models where such couplings are present, which could lead to $Z'$ signals at the LHC.
Fixing the Wilson coefficients $C_9$ and $C_{10}$ to their best fit values and assuming the flavour-changing coupling to have its maximal value (\ref{eq:Bsbound}) allowed by $B_s$ mixing, we find a lower bound on the muon coupling,
\begin{equation}
\Delta_{L}^{\mu\mu}\gtrsim 0.3 \left[\frac{M_{Z'}}{\text{TeV}}\right]\,.
\label{eq:deltal}
\end{equation}
Adopting the lower end of this range,
ATLAS and CMS searches for quark-lepton contact interactions~\cite{Aad:2014wca,CMS-PAS-EXO-12-020} can be used to put an upper bound on the $Z'$ coupling to the left-handed first-generation quark doublet. Using the CMS results~\cite{CMS-PAS-EXO-12-020}, we find 
\begin{equation}
\frac{M_{Z'}}{|\Delta^{qq}_L|}  \gtrsim 11~\text{TeV} ~(7~\text{TeV})
\end{equation}
for constructive (destructive) interference with the SM $q_L \bar q_L \to \mu^+\mu^-$ amplitude. Comparing this to (\ref{eq:Bsbound}), we conclude that models with a rough scaling $|\Delta^{bs}_L|\sim|V_{tb}V_{ts}^*\Delta^{qq}_L|$ are compatible with these bounds.

\begin{figure}[tb]
\centering\includegraphics[width=0.6\textwidth]{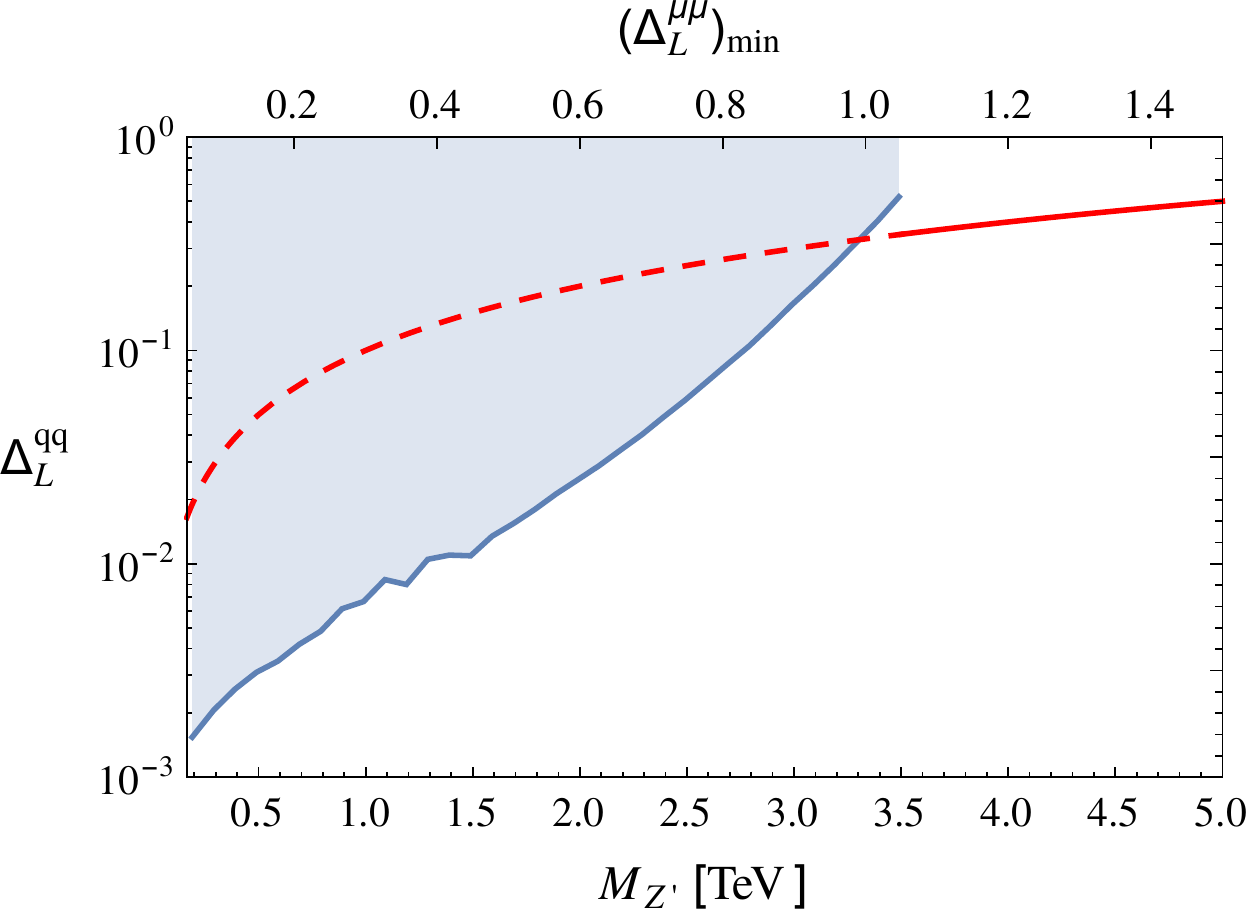}
\caption{Bounds in the plane of $Z'$ mass and the $Z'$ coupling to the left-handed first-generation quark doublet. The blue shaded region is excluded by searches for resonances in the dimuon invariant mass spectrum~\cite{Aad:2014cka}, assuming a $Z' \to \mu^+\mu^-$ branching ratio of 50\%. The region above the red curve is excluded by searches for quark-lepton contact interactions~\cite{CMS-PAS-EXO-12-020}.
The upper plot axis shows the minimal value of the $Z'$ coupling to left-handed muons~(\ref{eq:deltal}), required to obtain the best fit values for $C_9$ and $C_{10}$.
}
\label{fig:MZbound}
\end{figure}

For a $Z'$ mass between 200~GeV and 3.5~TeV, also LHC searches for resonances~\cite{CMS-PAS-EXO-12-061,Aad:2014cka} in the dimuon mass spectrum can be used to put an upper bound on the $Z'$ coupling to first-generation quarks as a function of $M_{Z'}$. In fig.~\ref{fig:MZbound} we show the bound on $\Delta_L^{qq}$ using the results from the ATLAS search~\cite{Aad:2014cka} (shaded blue region).
For the branching ratios of the $Z'$ we assume $\text{BR}(Z'\to \mu^+\mu^-)=\text{BR}(Z'\to \nu_\mu\bar\nu_\mu)=\frac{1}{2}$, which approximately holds as long as the $\Delta_L^{\mu\mu}$ coupling is sufficiently large compared to couplings to other states. The bound from resonance searches could be weaker if the $Z'$ has e.g. a sizable branching ratio into a dark sector.
In the same plot, we also show the bound from quark-lepton contact interaction searches from CMS~\cite{CMS-PAS-EXO-12-020}, assuming (\ref{eq:deltal}) (red line).
Below $3.5$~TeV, we show this bound as a dashed line, because for such light $Z'$ masses the contact interaction approximation becomes invalid.

We conclude that, in order to lead to visible effects in $b\to s\mu^+\mu^-$ transitions, a heavy $Z'$ with $M_{Z'} \gtrsim 3~\text{TeV}$ can have weak-interaction strength couplings to first-generation quarks without being in conflict with the bounds from contact interactions. Such a heavy $Z'$ must have strong couplings to muons ($\Delta_L^{\mu\mu} \gtrsim 1$). A lighter $Z'$ can be weakly coupled to muons, but requires a suppression of the coupling to first-generation quarks by roughly two orders of magnitude to avoid the bounds from direct searches.

\subsubsection{\texorpdfstring{$Z'$}{Z'} with vector-like coupling to muons}

If the couplings of the $Z'$ to muons are purely vector-like we can define $\Delta_L^{\mu\mu} = \Delta_R^{\mu\mu} \equiv \Delta_V^{\mu\mu}/2$. 
In this case, the correction to the neutrino trident cross section reads
\begin{equation}
 \frac{\sigma}{\sigma_\text{SM}} = \frac{1}{1+(1+4s_W^2)^2}\left[  1 + \left( 1 + 4s_W^2 + \frac{%2
 v^2 (\Delta_V^{\mu\mu})^2}{2 M_{Z^\prime}^2} \right)^2 \right] ~,
\end{equation} 
and we obtain the following bound using the CCFR measurement
\begin{equation} \label{eq:vectormuon}
\frac{M_{Z'}}{|\Delta^{\mu\mu}_V|}
\gtrsim 0.27~\text{TeV}
\,.
\end{equation}
Now the NP contribution to the Wilson coefficient $C_{10}$ vanishes, while for $C_9$ one has
\begin{equation}
C_9^\text{NP} = -\frac{\Delta^{bs}_L\Delta^{\mu\mu}_V}{V_{tb}V_{ts}^*}\left[\frac{\Lambda_v}{M_{Z'}}\right]^2\,.
\end{equation}
Again, one finds that sizable effects are possible: adopting the maximum allowed values for the couplings~(\ref{eq:vectormuon}) and~(\ref{eq:Bsbound}), we find $|C_9^\text{NP}|<9.3$.
The bounds on first-generation quark couplings from contact interaction and dimuon resonance searches are qualitatively similar to the left-handed case. 

\subsubsection{\texorpdfstring{$Z'$}{Z'} with universal coupling to leptons}

If the $Z'$ coupling to leptons is flavour-universal, stringent bounds on $\Delta^{\ell\ell}$ can be obtained from 
LEP2 searches for four lepton contact interactions~\cite{Schael:2013ita}. Depending on whether the coupling is to left-handed leptons only or is vector-like, we find
\begin{align}
\frac{M_{Z'}}{|\Delta^{\ell\ell}_L|} &\gtrsim 3.9~\text{TeV} \,,
&
\frac{M_{Z'}}{|\Delta^{\ell\ell}_V|} &\gtrsim 3.5~\text{TeV}\,,
\\
C_9^\text{NP} &= -C_{10}^\text{NP} = -\frac{\Delta^{bs}_L\Delta^{\ell\ell}_L}{V_{tb}V_{ts}^*}\left[\frac{\Lambda_v}{M_{Z'}}\right]^2
\,,&
C_9^\text{NP} &= -\frac{\Delta^{bs}_L\Delta^{\ell\ell}_V}{V_{tb}V_{ts}^*}\left[\frac{\Lambda_v}{M_{Z'}}\right]^2
\,,\\
\Rightarrow|C_9^\text{NP}|&=|C_{10}^\text{NP}|<0.64
\,,&
\Rightarrow|C_9^\text{NP}|&<0.72
\,.
\end{align}
where, for the last step, the flavour-changing coupling has been assumed to saturate the upper bound in~(\ref{eq:Bsbound}) coming from $B_s$ mixing.
We observe that the effects in $b\to s\mu^+\mu^-$ transitions are now much more limited, but, in particular for left-handed couplings, can still come close to the best-fit values in section~\ref{sec:1WC} (of course, the anomaly in $R_K$ cannot be explained in this scenario.)
Interestingly, this also implies that the effect in $B_s$ mixing is necessarily close to the current experimental bounds. Future improvements of the $B_s$ mixing constraints will then allow to test the lepton flavour universal scenario.

Concerning collider searches, the new feature of the lepton universal case is that there is an absolute lower bound on the $Z'$ mass from LEP2, $M_{Z'}>209$~GeV. LHC bounds on the coupling to first generation quarks, on the other hand, are qualitatively similar to the non-universal case discussed above.

\section{Summary and conclusions}\label{sec:concl}

Several recent results on rare B decays by the LHCb collaboration show tensions with standard model predictions. 
Those include discrepancies in angular observables in the $B \to K^* \mu^+\mu^-$ decay, a suppression in the branching ratios of $B \to K^* \mu^+\mu^-$ and $B_s \to \phi \mu^+\mu^-$, as well as a hint for the violation of lepton flavour universality in the form of a $B \to K \mu^+\mu^-$ branching ratio that is suppressed not only with respect to the SM prediction but also with respect to $B \to K e^+e^-$.
In this paper we performed global fits of the experimental data within the SM and in the context of new physics.

For our SM predictions we use state-of-the-art $B \to K$, $B \to K^*$ and $B_s \to \phi$ form factors taking into account results from lattice and light cone sum rule calculations.  
All relevant non-factorizable corrections to the $B \to K \mu^+\mu^-$, $B \to K^* \mu^+\mu^-$ and $B_s \to \phi \mu^+\mu^-$ amplitudes that are known are included in our analysis. Additional unknown contributions are parametrized in a conservative manner, such that existing estimates of their size are within the $1\sigma$ range of our parametrization.
We take into account all the correlations of theoretical uncertainties between different observables and between different bins of dilepton invariant mass. As experimental data is available for more and more observables in finer and finer bins, the theory error correlations have a strong impact on the result of the fits.\footnote{To quantify this statement: when all correlations of theory uncertainties are set to zero, the $\Delta\chi^2$ of the fit with NP in $C_9$ only increases from $13.7$ to $38.9$. This huge overestimate of the significance is easy to understand, as tensions in the same direction in adjacent bins are less significant if one knows that they are highly correlated.}

Making use of all relevant experimental data on radiative, leptonic and semi-leptonic $b \to s$ decays we find that there is on overall tension between the SM predictions and the experimental results.
Assuming the absence of new physics, we investigated to which extent non-perturbative QCD effects can be responsible for the apparent disagreement. 
We find that large non-factorizable corrections, a factor of 4 above our error estimate, could improve the agreement for the $B \to K^* \mu^+\mu^-$ angular observables and the branching ratios considerably.
Alternatively, the branching ratio predictions could also be brought into better agreement with the experimental data, if the involved form factors were all systematically below the theoretical determinations from the lattice and from LCSR. On the other hand, we find that non-standard values of the form factors could at most lead to a modest improvement of $B \to K^* \mu^+\mu^-$ angular observables. 
In both cases however, the hint for violation of lepton flavour universality cannot be explained.

Assuming that in our global fits the hadronic uncertainties are estimated in a sufficiently conservative way, we discussed the implications of the experimental results on new physics.
Effects from new physics at short distances can be described model independently by an effective Hamiltonian and the experimental data can be used to obtain allowed regions for the new physics contributions to the Wilson coefficients.
We find that the by far largest decrease in the $\chi^2$ can be obtained either by a negative new physics contribution to $C_9$ (with $C_9^\text{NP} \sim -25\% \times C_9^\text{SM}$), or by new physics in the $SU(2)_L$ invariant direction $C_9^\text{NP}=-C_{10}^\text{NP}$, (with $C_9^\text{NP} \sim -12\% \times C_9^\text{SM}$). A positive NP contribution to $C_{10}$ alone would also improve the fit, although to a lesser extent.

Concerning the hint for violation of lepton flavour universality, we observe that new physics exclusively in the muonic decay modes leads to an excellent description of the data. We do not find any preference for new physics in the electron modes. We provide predictions for other lepton flavour universality tests. We find that the ratio $R_{A_\text{FB}}$ of the forward-backward asymmetries in $B \to K^* \mu^+\mu^-$ and $B \to K^* e^+e^-$ at low dilepton invariant mass is a particularly sensitive probe of new physics in $C_9^\mu$. A precise measurement of $R_{A_\text{FB}}$ would allow to distinguish the new physics scenarios that give the best description of the current data.

Finally we also discussed the implications of the model independent fits for the minimal supersymmetric standard model and models that contain $Z^\prime$ gauge bosons with flavour-changing couplings. In the MSSM, large flavour changing trilinear couplings in the up-squark sector can give sizable contributions to the Wilson coefficient $C_{10}$ and we identified regions of MSSM parameter space that are favoured or disfavoured by the current experimental data.
Heavy $Z'$ bosons can have the required properties to explain the discrepancies observed in the $b \to s \ell \ell$ data. If the $Z'$ couples to muons but not to electrons (as preferred by the data), it is only weakly constrained by indirect probes. On the other hand, if the $Z'$ couplings to leptons are flavour universal, LEP constraints on 4 lepton contact interactions imply that an explanation of the $b \to s \ell \ell$ discrepancies results in new physics effects in $B_s$ mixing of at least $\sim 10\%$.  
In all scenarios, the couplings of the $Z'$ to first generation quarks are strongly constrained by ATLAS and CMS measurements of dilepton production. 

We look forward to the updated experimental results using the full LHCb data set, which will be crucial in helping to establish or to refute the exciting possibility of new physics in $b \to s$ transitions.

%###################################################################
\section*{Acknowledgements}
We thank
Martin Beneke,
Aoife Bharucha,
Christoph Bobeth,
Gerhard Buchalla,
Danny van Dyk,
Thorsten Feldmann,
Christoph Niehoff,
Yuming Wang,
Roman Zwicky,
and all the participants of the 
``Workshop on $b \to s ll$ processes'' at Imperial College in April 2014
for useful discussions.
We also thank the National Science Foundation for partial support (under 
Grant No. PHYS-1066293), the  Aspen Center for Physics for hospitality
and the German national football team for moral support during the
workshop ``Connecting Flavor Physics with Naturalness: from Theory to Experiment''
in July 2014. 
Research at Perimeter Institute is supported by the Government of Canada through Industry Canada and by the Province of Ontario through the Ministry of Economic Development \& Innovation. 
The research of D.S.\ was supported by the DFG cluster of excellence ``Origin and Structure of the Universe''.%###################################################################

\appendix

\section{\texorpdfstring{$B\to K$}{B --> K} form factors}\label{sec:BKFFs}

\begin{figure}[tbp]
\centering
\includegraphics[width=0.47\textwidth]{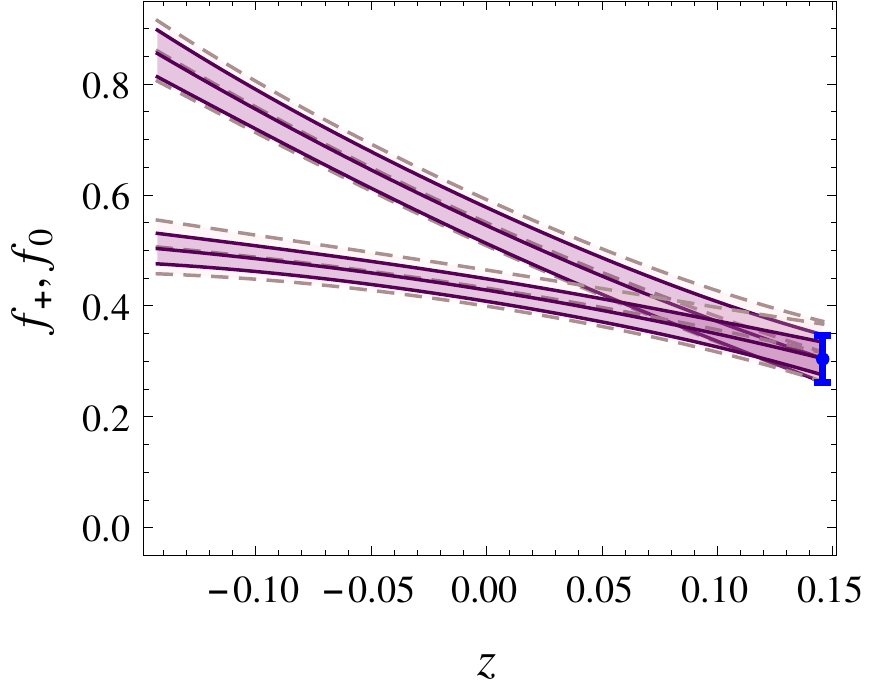}%
\hspace{0.04\textwidth}
\includegraphics[width=0.47\textwidth]{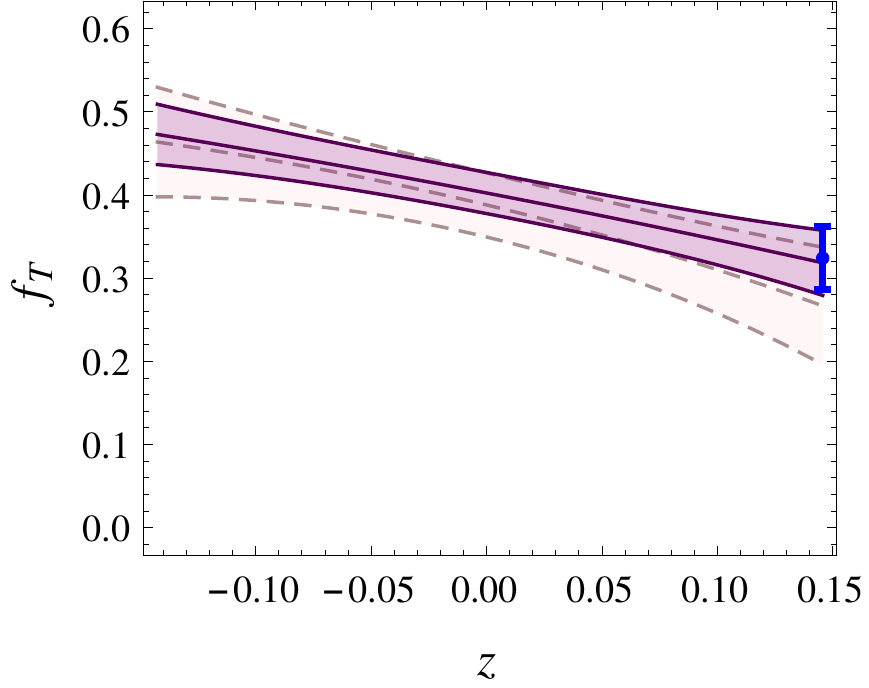}%
\caption{$B\to K$ form factors from a combined fit to lattice and LCSR calculations as a function of the series expansion parameter $z(q^2)$.
Solid: fit result with uncertainties. Blue: LCSR prediction at $q^2=0$. Dashed: extrapolated lattice prediction.}
\label{fig:FFK}
\end{figure}

For the $B\to K$ form factors, we perform a combined fit to the recent lattice computation \cite{Bouchard:2013eph} as well as LCSR predictions at $q^2=0$ \cite{Ball:2004ye,Bartsch:2009qp}, using the parametrization and conventions of \cite{Bouchard:2013eph}. The method was already described in appendix~A of \cite{Buras:2014fpa} and here we simply provide our results for the series expansion coefficients of all three $B\to K$ form factors.
For the central values and uncertainties, we find
\begin{align}
a_0^0 &= 0.54 \pm 0.03 \,,&
a_1^0 &= -1.91 \pm 0.1 \,,&
a_2^0 &= 1.83 \pm 1.07 \,,&
a_3^0 &= -0.02 \pm 2.74 \,,\\
a_0^+ &= 0.43 \pm 0.02 \,,&
a_1^+ &= -0.67 \pm 0.09 \,,&
a_2^+ &= -1.12 \pm 0.76 \,,\\
a_0^T &= 0.4 \pm 0.02 \,,&
a_1^T &= -0.53 \pm 0.13 \,,&
a_2^T &= -0.29 \pm 1.00 \,,
\end{align}
while the correlation matrix for the set $(a_0^0,a_1^0,a_2^0,a_3^0,a_0^+,a_1^+,a_2^+,a_0^T,a_1^T,a_2^T)$
reads
\begin{align}
\begin{pmatrix}
 1.0 & -0.39 & -0.71 & -0.63 & 0.49 & -0.03 & -0.22 & 0.16 & -0.08 & -0.09 \\
 -0.39 & 1.0 & 0.66 & 0.26 & 0.05 & 0.72 & 0.48 & -0.08 & 0.03 & 0.01 \\
 -0.71 & 0.66 & 1.0 & 0.54 & -0.17 & 0.51 & 0.59 & -0.16 & 0.05 & 0.09 \\
 -0.63 & 0.26 & 0.54 & 1.0 & 0.05 & 0.14 & 0.05 & 0.0 & 0.03 & -0.01 \\
 0.49 & 0.05 & -0.17 & 0.05 & 1.0 & 0.09 & -0.47 & 0.34 & -0.06 & -0.28 \\
 -0.03 & 0.72 & 0.51 & 0.14 & 0.09 & 1.0 & 0.43 & -0.06 & 0.11 & -0.04 \\
 -0.22 & 0.48 & 0.59 & 0.05 & -0.47 & 0.43 & 1.0 & -0.32 & -0.05 & 0.29 \\
 0.16 & -0.08 & -0.16 & 0.0 & 0.34 & -0.06 & -0.32 & 1.0 & 0.0 & -0.35 \\
 -0.08 & 0.03 & 0.05 & 0.03 & -0.06 & 0.11 & -0.05 & 0.0 & 1.0 & 0.21 \\
 -0.09 & 0.01 & 0.09 & -0.01 & -0.28 & -0.04 & 0.29 & -0.35 & 0.21 & 1.0 \\
\end{pmatrix}.
\end{align}
The form factors are plotted as a functions of $z$ in fig.~\ref{fig:FFK}. The solid curves show our central value and  $1\sigma$ error band, the blue error bar shows the LCSR prediction used as input to the fit, and the dashed lines show the extrapolation of the lattice result in \cite{Bouchard:2013eph} for comparison.

\section{Theory predictions vs.\ experimental data}\label{sec:obstables}

In this appendix, we give all the SM predictions for the relevant observables 
as well as the corresponding experimental measurements. Differential branching 
ratios are given in units of GeV$^{-2}$.

\renewcommand{\arraystretch}{1.2}
\subsection*{\texorpdfstring{$\bar B^0\to\bar K^{*0}\mu^+\mu^-$}{}}

\begin{longtable}[l]{cccccc}
\hline
Obs. & $q^2$ bin & SM pred. & \multicolumn{2}{c}{measurement} & pull\\
\hline
\multirow{9}{*}{$10^{7}~\frac{d\text{BR}}{dq^2}$} & \multirow{1}{*}{$[0,2]$} & \multirow{1}{*}{$0.82 \pm 0.11$} & $0.91 \pm 0.18$ & CDF & {$-0.4$}\\
 & \multirow{1}{*}{$[0.1,2]$} & \multirow{1}{*}{$0.76 \pm 0.11$} & $0.58 \pm 0.09$ & LHCb & {$+1.3$}\\
 & \multirow{1}{*}{$[1,2]$} & \multirow{1}{*}{$0.49 \pm 0.08$} & $0.49 \pm 0.14$ & CMS & {$+0.0$}\\
\arrayrulecolor{black!20}\cline{3-3}\arrayrulecolor{black}
 & \multirow{3}{*}{$[2,4.3]$} & \multirow{3}{*}{$0.44 \pm 0.07$} & $0.46 \pm 0.12$ & CDF & {$-0.1$}\\
 &  &  & $0.38 \pm 0.08$ & CMS & {$+0.6$}\\
 &  &  & $0.29 \pm 0.05$ & LHCb & {$+1.8$}\\
\arrayrulecolor{black!20}\cline{3-3}\arrayrulecolor{black}
 & \multirow{2}{*}{$[16,19]$} & \multirow{2}{*}{$0.49 \pm 0.06$} & $0.52 \pm 0.08$ & CMS & {$-0.3$}\\
 &  &  & $0.40 \pm 0.07$ & LHCb & {$+1.1$}\\
\arrayrulecolor{black!20}\cline{3-3}\arrayrulecolor{black}
 & \multirow{1}{*}{$[16,19.25]$} & \multirow{1}{*}{$0.47 \pm 0.06$} & $0.31 \pm 0.07$ & CDF & {$+1.8$}\\
\hline
\multirow{6}{*}{$A_9$} & \multirow{1}{*}{$[0,2]$} & \multirow{1}{*}{$0.00 \pm 0.00$} & $0.30 \pm 0.25$ & CDF & {$-1.2$}\\
 & \multirow{1}{*}{$[0.1,2]$} & \multirow{1}{*}{$0.00 \pm 0.00$} & $0.14 \pm 0.11$ & LHCb & {$-1.3$}\\
\arrayrulecolor{black!20}\cline{3-3}\arrayrulecolor{black}
 & \multirow{2}{*}{$[2,4.3]$} & \multirow{2}{*}{$0.00 \pm 0.00$} & $-0.08 \pm 0.37$ & CDF & {$+0.2$}\\
 &  &  & $0.08 \pm 0.10$ & LHCb & {$-0.8$}\\
\arrayrulecolor{black!20}\cline{3-3}\arrayrulecolor{black}
 & \multirow{1}{*}{$[16,19]$} & \multirow{1}{*}{$0.00 \pm 0.00$} & $0.00 \pm 0.11$ & LHCb & {$+0.0$}\\
 & \multirow{1}{*}{$[16,19.25]$} & \multirow{1}{*}{$0.00 \pm 0.00$} & $-0.01 \pm 0.25$ & CDF & {$+0.0$}\\
\hline
\multirow{13}{*}{$A_\text{FB}$} & \multirow{1}{*}{$[0,2]$} & \multirow{1}{*}{$-0.10 \pm 0.01$} & $0.06 \pm 0.29$ & CDF & {$-0.5$}\\
 & \multirow{1}{*}{$[0.1,1]$} & \multirow{1}{*}{$-0.09 \pm 0.01$} & $-0.00 \pm 0.06$ & LHCb & {$-1.5$}\\
 & \multirow{1}{*}{$[1,2]$} & \multirow{1}{*}{$-0.15 \pm 0.02$} & $-0.11 \pm 0.26$ & CMS & {$-0.2$}\\
 & \multirow{1}{*}{$[1.1,2.5]$} & \multirow{1}{*}{$-0.13 \pm 0.02$} & $-0.20 \pm 0.07$ & LHCb & {$+0.8$}\\
\arrayrulecolor{black!20}\cline{3-3}\arrayrulecolor{black}
 & \multirow{3}{*}{$[2,4.3]$} & \multirow{3}{*}{$-0.03 \pm 0.03$} & $0.22 \pm 0.31$ & ATLAS & {$-0.8$}\\
 &  &  & $-0.15 \pm 0.41$ & CDF & {$+0.3$}\\
 &  &  & $-0.07 \pm 0.20$ & CMS & {$+0.2$}\\
\arrayrulecolor{black!20}\cline{3-3}\arrayrulecolor{black}
 & \multirow{1}{*}{$[2.5,4]$} & \multirow{1}{*}{$-0.02 \pm 0.03$} & $-0.12 \pm 0.08$ & LHCb & {$+1.2$}\\
 & \multirow{1}{*}{$[4,6]$} & \multirow{1}{*}{$0.12 \pm 0.04$} & $0.03 \pm 0.05$ & LHCb & {$+1.4$}\\
 & \multirow{1}{*}{$[15,19]$} & \multirow{1}{*}{$0.37 \pm 0.03$} & $0.36 \pm 0.03$ & LHCb & {$+0.3$}\\
\arrayrulecolor{black!20}\cline{3-3}\arrayrulecolor{black}
 & \multirow{2}{*}{$[16,19]$} & \multirow{2}{*}{$0.35 \pm 0.03$} & $0.16 \pm 0.10$ & ATLAS & {$+1.7$}\\
 &  &  & $0.41 \pm 0.06$ & CMS & {$-0.9$}\\
\arrayrulecolor{black!20}\cline{3-3}\arrayrulecolor{black}
 & \multirow{1}{*}{$[16,19.25]$} & \multirow{1}{*}{$0.35 \pm 0.03$} & $0.34 \pm 0.19$ & CDF & {$+0.0$}\\
\hline
\multirow{13}{*}{$F_L$} & \multirow{1}{*}{$[0,2]$} & \multirow{1}{*}{$0.39 \pm 0.04$} & $0.26 \pm 0.14$ & CDF & {$+0.9$}\\
 & \multirow{1}{*}{$[0.1,1]$} & \multirow{1}{*}{$0.30 \pm 0.04$} & $0.26 \pm 0.05$ & LHCb & {$+0.5$}\\
 & \multirow{1}{*}{$[1,2]$} & \multirow{1}{*}{$0.73 \pm 0.04$} & $0.46 \pm 0.24$ & CMS & {$+1.1$}\\
 & \multirow{1}{*}{$[1.1,2.5]$} & \multirow{1}{*}{$0.77 \pm 0.03$} & $0.67 \pm 0.08$ & LHCb & {$+1.1$}\\
\arrayrulecolor{black!20}\cline{3-3}\arrayrulecolor{black}
 & \multirow{3}{*}{$[2,4.3]$} & \multirow{3}{*}{$0.81 \pm 0.02$} & $0.26 \pm 0.19$ & ATLAS & {\boldmath$+2.9$}\\
 &  &  & $0.70 \pm 0.17$ & CDF & {$+0.6$}\\
 &  &  & $0.65 \pm 0.17$ & CMS & {$+0.9$}\\
\arrayrulecolor{black!20}\cline{3-3}\arrayrulecolor{black}
 & \multirow{1}{*}{$[2.5,4]$} & \multirow{1}{*}{$0.82 \pm 0.02$} & $0.87 \pm 0.09$ & LHCb & {$-0.6$}\\
 & \multirow{1}{*}{$[4,6]$} & \multirow{1}{*}{$0.74 \pm 0.04$} & $0.61 \pm 0.06$ & LHCb & {$+1.9$}\\
 & \multirow{1}{*}{$[15,19]$} & \multirow{1}{*}{$0.34 \pm 0.04$} & $0.34 \pm 0.03$ & LHCb & {$-0.1$}\\
\arrayrulecolor{black!20}\cline{3-3}\arrayrulecolor{black}
 & \multirow{2}{*}{$[16,19]$} & \multirow{2}{*}{$0.33 \pm 0.04$} & $0.35 \pm 0.08$ & ATLAS & {$-0.2$}\\
 &  &  & $0.44 \pm 0.08$ & CMS & {$-1.3$}\\
\arrayrulecolor{black!20}\cline{3-3}\arrayrulecolor{black}
 & \multirow{1}{*}{$[16,19.25]$} & \multirow{1}{*}{$0.33 \pm 0.04$} & $0.20 \pm 0.13$ & CDF & {$+1.0$}\\
\hline
\multirow{5}{*}{$S_3$} & \multirow{1}{*}{$[0.1,1]$} & \multirow{1}{*}{$0.01 \pm 0.00$} & $-0.04 \pm 0.06$ & LHCb & {$+0.7$}\\
 & \multirow{1}{*}{$[1.1,2.5]$} & \multirow{1}{*}{$0.00 \pm 0.00$} & $-0.08 \pm 0.10$ & LHCb & {$+0.9$}\\
 & \multirow{1}{*}{$[2.5,4]$} & \multirow{1}{*}{$-0.01 \pm 0.00$} & $0.04 \pm 0.09$ & LHCb & {$-0.6$}\\
 & \multirow{1}{*}{$[4,6]$} & \multirow{1}{*}{$-0.02 \pm 0.01$} & $0.04 \pm 0.07$ & LHCb & {$-0.9$}\\
 & \multirow{1}{*}{$[15,19]$} & \multirow{1}{*}{$-0.21 \pm 0.02$} & $-0.18 \pm 0.02$ & LHCb & {$-1.0$}\\
\hline
\multirow{5}{*}{$S_4$} & \multirow{1}{*}{$[0.1,1]$} & \multirow{1}{*}{$0.10 \pm 0.00$} & $0.08 \pm 0.07$ & LHCb & {$+0.2$}\\
 & \multirow{1}{*}{$[1.1,2.5]$} & \multirow{1}{*}{$-0.01 \pm 0.01$} & $-0.08 \pm 0.11$ & LHCb & {$+0.6$}\\
 & \multirow{1}{*}{$[2.5,4]$} & \multirow{1}{*}{$-0.14 \pm 0.02$} & $-0.24 \pm 0.14$ & LHCb & {$+0.7$}\\
 & \multirow{1}{*}{$[4,6]$} & \multirow{1}{*}{$-0.22 \pm 0.02$} & $-0.22 \pm 0.09$ & LHCb & {$+0.0$}\\
 & \multirow{1}{*}{$[15,19]$} & \multirow{1}{*}{$-0.30 \pm 0.01$} & $-0.29 \pm 0.04$ & LHCb & {$-0.4$}\\
\hline
\multirow{5}{*}{$S_5$} & \multirow{1}{*}{$[0.1,1]$} & \multirow{1}{*}{$0.24 \pm 0.01$} & $0.17 \pm 0.06$ & LHCb & {$+1.2$}\\
 & \multirow{1}{*}{$[1.1,2.5]$} & \multirow{1}{*}{$0.06 \pm 0.03$} & $0.14 \pm 0.10$ & LHCb & {$-0.7$}\\
 & \multirow{1}{*}{$[2.5,4]$} & \multirow{1}{*}{$-0.18 \pm 0.04$} & $-0.02 \pm 0.11$ & LHCb & {$-1.5$}\\
 & \multirow{1}{*}{$[4,6]$} & \multirow{1}{*}{$-0.33 \pm 0.03$} & $-0.15 \pm 0.08$ & LHCb & {\boldmath$-2.2$}\\
 & \multirow{1}{*}{$[15,19]$} & \multirow{1}{*}{$-0.28 \pm 0.02$} & $-0.33 \pm 0.04$ & LHCb & {$+1.0$}\\
\hline
\end{longtable}

\subsection*{\texorpdfstring{$B^-\to K^{*-}\mu^+\mu^-$}{}}

\begin{longtable}[l]{cccccc}
\hline
Obs. & $q^2$ bin & SM pred. & \multicolumn{2}{c}{measurement} & pull\\
\hline
\multirow{4}{*}{$10^{7}~\frac{d\text{BR}}{dq^2}$} & \multirow{1}{*}{$[0.1,2]$} & \multirow{1}{*}{$0.81 \pm 0.11$} & $0.60 \pm 0.14$ & LHCb & {$+1.2$}\\
 & \multirow{1}{*}{$[2,4]$} & \multirow{1}{*}{$0.48 \pm 0.08$} & $0.57 \pm 0.16$ & LHCb & {$-0.5$}\\
 & \multirow{1}{*}{$[4,6]$} & \multirow{1}{*}{$0.54 \pm 0.08$} & $0.26 \pm 0.10$ & LHCb & {\boldmath$+2.1$}\\
 & \multirow{1}{*}{$[15,19]$} & \multirow{1}{*}{$0.58 \pm 0.07$} & $0.40 \pm 0.08$ & LHCb & {$+1.7$}\\
\hline
\end{longtable}

\subsection*{\texorpdfstring{$\bar B^0\to\bar K^{0}\mu^+\mu^-$}{}}

\begin{longtable}[l]{cccccc}
\hline
Obs. & $q^2$ bin & SM pred. & \multicolumn{2}{c}{measurement} & pull\\
\hline
\multirow{7}{*}{$10^{8}~\frac{d\text{BR}}{dq^2}$} & \multirow{1}{*}{$[0,2]$} & \multirow{1}{*}{$2.63 \pm 0.49$} & $2.45 \pm 1.60$ & CDF & {$+0.1$}\\
 & \multirow{1}{*}{$[0.1,2]$} & \multirow{1}{*}{$2.71 \pm 0.50$} & $1.26 \pm 0.56$ & LHCb & {$+1.9$}\\
 & \multirow{1}{*}{$[2,4]$} & \multirow{1}{*}{$2.76 \pm 0.47$} & $1.90 \pm 0.53$ & LHCb & {$+1.2$}\\
 & \multirow{1}{*}{$[2,4.3]$} & \multirow{1}{*}{$2.77 \pm 0.47$} & $2.55 \pm 1.74$ & CDF & {$+0.1$}\\
 & \multirow{1}{*}{$[4,6]$} & \multirow{1}{*}{$2.81 \pm 0.46$} & $1.76 \pm 0.51$ & LHCb & {$+1.5$}\\
 & \multirow{1}{*}{$[15,22]$} & \multirow{1}{*}{$1.19 \pm 0.15$} & $0.96 \pm 0.16$ & LHCb & {$+1.1$}\\
 & \multirow{1}{*}{$[16,23]$} & \multirow{1}{*}{$0.93 \pm 0.12$} & $0.37 \pm 0.22$ & CDF & {\boldmath$+2.2$}\\
\hline
\end{longtable}

\subsection*{\texorpdfstring{$B^+\to K^+\mu^+\mu^-$}{}}

\begin{longtable}[l]{cccccc}
\hline
Obs. & $q^2$ bin & SM pred. & \multicolumn{2}{c}{measurement} & pull\\
\hline
\multirow{10}{*}{$10^{8}~\frac{d\text{BR}}{dq^2}$} & \multirow{1}{*}{$[0,2]$} & \multirow{1}{*}{$2.84 \pm 0.53$} & $1.80 \pm 0.53$ & CDF & {$+1.4$}\\
 & \multirow{1}{*}{$[0.1,1]$} & \multirow{1}{*}{$2.90 \pm 0.56$} & $3.32 \pm 0.25$ & LHCb & {$-0.7$}\\
 & \multirow{1}{*}{$[1.1,2]$} & \multirow{1}{*}{$2.94 \pm 0.53$} & $2.33 \pm 0.19$ & LHCb & {$+1.1$}\\
 & \multirow{1}{*}{$[2,3]$} & \multirow{1}{*}{$2.97 \pm 0.51$} & $2.82 \pm 0.21$ & LHCb & {$+0.3$}\\
 & \multirow{1}{*}{$[2,4.3]$} & \multirow{1}{*}{$2.99 \pm 0.50$} & $3.16 \pm 0.57$ & CDF & {$-0.2$}\\
 & \multirow{1}{*}{$[3,4]$} & \multirow{1}{*}{$3.00 \pm 0.50$} & $2.54 \pm 0.20$ & LHCb & {$+0.8$}\\
 & \multirow{1}{*}{$[4,5]$} & \multirow{1}{*}{$3.02 \pm 0.50$} & $2.21 \pm 0.18$ & LHCb & {$+1.5$}\\
 & \multirow{1}{*}{$[5,6]$} & \multirow{1}{*}{$3.05 \pm 0.50$} & $2.31 \pm 0.18$ & LHCb & {$+1.4$}\\
 & \multirow{1}{*}{$[15,22]$} & \multirow{1}{*}{$1.29 \pm 0.17$} & $1.21 \pm 0.07$ & LHCb & {$+0.4$}\\
 & \multirow{1}{*}{$[16,23]$} & \multirow{1}{*}{$1.01 \pm 0.13$} & $0.72 \pm 0.15$ & CDF & {$+1.5$}\\
\hline
\end{longtable}

\subsection*{\texorpdfstring{$\bar B^0\to\bar K^{*0}\gamma$}{}}

\begin{longtable}[l]{cccccc}
\hline
Obs. & SM pred. & \multicolumn{2}{c}{measurement} & pull \\
\hline
\multirow{1}{*}{$10^{5}~\text{BR}$}  & \multirow{1}{*}{$4.21 \pm 0.68$} & $4.33 \pm 0.15$ & HFAG & {$-0.2$}\\
\hline
\multirow{1}{*}{$S$}  & \multirow{1}{*}{$-0.02 \pm 0.00$} & $-0.16 \pm 0.22$ & HFAG & {$+0.6$}\\
\hline
\end{longtable}

\subsection*{\texorpdfstring{$B^-\to K^{*-}\gamma$}{}}

\begin{longtable}[l]{cccccc}
\hline
Obs. & SM pred. & \multicolumn{2}{c}{measurement} & pull \\
\hline
\multirow{1}{*}{$10^{5}~\text{BR}$}  & \multirow{1}{*}{$4.42 \pm 0.73$} & $4.21 \pm 0.18$ & HFAG & {$+0.3$}\\
\hline
\end{longtable}

\subsection*{\texorpdfstring{$B\to X_s\gamma$}{}}

\begin{longtable}[l]{cccccc}
\hline
Obs. & SM pred. & \multicolumn{2}{c}{measurement} & pull \\
\hline
\multirow{1}{*}{$10^{4}~\text{BR}$}  & \multirow{1}{*}{$3.36 \pm 0.23$} & $3.43 \pm 0.22$ & HFAG & {$-0.2$}\\
\hline
\end{longtable}

\subsection*{\texorpdfstring{$B_s\to\mu^+\mu^-$}{}}

\begin{longtable}[l]{cccccc}
\hline
Obs. & SM pred. & \multicolumn{2}{c}{measurement} & pull \\
\hline
\multirow{1}{*}{$10^{9}~\text{BR}$}  & \multirow{1}{*}{$3.40 \pm 0.21$} & $2.90 \pm 0.70$ & LHCb+CMS & {$+0.7$}\\
\hline
\end{longtable}

\subsection*{\texorpdfstring{$B\to X_s\mu^+\mu^-$}{}}

\begin{longtable}[l]{cccccc}
\hline
Obs. & $q^2$ bin & SM pred. & \multicolumn{2}{c}{measurement} & pull\\
\hline
\multirow{2}{*}{$10^{6}~\text{BR}$} & \multirow{1}{*}{$[1,6]$} & \multirow{1}{*}{$1.59 \pm 0.11$} & $0.72 \pm 0.84$ & BaBar & {$+1.0$}\\
 & \multirow{1}{*}{$[14.2,25]$} & \multirow{1}{*}{$0.24 \pm 0.07$} & $0.62 \pm 0.30$ & BaBar & {$-1.2$}\\
\hline
\end{longtable}

\subsection*{\texorpdfstring{$B_s\to\phi\mu^+\mu^-$}{}}

\begin{longtable}[l]{cccccc}
\hline
Obs. & $q^2$ bin & SM pred. & \multicolumn{2}{c}{measurement} & pull\\
\hline
\multirow{4}{*}{$10^{7}~\frac{d\text{BR}}{dq^2}$} & \multirow{2}{*}{$[1,6]$} & \multirow{2}{*}{$0.48 \pm 0.06$} & $0.21 \pm 0.15$ & CDF & {$+1.7$}\\
 &  &  & $0.23 \pm 0.05$ & LHCb & {\boldmath$+3.1$}\\
\arrayrulecolor{black!20}\cline{3-3}\arrayrulecolor{black}
 & \multirow{2}{*}{$[16,19]$} & \multirow{2}{*}{$0.41 \pm 0.05$} & $0.80 \pm 0.32$ & CDF & {$-1.2$}\\
 &  &  & $0.36 \pm 0.08$ & LHCb & {$+0.6$}\\
\hline
\end{longtable}

\subsection*{\texorpdfstring{$B^+\to K^+e^+e^-$}{}}

\begin{longtable}[l]{cccccc}
\hline
Obs. & $q^2$ bin & SM pred. & \multicolumn{2}{c}{measurement} & pull\\
\hline
\multirow{1}{*}{$10^{8}~\frac{d\text{BR}}{dq^2}$} & \multirow{1}{*}{$[1,6]$} & \multirow{1}{*}{$2.99 \pm 0.50$} & $3.18 \pm 0.35$ & LHCb & {$-0.3$}\\
\hline
\end{longtable}

\subsection*{\texorpdfstring{$B\to X_se^+e^-$}{}}

\begin{longtable}[l]{cccccc}
\hline
Obs. & $q^2$ bin & SM pred. & \multicolumn{2}{c}{measurement} & pull\\
\hline
\multirow{2}{*}{$10^{6}~\text{BR}$} &  & \multirow{1}{*}{$1.64 \pm 0.11$} & $1.97 \pm 0.53$ & BaBar & {$-0.6$}\\
 & \multirow{1}{*}{$[14.2,25]$} & \multirow{1}{*}{$0.21 \pm 0.07$} & $0.57 \pm 0.19$ & BaBar & {$-1.8$}\\
\hline
\end{longtable}
\renewcommand{\arraystretch}{1.0}

\section{Constraints on pairs of Wilson coefficients}\label{sec:2Dplots}

\begin{figure}[p]
\includegraphics[width=0.43\textwidth]{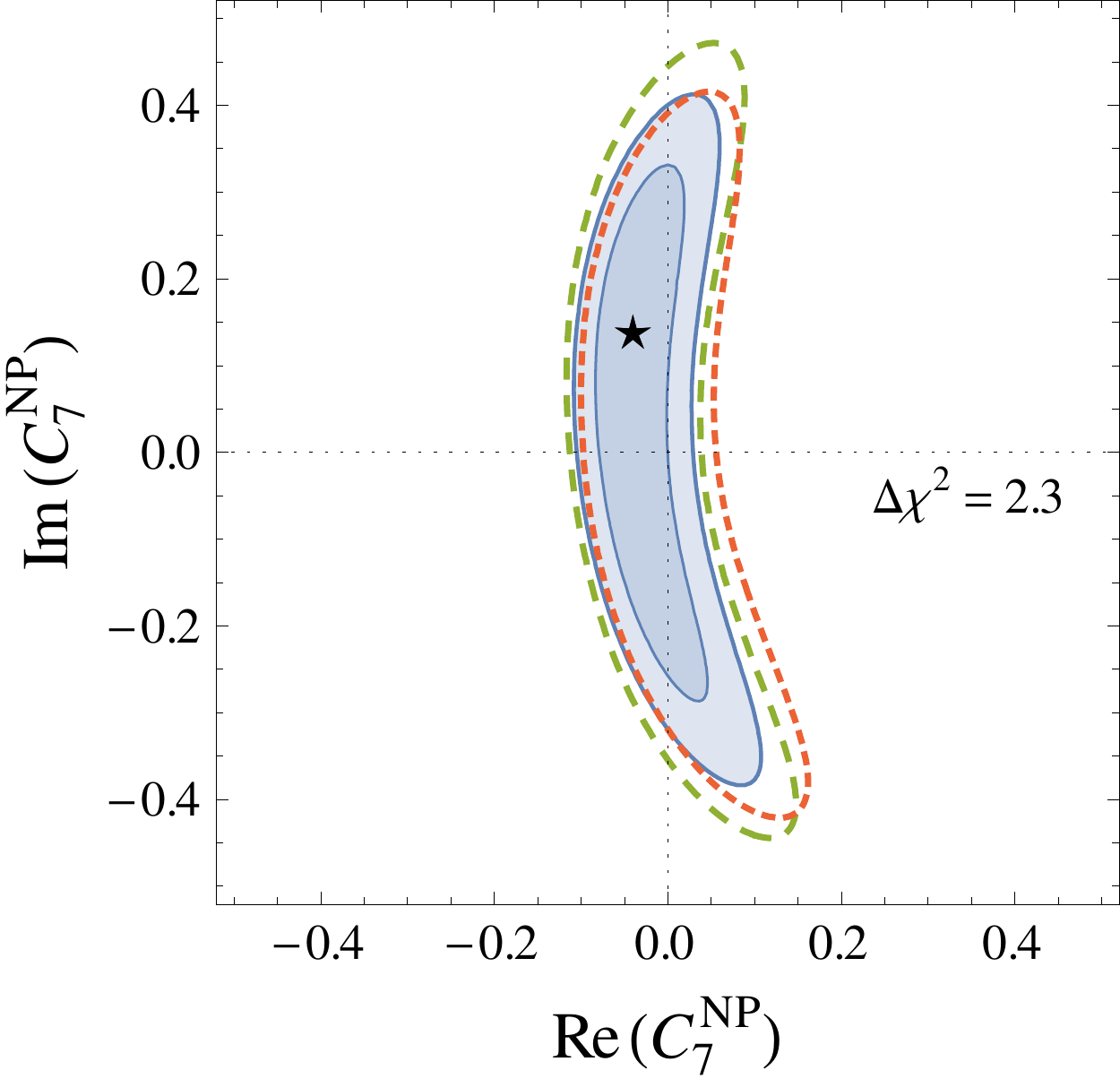} ~~~~%
\includegraphics[width=0.43\textwidth]{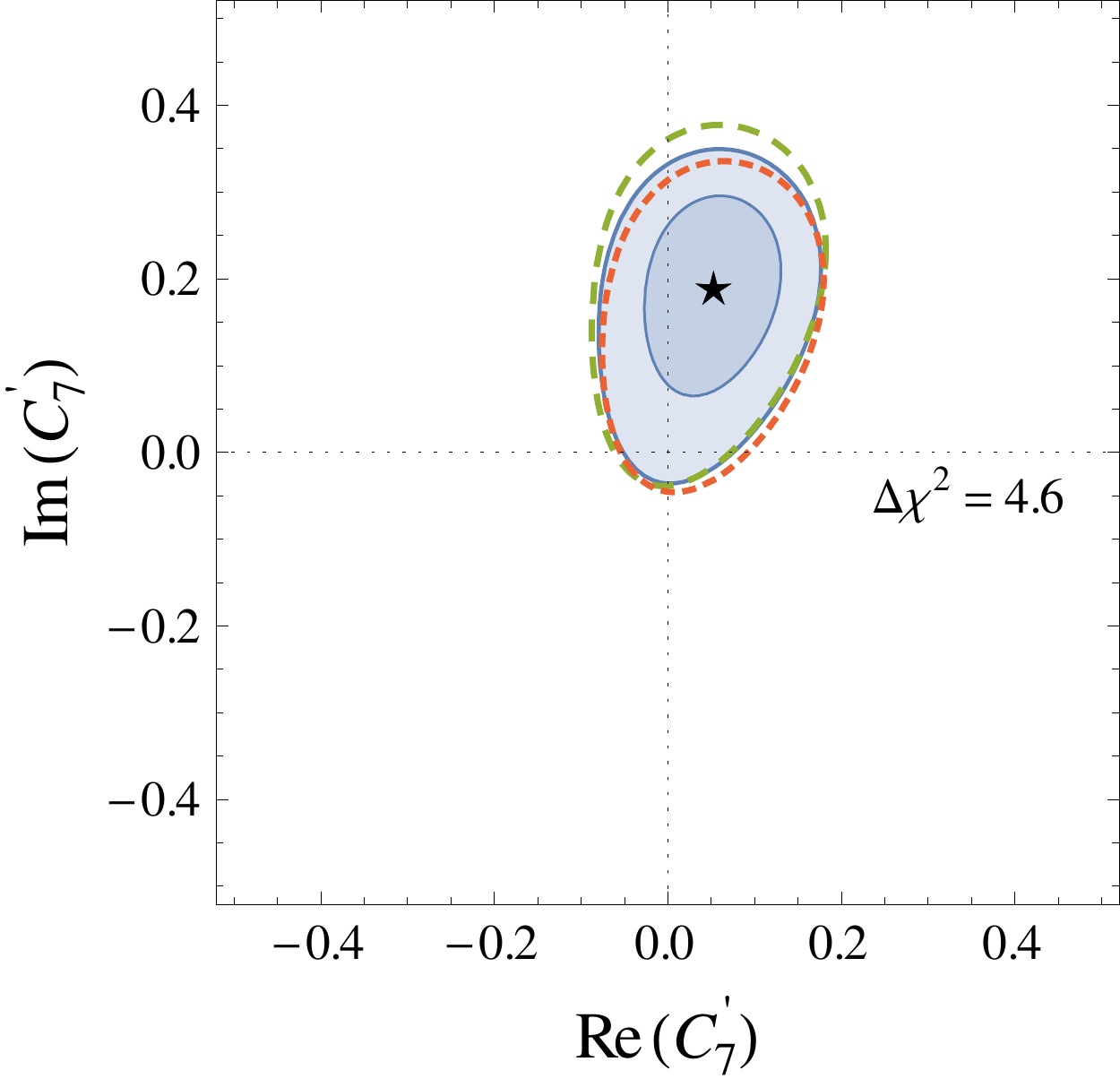} \\[8pt]%

~~\includegraphics[width=0.42\textwidth]{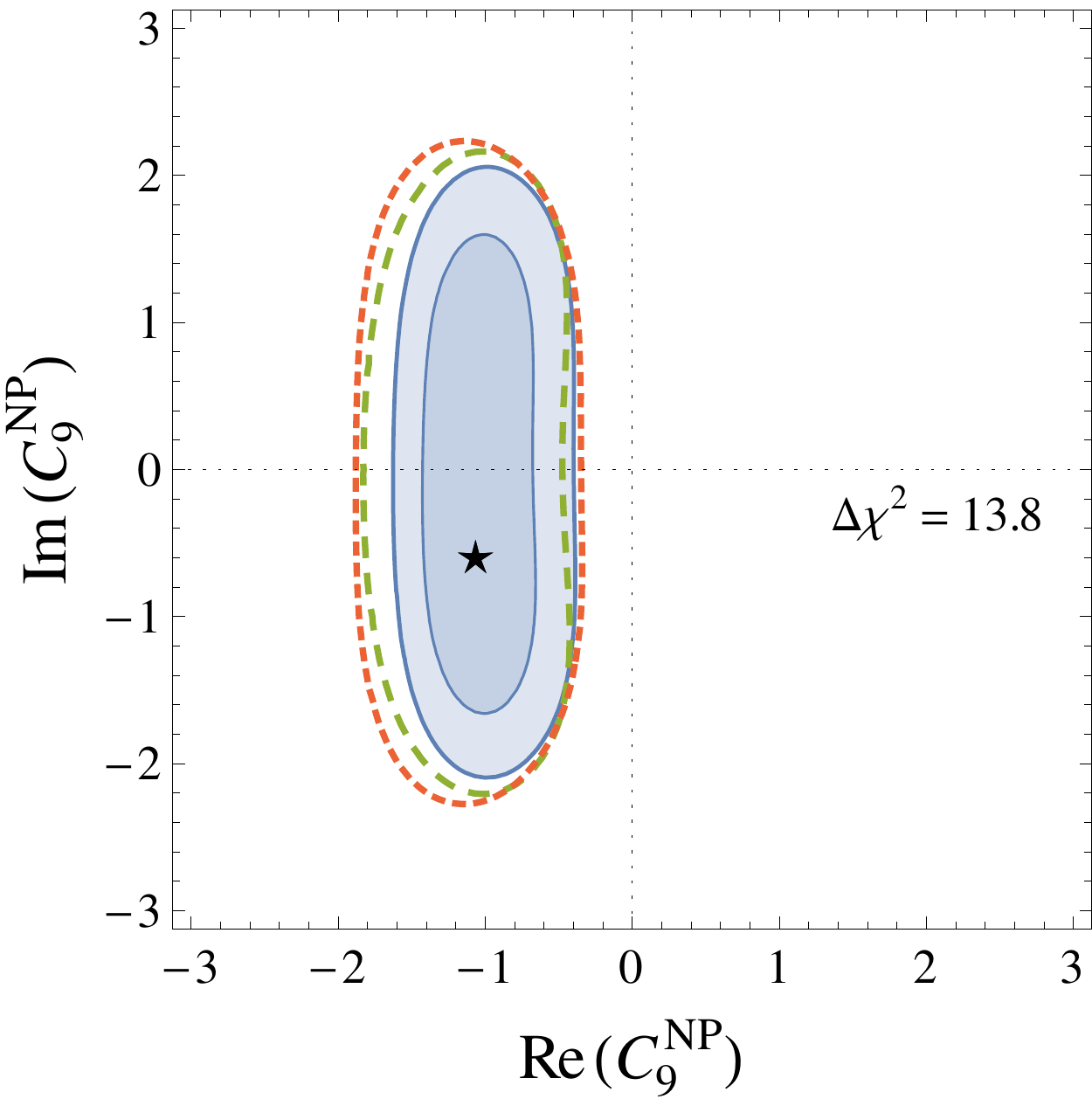} ~~~~~%
\includegraphics[width=0.42\textwidth]{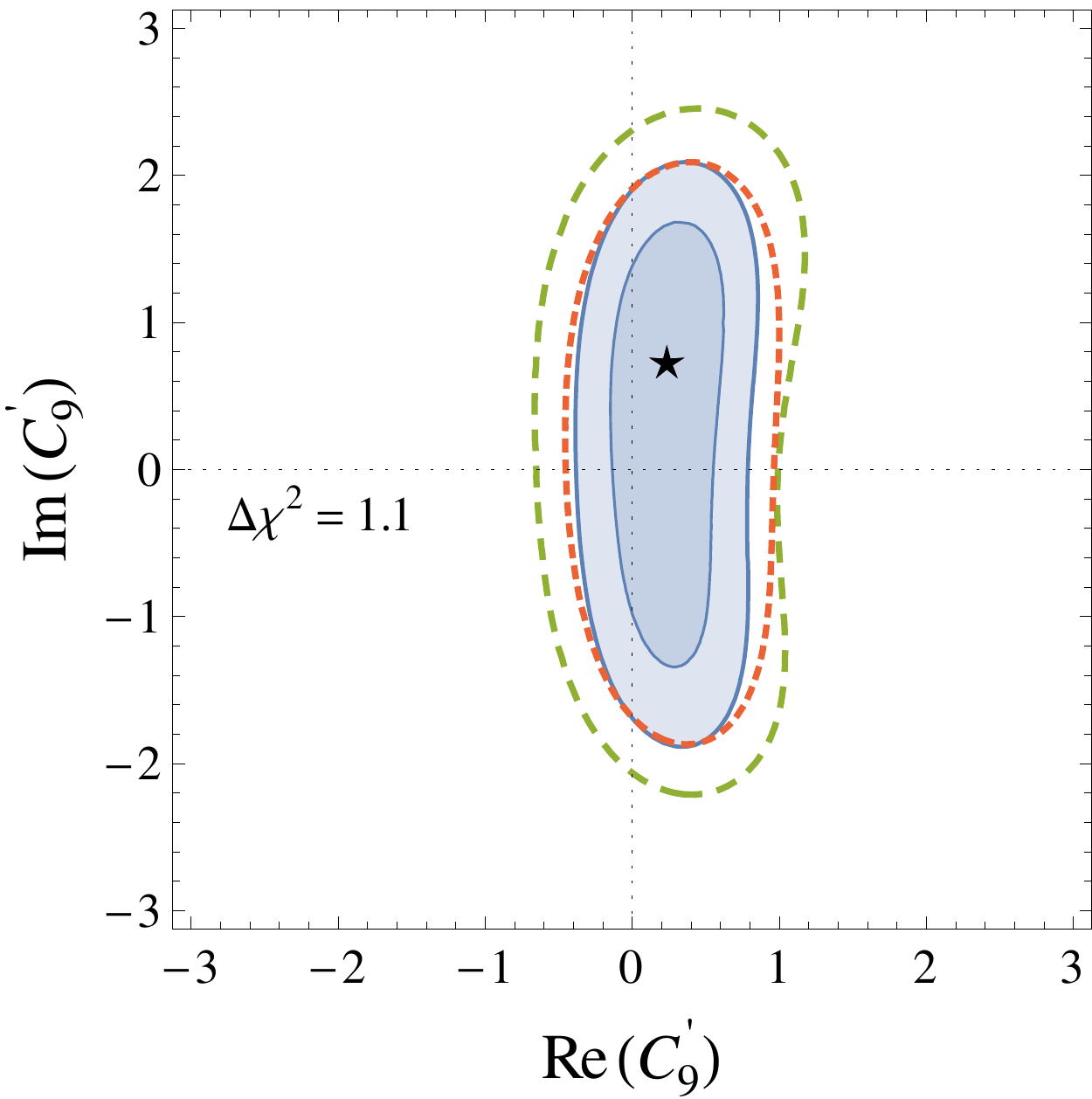} \\[8pt]%

~~\includegraphics[width=0.42\textwidth]{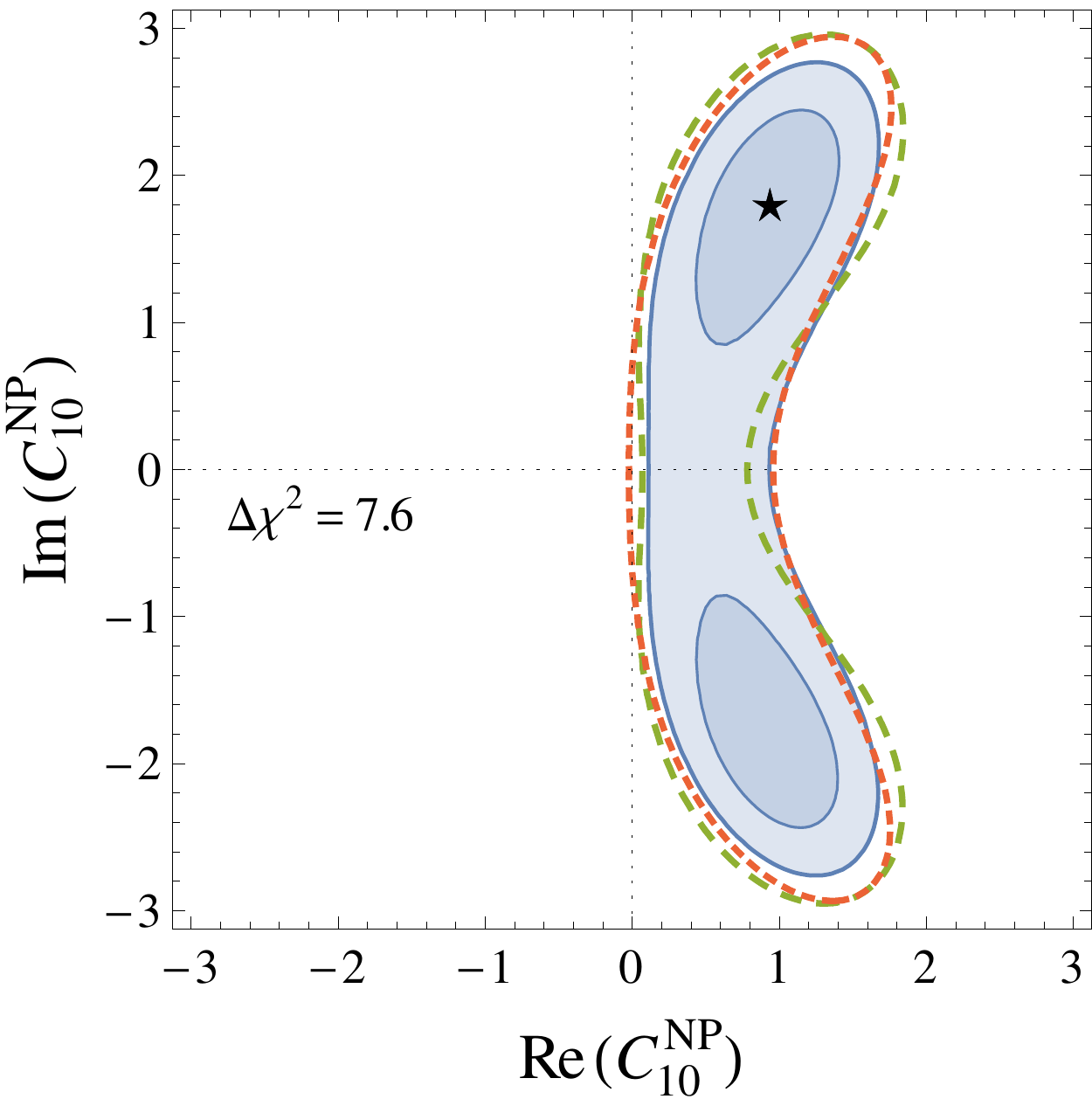} ~~~~~%
\includegraphics[width=0.42\textwidth]{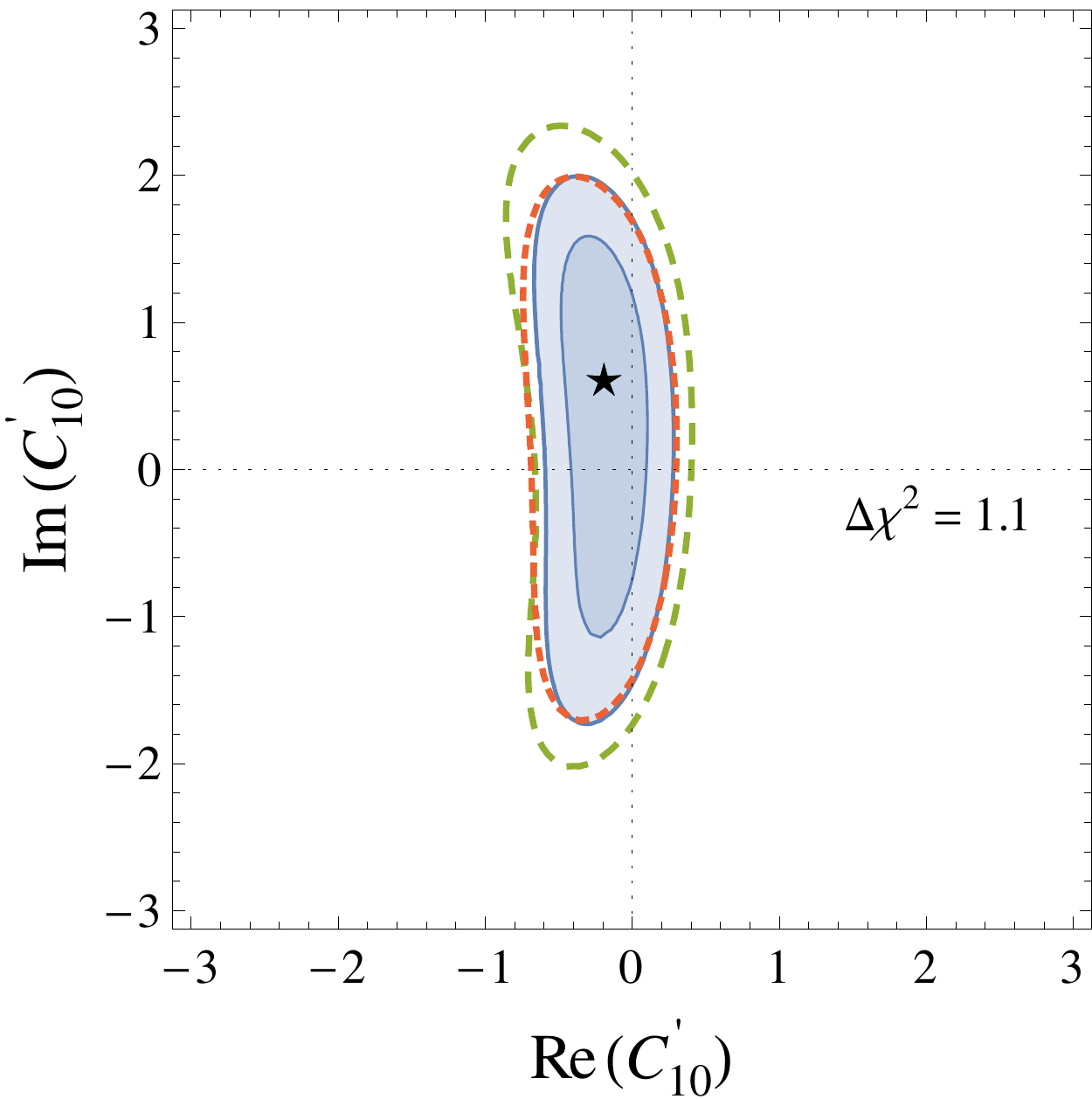}%
\caption{Constraints on complex Wilson coefficients. Contours are as in fig.~\ref{fig:2D-zoom}}
\label{fig:2Dconstraints1}
\end{figure}

\begin{figure}[p]
\centering
\includegraphics[width=0.43\textwidth]{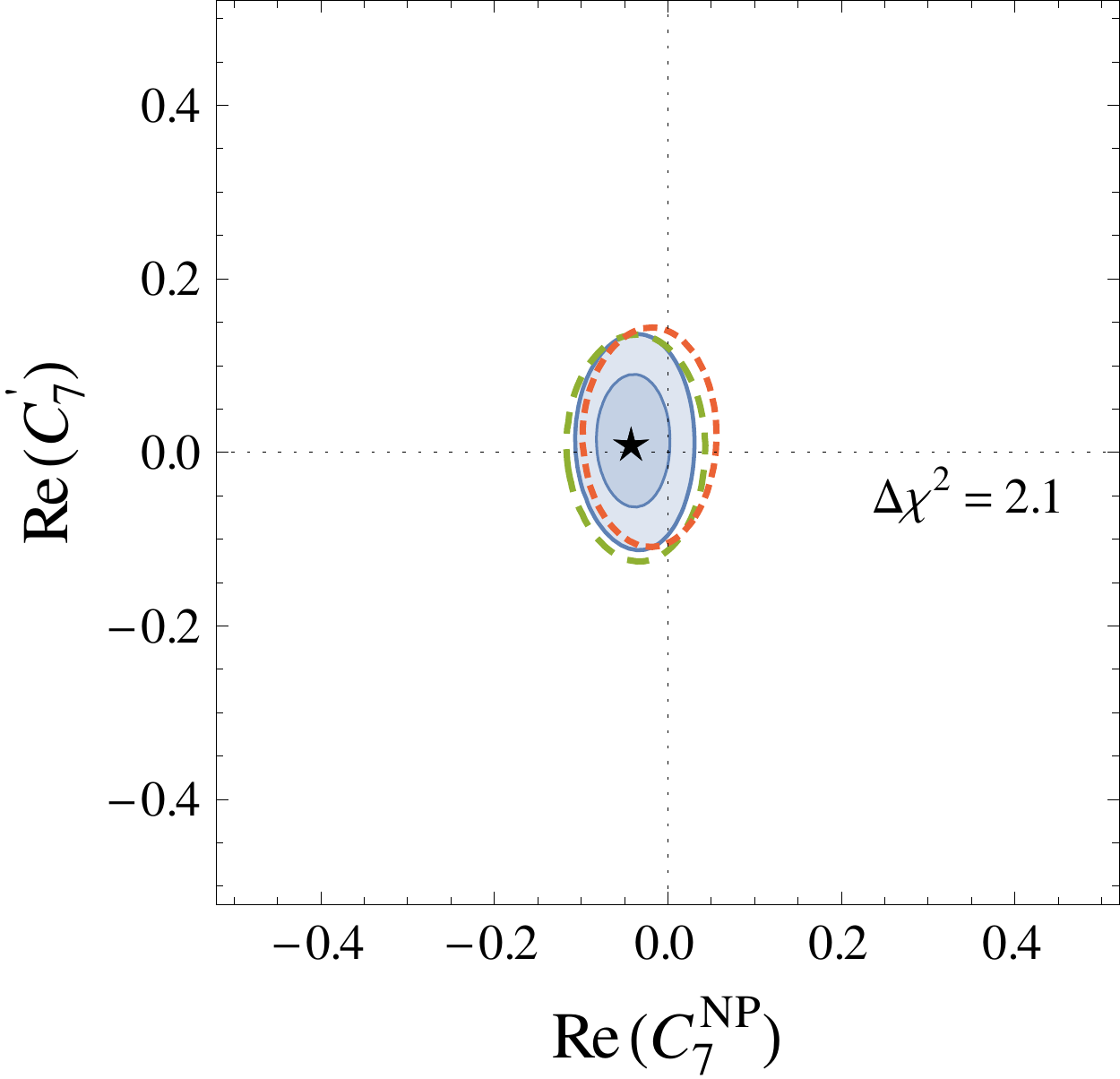} ~~~~~%
\includegraphics[width=0.42\textwidth]{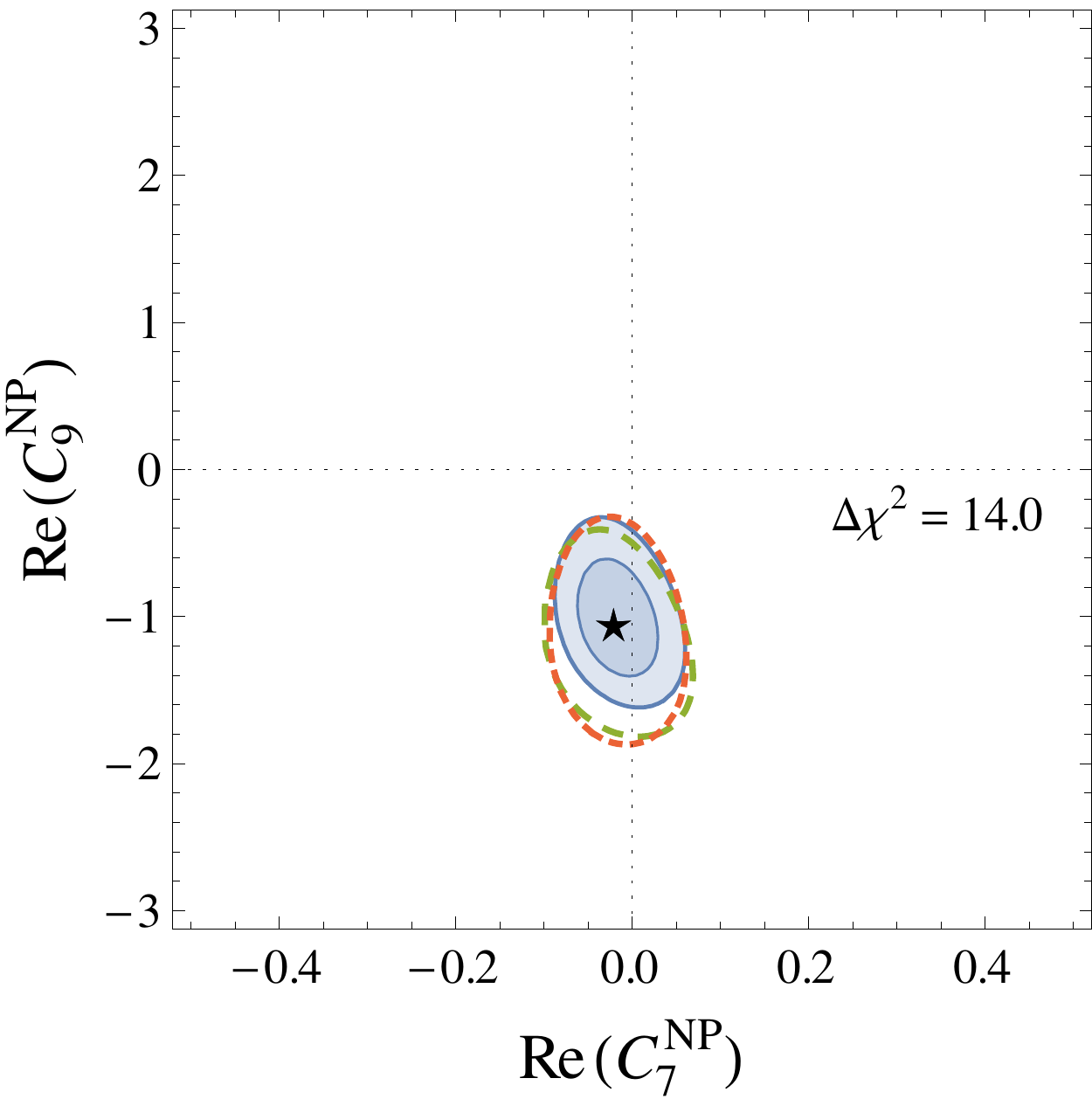} \\[8pt]%

~~\includegraphics[width=0.42\textwidth]{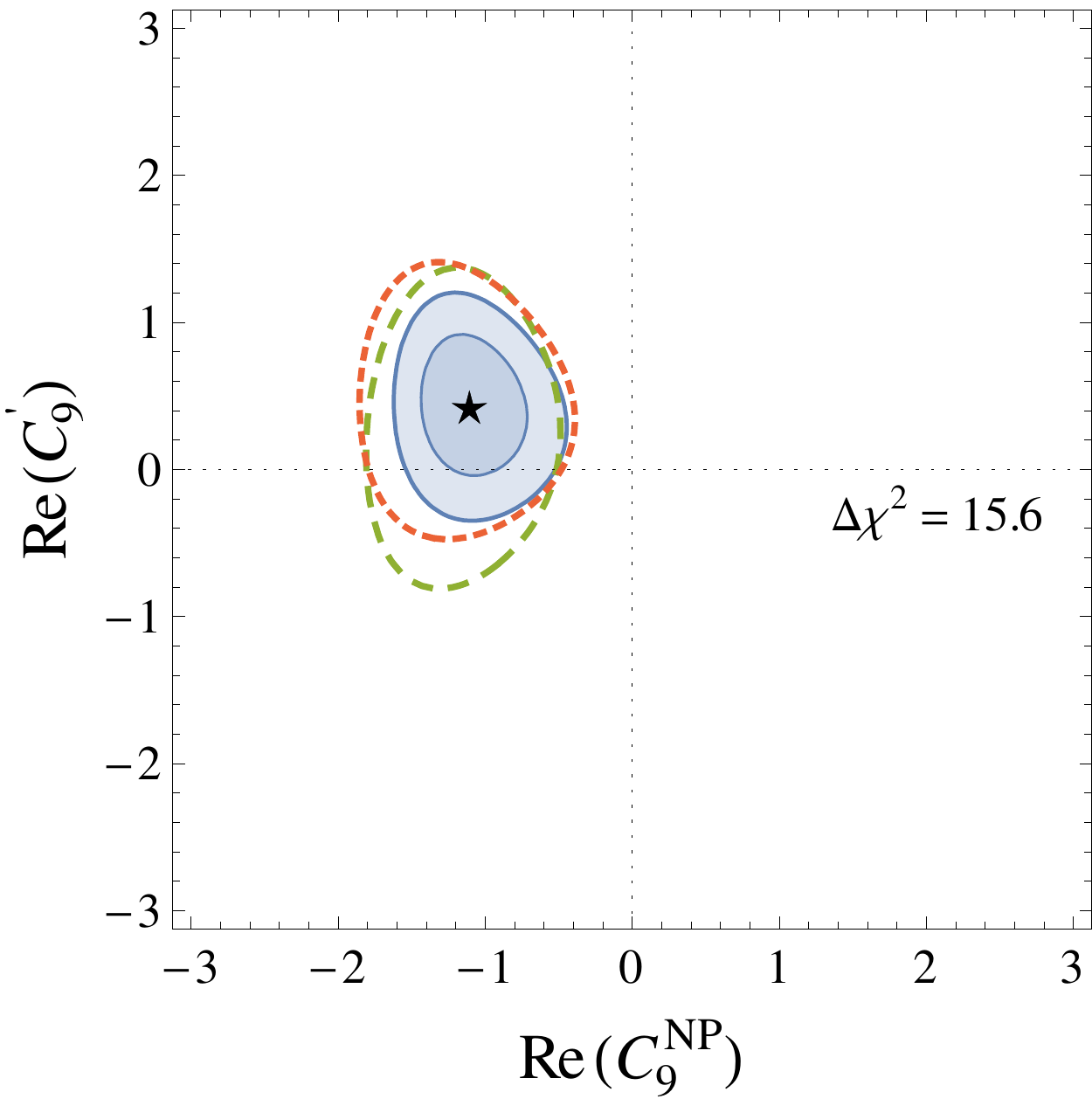} ~~~~~%
\includegraphics[width=0.42\textwidth]{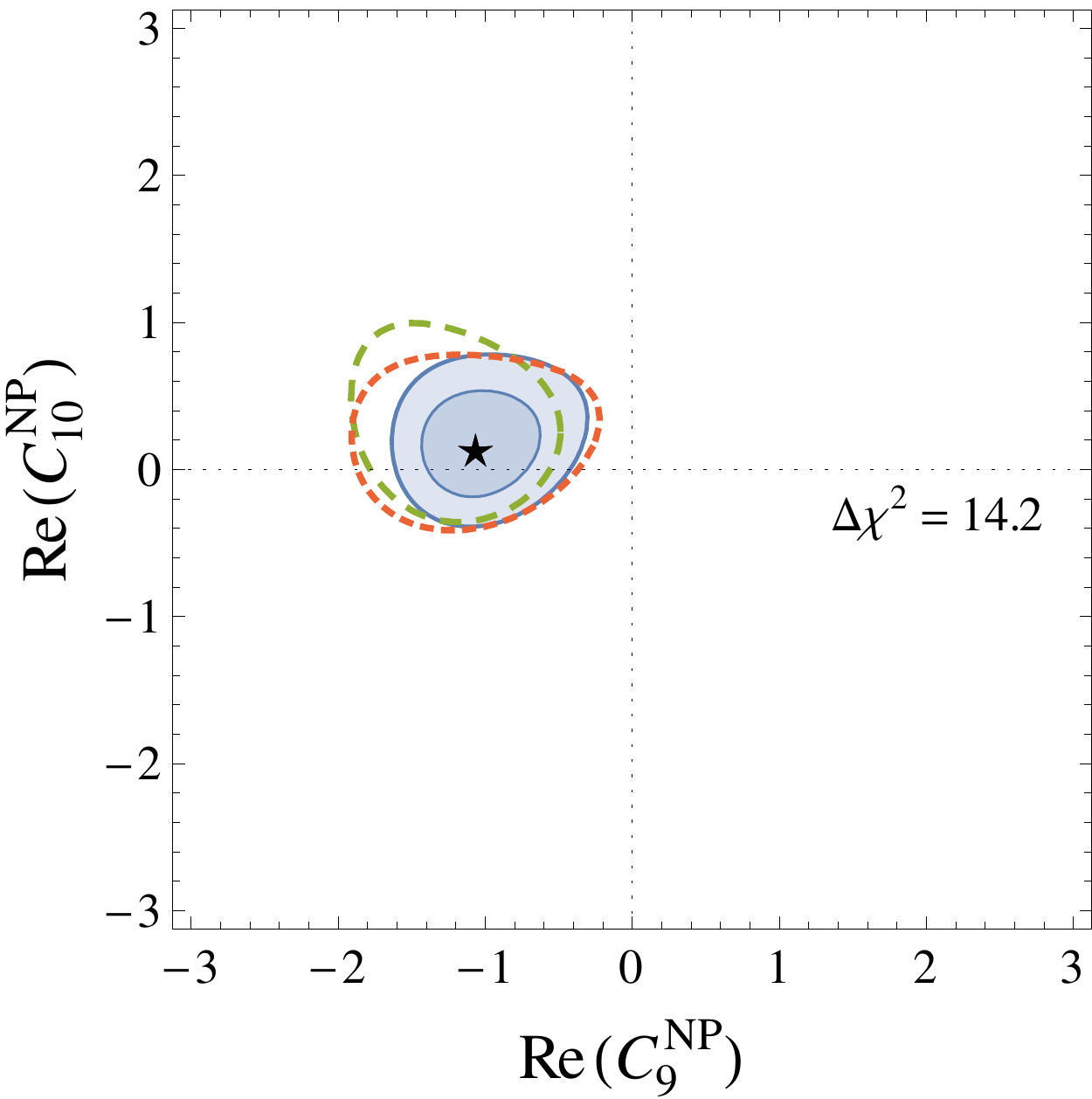} \\[8pt]%

~~\includegraphics[width=0.42\textwidth]{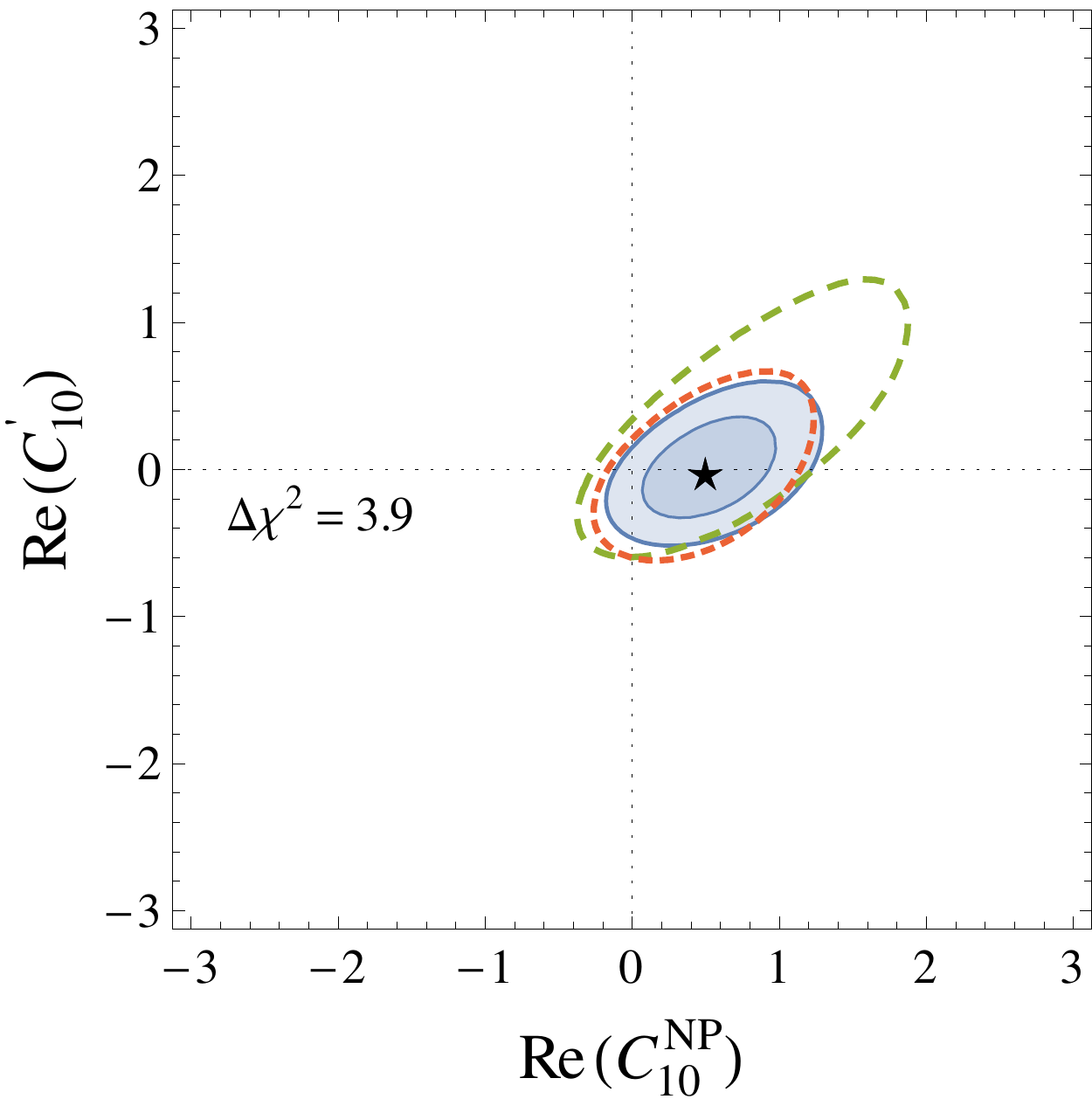} ~~~~~%
\includegraphics[width=0.42\textwidth]{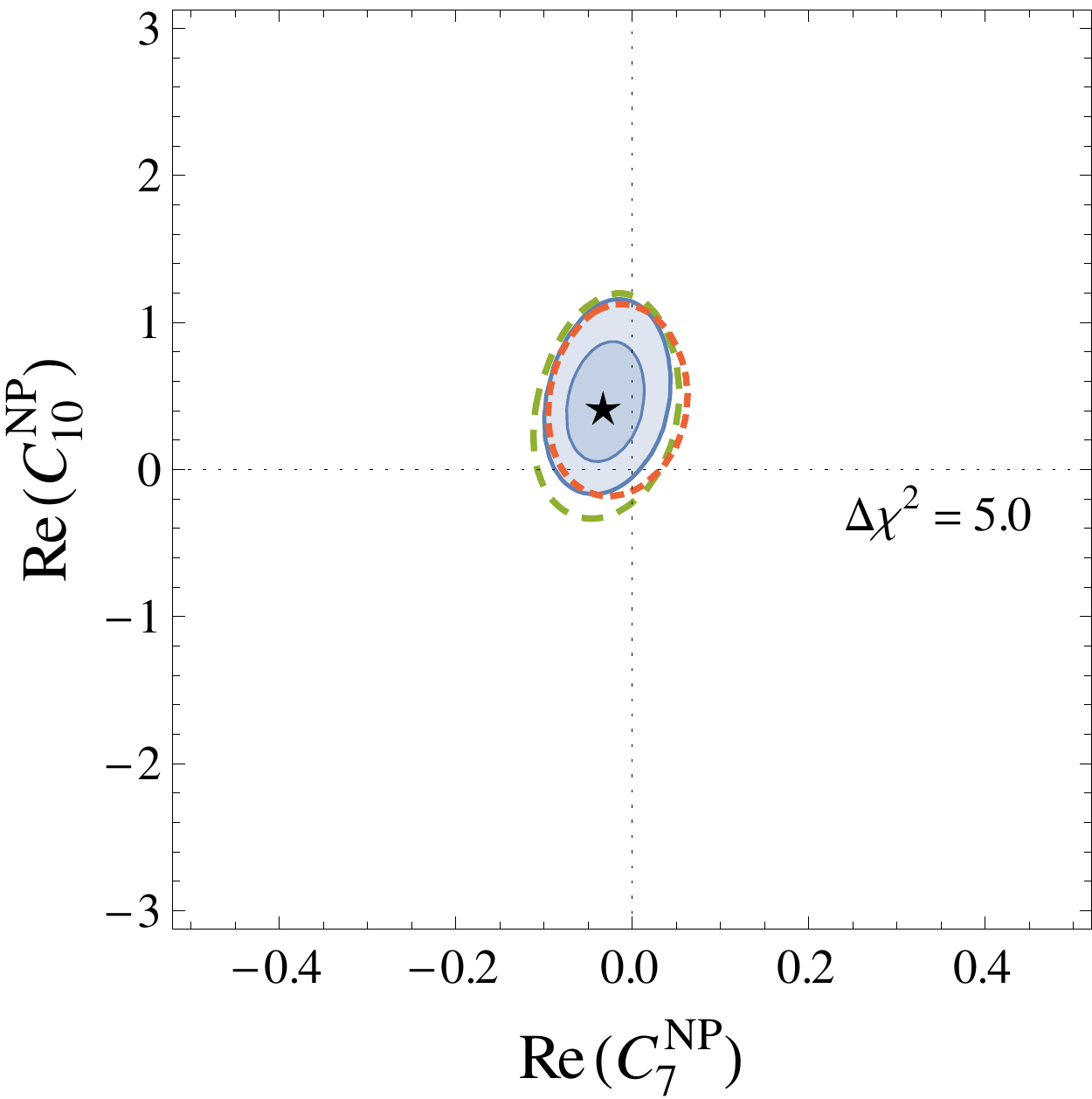}%
\caption{Constraints on pairs of real Wilson coefficients. Contours are as in fig.~\ref{fig:2D-zoom}}
\label{fig:2Dconstraints2}
\end{figure}

Figs.~\ref{fig:2Dconstraints1} and~\ref{fig:2Dconstraints2} shows the constraints in the planes of the complex Wilson coefficients or of various pairs of real Wilson coefficients. The blue contours correspond to the 1 and $2\sigma$ regions allowed by the global fit. The green short-dashed and the red short-dashed contours correspond to the $2\sigma$ allowed regions for scenarios with doubled form factor uncertainties and doubled uncertainties related to sub-leading non-factorizable corrections, respectively. The $\Delta\chi^2$ of the best fit point with respect to the SM is also given in the plots.

%#####################################################%
\bibliographystyle{utphys}
\bibliography{bsll}
%#####################################################%

\end{document}